\documentclass[twocolumn]{aastex63}
\usepackage{float,graphicx,amsmath,multirow,hyperref}
\usepackage{color}
\usepackage[version=4]{mhchem}
\usepackage{booktabs}
\usepackage{gensymb}

\begin{document}

\title{An Investigation of Spectral Line Stacking Techniques and Application to the Detection of \ce{HC11N} }
\author{Ryan A. Loomis}
\affiliation{National Radio Astronomy Observatory, Charlottesville, VA 22903, USA}
\author{Andrew M. Burkhardt}
\affiliation{Center for Astrophysics $\mid$ Harvard~\&~Smithsonian, Cambridge, MA 02138, USA}
\author{Christopher N. Shingledecker}
\affiliation{Department of Physics and Astronomy, Benedictine College, Atchison, KS 66002, USA}
\affiliation{Center for Astrochemical Studies, Max Planck Institute for Extraterrestrial Physics, Garching, Germany}
\affiliation{Institute for Theoretical Chemistry, University of Stuttgart, Stuttgart, Germany}
\author{Steven B. Charnley}
\affiliation{Astrochemistry Laboratory and the Goddard Center for Astrobiology, NASA Goddard Space Flight Center, Greenbelt, MD 20771, USA}
\author{Martin A. Cordiner}
\affiliation{Astrochemistry Laboratory and the Goddard Center for Astrobiology, NASA Goddard Space Flight Center, Greenbelt, MD 20771, USA}
\affiliation{Institute for Astrophysics and Computational Sciences, The Catholic University of America, Washington, DC 20064, USA}
\author{Eric Herbst}
\affiliation{Department of Chemistry, University of Virginia, Charlottesville, VA 22904, USA}
\affiliation{Department of Astronomy, University of Virginia, Charlottesville, VA 22904, USA}
\author{Sergei Kalenskii}
\affiliation{Astro Space Center, Lebedev Physical Institute, Russian Academy of Sciences, Moscow, Russia}
\author{Kin Long Kelvin Lee}
\affiliation{Department of Chemistry, Massachusetts Institute of Technology, Cambridge, MA 02139, USA}
\affiliation{Center for Astrophysics $\mid$ Harvard~\&~Smithsonian, Cambridge, MA 02138, USA}
\author{Eric R. Willis}
\affiliation{Department of Chemistry, University of Virginia, Charlottesville, VA 22904, USA}
\author{Ci Xue}
\affiliation{Department of Chemistry, University of Virginia, Charlottesville, VA 22904, USA}
\author{Anthony J. Remijan}
\affiliation{National Radio Astronomy Observatory, Charlottesville, VA 22903, USA}
\author{Michael C. McCarthy}
\affiliation{Center for Astrophysics $\mid$ Harvard~\&~Smithsonian, Cambridge, MA 02138, USA}
\author{Brett A. McGuire}
\affiliation{Department of Chemistry, Massachusetts Institute of Technology, Cambridge, MA 02139, USA}
\affiliation{National Radio Astronomy Observatory, Charlottesville, VA 22903, USA}
\affiliation{Center for Astrophysics $\mid$ Harvard~\&~Smithsonian, Cambridge, MA 02138, USA}

\correspondingauthor{Ryan A. Loomis, Brett A. McGuire}
\email{rloomis@nrao.edu, brettmc@mit.edu}

\begin{abstract}
 As the inventory of interstellar molecules continues to grow, the gulf between small species, whose individual rotational lines can be observed with radio telescopes, and large ones, such as polycyclic aromatic hydrocarbons (PAHs) best studied in bulk via infrared and optical observations, is slowly being bridged. Understanding the connection between these two molecular reservoirs is critical to understanding the interstellar carbon cycle, but will require pushing the boundaries of how far we can probe molecular complexity while still retaining observational specificity. Toward this end, we present a method for detecting and characterizing new molecular species in single-dish observations toward sources with sparse line spectra. We have applied this method to data from the ongoing GOTHAM (GBT Observations of TMC-1: Hunting Aromatic Molecules) Green Bank Telescope (GBT) large program, discovering six new interstellar species. In this paper we highlight the detection of HC$_{11}$N, the largest cyanopolyyne in the interstellar medium.
\end{abstract}

\section{Introduction}

As molecules increase in size, detection by rotational spectroscopy generally becomes more challenging. In large molecules, there are a substantially larger number of rotational energy levels over which population is distributed, reducing the emission between any two that give rise to an observable transition. Even at low temperatures the rotational partition function for such species can be high, with a large number of thermally populated rotational levels, diluting the intensity of any given transition. Additionally, larger species are generally less abundant than smaller species \citep{McGuire_2018_census}. Taken together, it is often far more difficult to detect individual rotational lines of a heavy species relative  to those of a light species even if both have identical dipole moments, rotational temperatures, and column densities. Even for small PAHs, for example, the total line intensity is diluted over potentially hundreds if not thousands of transitions, making it exceedingly difficult to detect any individual line in a reasonable amount of integration time.

\begin{figure*}[!htbp]
    \centering
    \includegraphics[width=1.\textwidth]{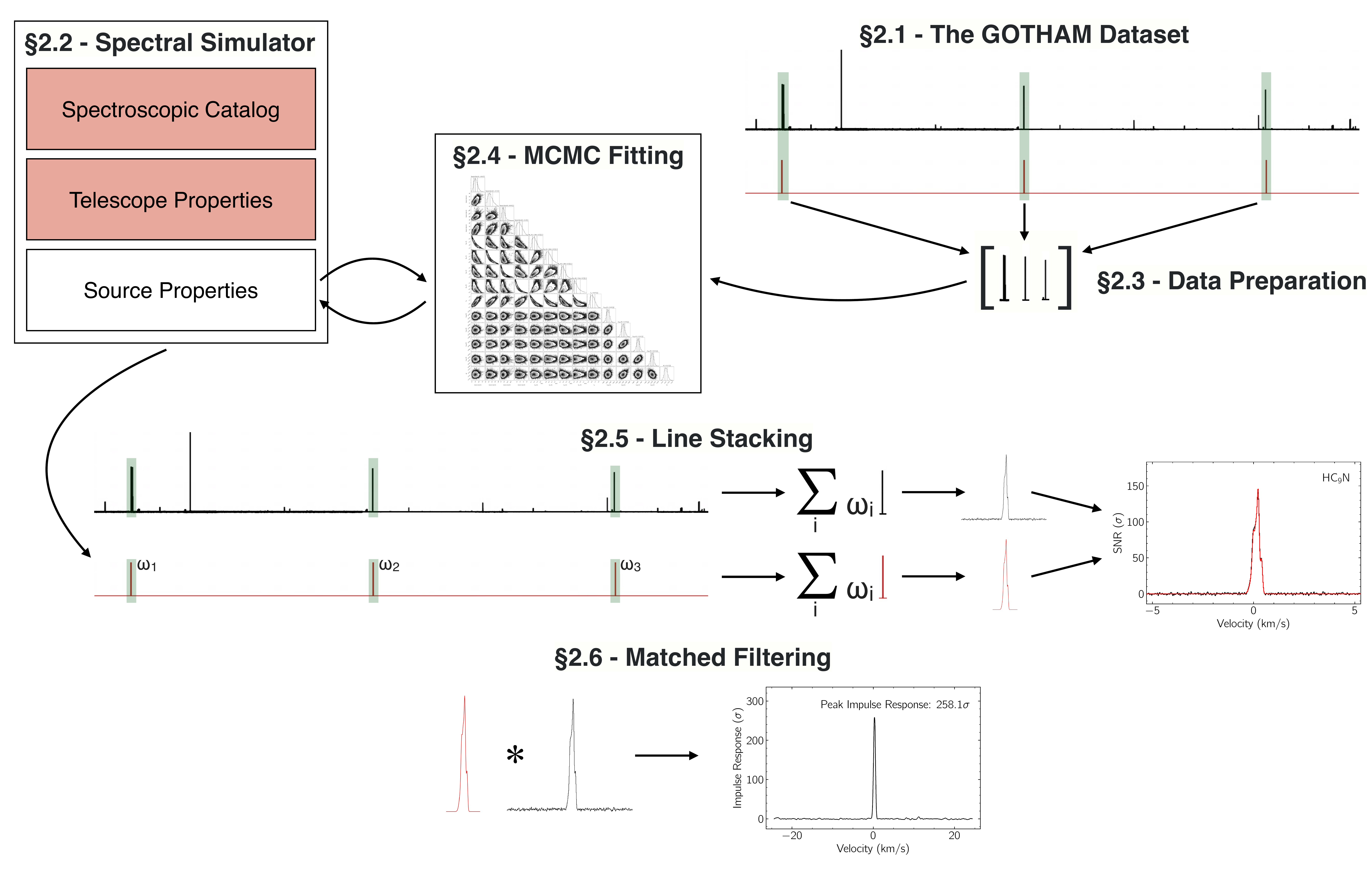}
    \caption{Schematic diagram of our method for molecular detection and characterization. In short, the GOTHAM dataset (\S\ref{Method_1}) and an initial spectral simulation (\S\ref{Method_2}) are used to select a relevant subsection of data (green shaded regions; \S\ref{Method_3}). A model is then fit to the data (\S\ref{Method_4}), with the source properties being varied while the telescope properties and spectral catalog are held fixed (shaded red). The best fit model is used to weight the data for stacking (\S\ref{Method_5}). To visualize the statistical significance of this detection, the stacked model is used as a matched filter and applied to the stacked data (\S\ref{Method_6}).}
    \label{Fig_1}
\end{figure*}

Here, we describe a new method that combines the techniques of Markov Chain Monte Carlo (MCMC) inference with spectral line stacking and matched filtering to counteract the effects of rotational dilution, improving detection efficiency and the characterization of weak emission from large molecules.

\subsection{Molecular detection technique}
\label{Method}
MCMC inference has grown in popularity in recent years in the astrochemical community as a tool for analyzing the properties of spectroscopic lines \citep{Czekala_2016, Gratier_2016, Loomis_2016}, allowing for straightforward characterization of parameter uncertainties and covariances. Similarly, line stacking and matched filtering techniques have regularly been applied to improve the signal-to-noise ratio (SNR) and detection efficiency of weak lines \citep{Walsh_2016, Loomis_2018, Loomis_2020}. Here, we present a hybrid combination of these techniques to robustly infer the presence of large astronomical molecules of interest in single dish spectra, as well as their emission parameters and associated uncertainties. In particular, this technique is ideal for identifying and characterizing species when no individual line is intense enough to be observed in a spectral line survey, but where many lines are present in the data itself, hidden under the noise. A flowchart providing an overview of our analysis method is shown in Fig.~\ref{Fig_1}, and we explain each step of the process in the following sub-sections.

\subsubsection{The GOTHAM dataset}
\label{Method_1}
Our method is best suited to a line-sparse single-dimensional spectral dataset, and here we investigate its application to data from the ongoing GBT large program GOTHAM. The details of these observations are presented in \citet{McGuire:2020bb}. In short, at the time of this analysis, the observations were $\sim$30\% complete, covering 13.1\,GHz of total bandwidth between 7.8 and 29.9\,GHz. With a frequency resolution of 1.4\,kHz (0.014-0.054\,km~s$^{-1}$), the dataset encompasses 9.3 million channels.

Despite the wide spectral range, the observations are relatively line-sparse. A total of 632 lines are detected above 5$\sigma$, yielding an effective average line density of 0.05 lines/MHz (1 line every 20\,MHz). In addition, the lines are quite narrow: $\sim$0.3\,km~s$^{-1}$ in aggregate, although we fit the contributions of several (2-4) $\sim$0.11\,km~s$^{-1}$ components to these features. The result is a spectrum which is sparse in `bright' channels: only 1 channel in every $\sim$1400 is $>$5$\sigma$ above the local noise level, which is equivalent to a filling factor of $<$0.1\%.  We discuss the importance of the line sparsity in more detail later.

\subsubsection{Spectral simulator}
\label{Method_2}
To infer the desired astrophysical properties (for example, excitation temperature and column density) of a given molecular species, we employ a forward modeling framework where spectra are iteratively simulated in a fashion similar to the observations themselves and then compared to the data. Our spectral simulator is based on the basic equations of molecular excitation and radiative transfer \citep{Liu_2001, Remijan_2005, Mangum_Shirley_2015}; the open source code can be found at \href{https://github.com/ryanaloomis/spectral_simulator}{https://github.com/ryanaloomis/spectral\_simulator}. The simulator has three main inputs: a spectroscopic catalog in SPCAT format from the \textsc{CALPGM} suite of programs, \citep{Pickett_1991, Drouin_2017} a collection of telescope properties, and a collection of source properties.

The most critical telescope property is the 100\,m dish size of the GBT, required for calculating an effective beam filling factor to account for beam dilution effects. Source properties for each source component are left as free parameters, and include effective source size (used for calculating filling factors and assumed to be a symmetric Gaussian), column density (\textit{N$_{\textrm{col}}$}), excitation temperature (\textit{T$_{\textrm{ex}}$}), source velocity (\textit{v$_{\textrm{LSRK}}$}), and linewidth (\textit{dv}).

In our modeling of TMC-1, we have found four distinct velocity components at similar velocities to those previously identified \citep{Dobashi_2018, Dobashi_2019} from which the majority of species emit from. Source size, column density, and source velocity were allowed to freely vary for each component, and the excitation temperature and linewidth are fit jointly across components. It is likely that excitation temperature and linewidth do vary slightly across the different cloud components, but our data is not sufficient to constrain these differences, which we discuss in more detail later. Several species in our analysis were best-fit by utilizing only a subset of three of these four cloud components, and their results are presented with a corresponding number of free parameters.

\begin{figure*}[!t]
    \centering
    \includegraphics[width=0.5\textwidth]{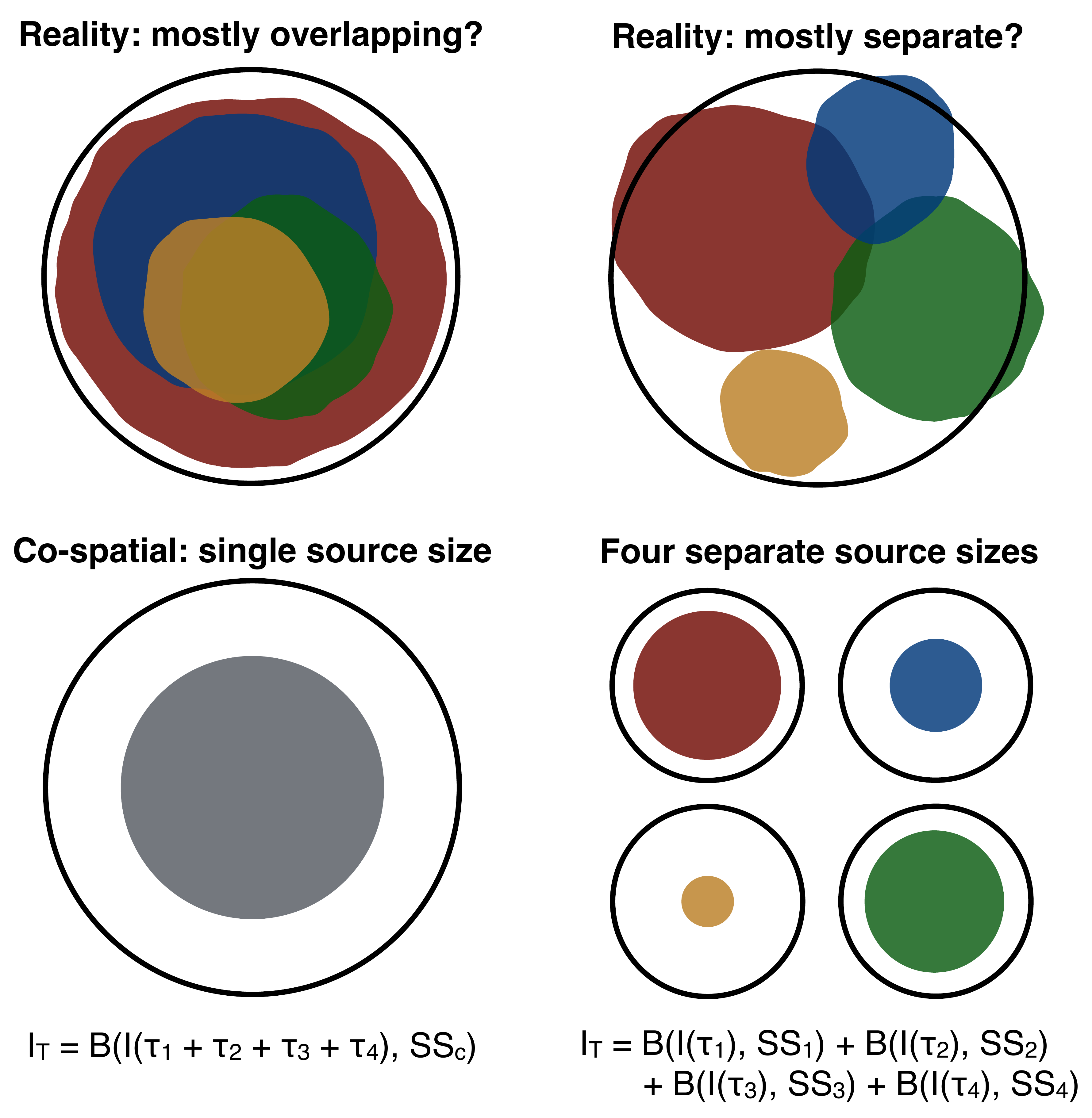}
    \caption{A schematic showing two spatial distribution regimes in which the emission from TMC-1 may fall into (upper panels) and the approximations we use in our analysis (lower panels). The FWHM primary beam of the GBT is denoted by the black circle. \textit{Upper left:} Emission may be mainly co-spatial, with significant overlap between velocity components. \textit{Upper right:} Emission may originate from spatially distinct velocity components, which are all still mostly within the primary beam. \textit{Lower left:} The `co-spatial' approximation, in which optical depths are added linearly before converting to intensity, and a common source size is fit. \textit{Lower right:} The `separate components' approximation, in which each component is separately beam diluted and then these intensities are added linearly.}
    \label{Fig_2}
\end{figure*}

The spatial orientation of the four cloud components on the sky (Fig. \ref{Fig_2}) has a pronounced impact on how their emission is measured by the telescope. First, since our dataset is only from a single pointing position and we do not have spatial information about these cloud components, we make the simplifying assumption that each component is centered in the beam. The Gaussian FWHM of the beam for the Green Bank Telescope is calculated for a given wavelength $\lambda$ and dish size $D$ as:

\begin{equation}
    \theta_{bm} [''] = \frac{206265\times1.22\lambda}{D}
\end{equation}

as documented in the GBT Proposer's Guide \citep{GBTObsGuide} and the corresponding beam dilution factor for a Gaussian source centered in the beam with FWHM $\theta_{source}$ is:

\begin{equation}
    \frac{\theta_{source}^{2}}{\theta_{bm}^{2} + \theta_{source}^{2}}.
\end{equation}

This beam dilution factor is applied at each frequency in the spectrum to all cloud components based on their given source sizes. In reality, the sources may be unequally distributed throughout the beam, leading to varying beam dilution effects at different frequencies, as the source begins to exit the beam. We discuss this point in more detail later.

In the optically thin limit, the spatial distribution of components does not strongly impact how their emission co-adds. Thus for species firmly within the optically thin limit, a beam diluted spectrum can be generated for each component and then summed. For species that may have lines which are more optically thick, however, the spatial distribution may have a more pronounced effect on how the emission co-adds.

If two optically thick lines lie at different velocities, and the linewidths are smaller than the separation between the central velocity of the components, then the components are radiative decoupled and can be added as in the optically thin case. This is the main assumption of the Large Velocity Gradient approximation. Additionally, if two optically thick components are spatially distinct, they will add linearly in measured intensity. If there are tow co-spatial optically thick lines that overlap in velocity, however, they need to be added in $\tau$ space before converting to intensity. We refer to these two limiting cases as `separate components' and `co-spatial'. As we lack the spatial information to disentangle the more complicated (and more likely) scenario of a situation between these two limiting cases, we instead present results from the two limits and discuss both when relevant. For the `co-spatial' case, it makes more sense to fit a common source-size across components (Fig. \ref{Fig_2}, bottom left), so the total number of model parameters is shrunk by three. As shown in the Supplementary Materials and discussed in more detail later, a co-spatial model does a much better job of describing the smaller and more optically thick cyanopolyynes.

\subsubsection{Initial data preparation}
\label{Method_3}
To begin the fitting process, it is first necessary to reduce the size of the dataset that will be simulated, as generating 9.3 million channels in every step of the MCMC process would not be computationally tractable. A dataset of much more manageable size would consist of only the small number of channels which are near lines of interest for a given species.

Using our spectral simulator and a nominal set of telescope and source properties, we generate an initial simulation for the target species across the full bandwidth of the GOTHAM observations. A dish size of 100~m, source size of 100\arcsec, excitation temperature of 8~K, column density of 10$^{12}$~cm$^{-2}$, and linewidth of 0.37~km~s$^{-1}$ are assumed. As this initial simulation is only used for selecting regions of the spectrum to perform the fit on, relative line strengths need only be approximate, and knowing the exact source size, excitation temperature, column density, or linewidth is not necessary. The linewidth and excitation temperature are estimated based on previous observations of TMC-1 \citep{Remijan_2006, Dobashi_2018, Dobashi_2019}, with the linewidth being large enough to encompass all of the known multiple velocity components.

Nominally the method will work when including all lines in a catalog file that fall within the range of the observations.  Indeed, for this work all lines were used with simple linear species, but for the analysis of species such as 2-cyanonaphthalene where there are thousands of extremely weak lines, applying a threshold significantly improves the computational efficiency. In these cases a  threshold of 5\% of the peak intensity in the initial simulation was used, discarding all lines below this threshold as they will not contribute significantly to the final fit or stacked detection. For each remaining line, a window was generated at 5.8$\pm$0.5~km~s$^{-1}$ and applied to the GOTHAM spectrum, yielding a final sparse spectrum with a much smaller datasize, as shown in Fig. \ref{Fig_1}. Within each window, a local estimate of the noise was taken by calculating the standard deviation of all points less than 3.5$\sigma$ (where $\sigma$ is an initial standard deviation taken considering all points). This method reduces the impact of any strong lines on the estimate of the local noise. For the analysis of weaker species, a 6$\sigma$ threshold was then applied to block any interloping lines from other species, preventing them from contaminating the model fit or final stack. Interloping lines were removed from the windowed dataset.

The final output of this procedure is a small, sparse spectrum for each species being considered, as well as a noise spectrum of identical dimensionality.

\subsubsection{MCMC fitting}
\label{Method_4}
With a reasonably sized dataset now available for a given species, we then utilize an MCMC fitting method to derive posterior probability distributions and covariances for each free parameter in the spectral simulator model. This process is very similar to that described in \citet{Loomis_2016}.

The degrees of freedom for each model are set by the considerations described earlier, with a maximum of 14 free parameters. The affine-invariant MCMC implementation \texttt{emcee} \citep{Foreman-Mackey_2013} was used with 100 walkers run for up to 10,000 steps. Convergence was assessed using a Gelman-Rubin convergence diagnostic \citep{Gelman_1992}.

Parameter initialization and priors were determined using two well-characterized `template' species. While initially investigating the properties of species within the GOTHAM data, we found that as one might expect from chemical intuition, linear species appeared to share source properties with HC$_9$N, while cyclic species appeared to share source properties more similar to benzonitrile. These two species, which both have easily identified bright individual lines, were therefore fit first with very simple priors - source velocities were forced to be in a sequential order and all other values had physical bounds set on them (e.g. positivity constraints). An example corner plot of the HC$_9$N fit is shown in Fig. \ref{Fig_3}.

\begin{figure*}[!t]
    \centering
    \includegraphics[width=0.9\textwidth]{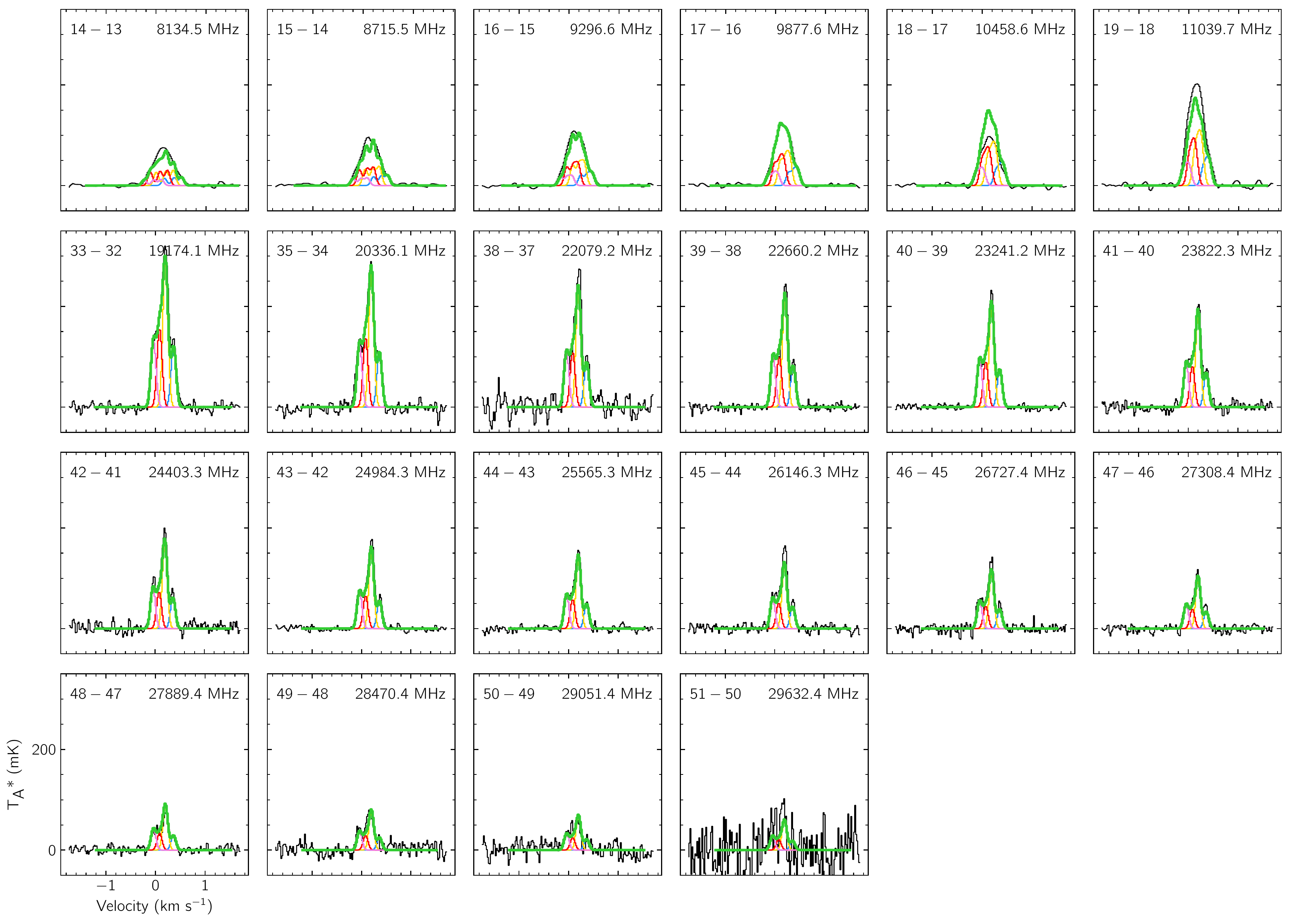}
    \caption{Individual line detections of \ce{HC9N} in the GOTHAM data.  The spectra (black) are displayed in velocity space relative to 5.8\,km\,s$^{-1}$, and using the rest frequency given in the top right of each panel. Quantum numbers are given in the top left of each panel, neglecting hyperfine splitting. The best-fit model to the data, including all velocity components, is overlaid in green.  Simulated spectra of the individual velocity components are shown in: blue (5.63\,km\,s$^{-1}$), gold (5.79\,km\,s$^{-1}$), red (5.91\,km\,s$^{-1}$), and violet (6.03\,km\,s$^{-1}$).  See Table~\ref{Table_1}.}
    \label{Fig_4}
\end{figure*}

The quality of these fits was then assessed visually, assuring the suitability of the model for the data. As seen in Fig. \ref{Fig_4}, these nominal fit parameters reproduced all observed lines within uncertainties. The HC$_9$N and benzonitrile posteriors were then used as priors for their respective template families for all values other than column density, and 50th percentile values were used to initialize walkers in a tight ball. Column densities were initialized via quick maximum likelihood fits, holding the other initialized values fixed.

From these fits for each species, we report parameters and their uncertainties using 16th, 50th, and 84th percentile intervals (e.g. Table \ref{Table_1} for HC$_9$N). These intervals are also denoted in the corner plots (e.g. Fig. \ref{Fig_3}). 50th percentile values are used for all stacking analyses.

\subsubsection{Line stacking}
\label{Method_5}
The posterior probability distributions from the MCMC fitting describe the range of parameter values consistent with the data, but are predicated on the assumption that our model does a good job of describing the underlying data. This is easily justified when individual lines can be detected and compared to the model predictions (Fig \ref{Fig_4}), but is less easy to visualize when individual lines are not seen above the noise level. Calculating a detection significance is therefore crucial to interpreting the MCMC constraints. To provide a visually intuitive interpretation of detection significance, we break this process down into two steps. First, we stack all of the windowed lines which have no interlopers, and second, we apply the stacked best fit simulation as a matched filter to the data stack.

The application of line stacking techniques to increase SNR in spectroscopic data is a well-known technique, particularly in an astrochemical context for the detection of new species \citep{Langston_2007, Walsh_2016, Loomis_2016}. Here we follow the normal prescription of SNR weighted stacking of each line (see Fig. \ref{Fig_1}), but with a minor modification for some species. When a species has a more complex spectrum where transitions are not always well separated (e.g. closely spaced hyperfine components), a naive stack of every transition will over-count the contributions from other nearby transitions, and may also contaminate the signal-free noise regions of the stack with signal from these nearby lines. To avoid these issues, we treat groups of transitions which are blended or closely spaced (typically $<$3 FWHM) as a single spectral line feature. This has the effect of slightly blurring the contribution to the total line stack, but avoids any over-counting. As the stacking procedure is performed identically for both the data spectrum and the predicted spectrum, the full signal is recovered during the matched filtering stage.

An example of this line stacking for the HC$_9$N lines in Fig. \ref{Fig_4} is shown in Fig. \ref{Fig_6}. Even though each of the individual lines were strongly detected, the overall significance of the detection is greatly enhanced, now with a peak value of $\sim$140$\sigma$. A similar stack of our best-fit model is overlaid in red, illustrating the quality of the fit. Demonstrations of the robustness of this line stacking method for our dataset are shown in the Supplementary Materials. We additionally discuss its limitations later, particularly with respect to source line density.

\subsubsection{Matched filtering}
\label{Method_6}
\begin{figure*}
    \centering
    \includegraphics[width=0.45\textwidth]{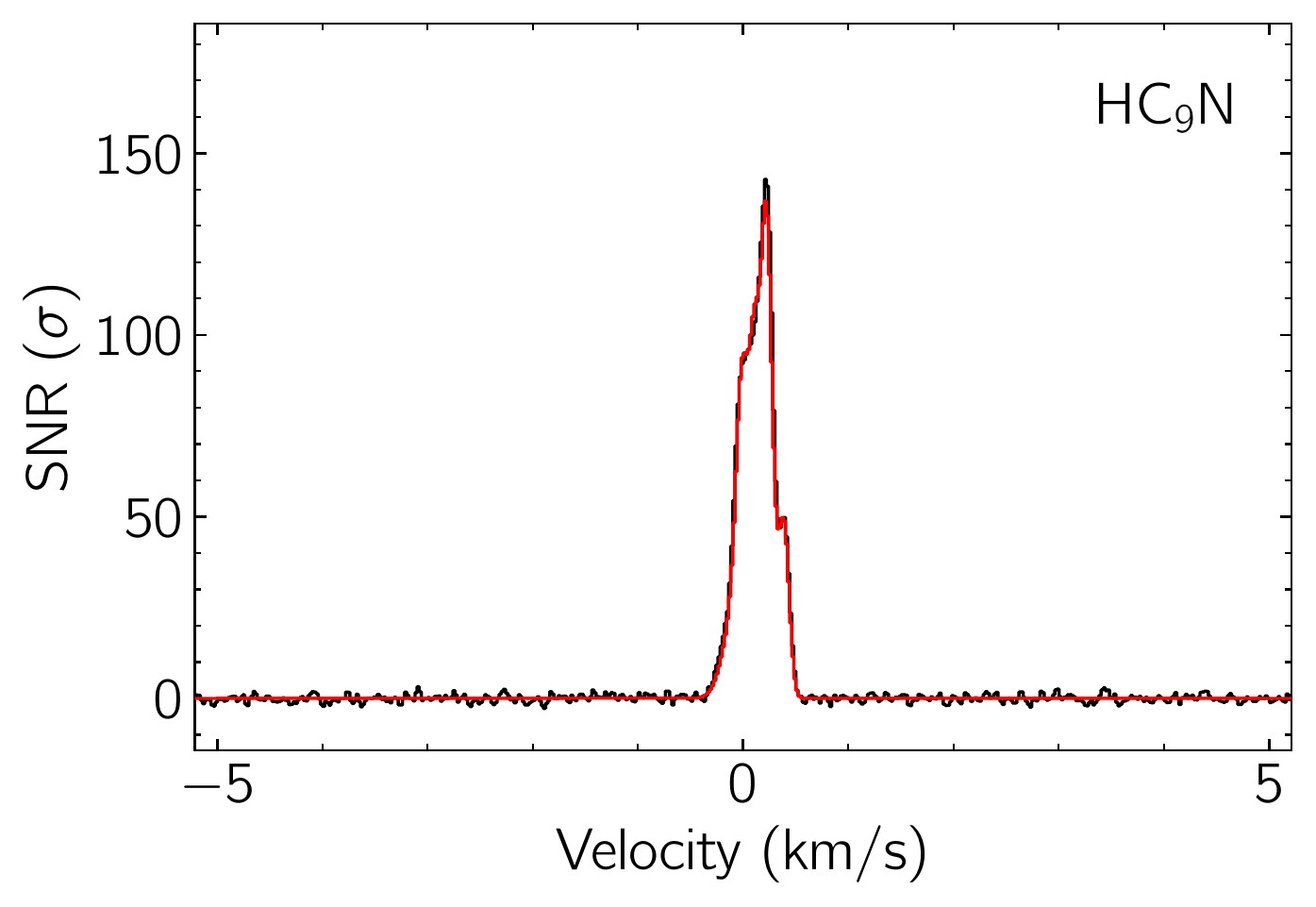}
    \includegraphics[width=0.45\textwidth]{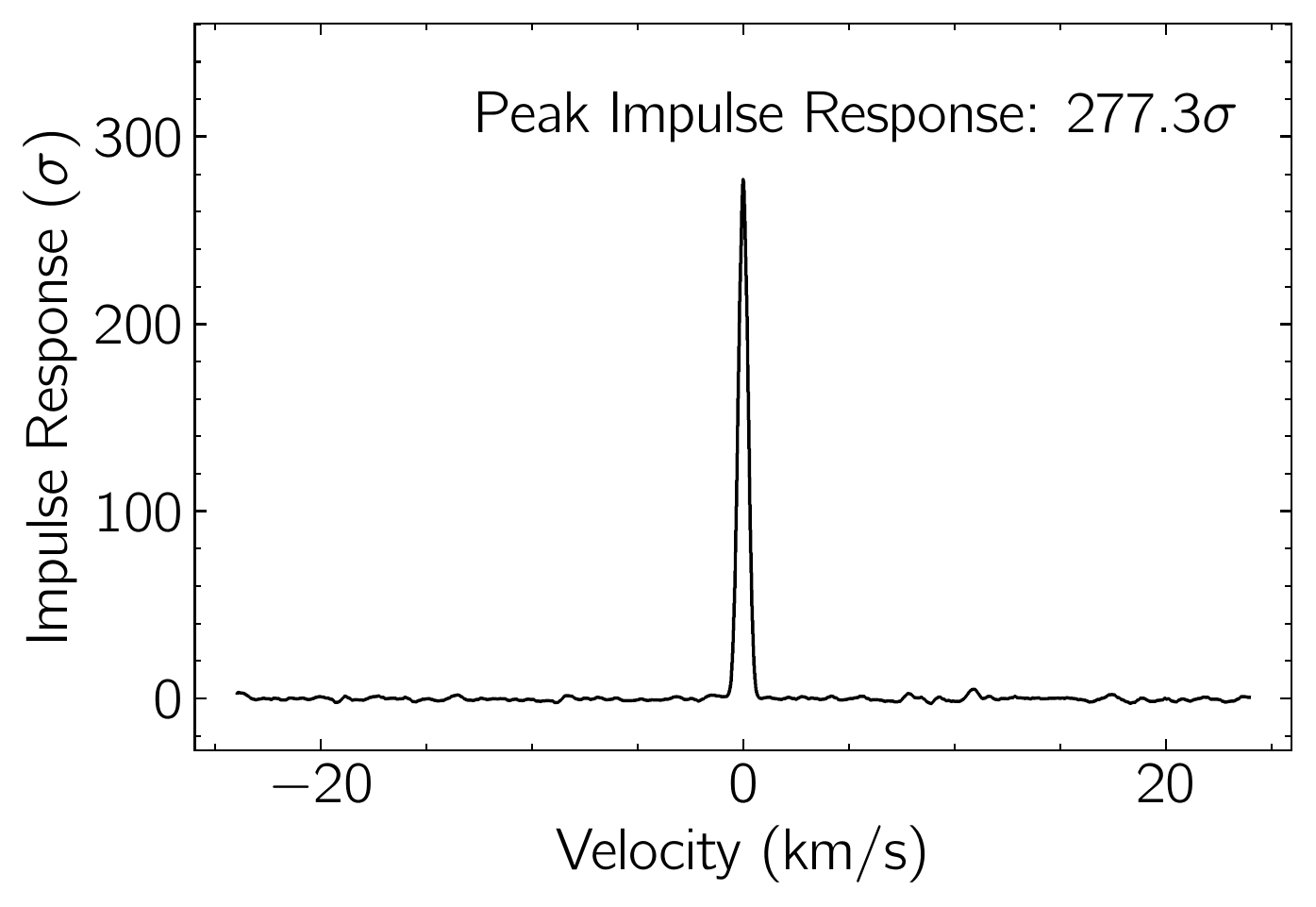}
    \caption{\textit{Left:} Velocity-stacked spectra of \ce{HC9N} in black, with the corresponding stack of the simulation using the best-fit parameters to the individual lines in red. The data have been uniformly sampled to a resolution of 0.02\,km\,s$^{-1}$. The intensity scale is the signal-to-noise ratio of the spectrum at any given velocity. \textit{Right:} Impulse response function of the stacked spectrum using the simulated line profile as a matched filter.  The intensity scale is the signal-to-noise ratio of the response function when centered at a given velocity.  The peak of the impulse response function provides a minimum significance for the detection of 277.3$\sigma$.}
    \label{Fig_6}
\end{figure*}

As described in \citet{Loomis_2018}, the technique of matched filtering first presented by \citet{Woodward_1953} and \citet{North_1963} can be used on astronomical spectroscopic data to optimally extract a detection significance when the shape of the signal is known. In our case, the stacked line signal still retains velocity structure, as seen in Fig. \ref{Fig_6}, and is thus not yet the maximum SNR attainable.

As shown in Fig. \ref{Fig_1}, we select a narrow region around the stacked predicted spectrum to use as the template filter, and then cross-correlate this filter with the stacked data spectrum, yielding an impulse response spectrum. The spectrum is then normalized by calculating the standard deviation of the spectrum (excluding the central region where we expect to see signal) and dividing by this standard deviation \citep{Loomis_2018}. The units of the impulse response are now $\sigma$, rather than a flux unit, and describe the SNR of the response. The peak response can therefore be thought of as a minimum detection significance for the species.  An example of this impulse response spectrum for HC$_9$N is shown in Fig. \ref{Fig_6}, where the peak detection significance is now almost doubled, at 258.1$\sigma$.

With a better model and hence a better matching filter, the significance of the detection could be improved, but it cannot be lower than the current peak response. We discuss this point in more detail later, along with an exploration of the effects of spectroscopic catalog accuracy on the recovered detection significance.

\subsubsection{Upper limits}
\label{Method_7}
In cases where our matched filtering analysis yields an impulse response with a significance not large enough to claim a detection (e.g. $<$4$\sigma$), we refit the data using a modified MCMC process to yield more useful posteriors on the column densities. Instead of letting all of the parameters described above run free, we instead fix the source sizes, velocities, and excitation temperature to the values reported for a similar molecule, as was done for the priors described earlier (for example, HC$_9$N for linear species and benzonitrile for cyclic species). From the resultant posterior distributions, 95th percentile confidence interval values are reported as 2$\sigma$ upper limit column densities. An example upper limit posterior is shown in Fig \ref{Fig_7} for HC$_{13}$N, which we do not currently detect above a 4$\sigma$ significance in the GOTHAM data.

\subsubsection{Broader applicability and limitations of method}
\label{Discussion_2}
Both the MCMC fitting and stacking analysis presented here are predicated on the assumption that signal (i.e. coherent information content) within the windowed data being fit or stacked is dominated by species of interest, rather than some red noise or contribution from competing species. In the context of well-calibrated single dish spectra, this can be more simply stated as a requirement of line sparsity. Analysis of interferometric data with this technique is possible, but beyond the scope of this paper. The degree of line sparsity necessary for a given analysis will be different for each species of interest. As discussed earlier, thresholding data is able to prevent the most egregious interloping lines from contaminating an analysis, but low level line confusion would prevent successful stacking of the thousands of lines necessary to detect a species such as 2-cyanonaphthalene. In contrast, the several dozen lines of HC$_{11}$N would be more tolerant to a low level of line confusion (as each individual line of interest would be brighter in comparison to the confusing lines).

Of the handful of astronomical sources that have yielded the vast majority of new interstellar molecular detections, TMC-1 has by far the most sparse spectra. Application of our technique to other sources, such as Sgr B2(N), IRAS 16293-2422, or Orion KL, is likely not as straightforward due to their higher line density. A more fruitful approach may be to take inspiration from other solutions to the analogous problems of detrending and source separation, where advancements in Bayesian methods such as probabilistic cataloging \citep{Portillo_2017} hold promise for the bulk analysis of large datasets \citep{Astrostats_2020}.

\begin{figure}[t]
    \centering
    \includegraphics[width=0.5\textwidth]{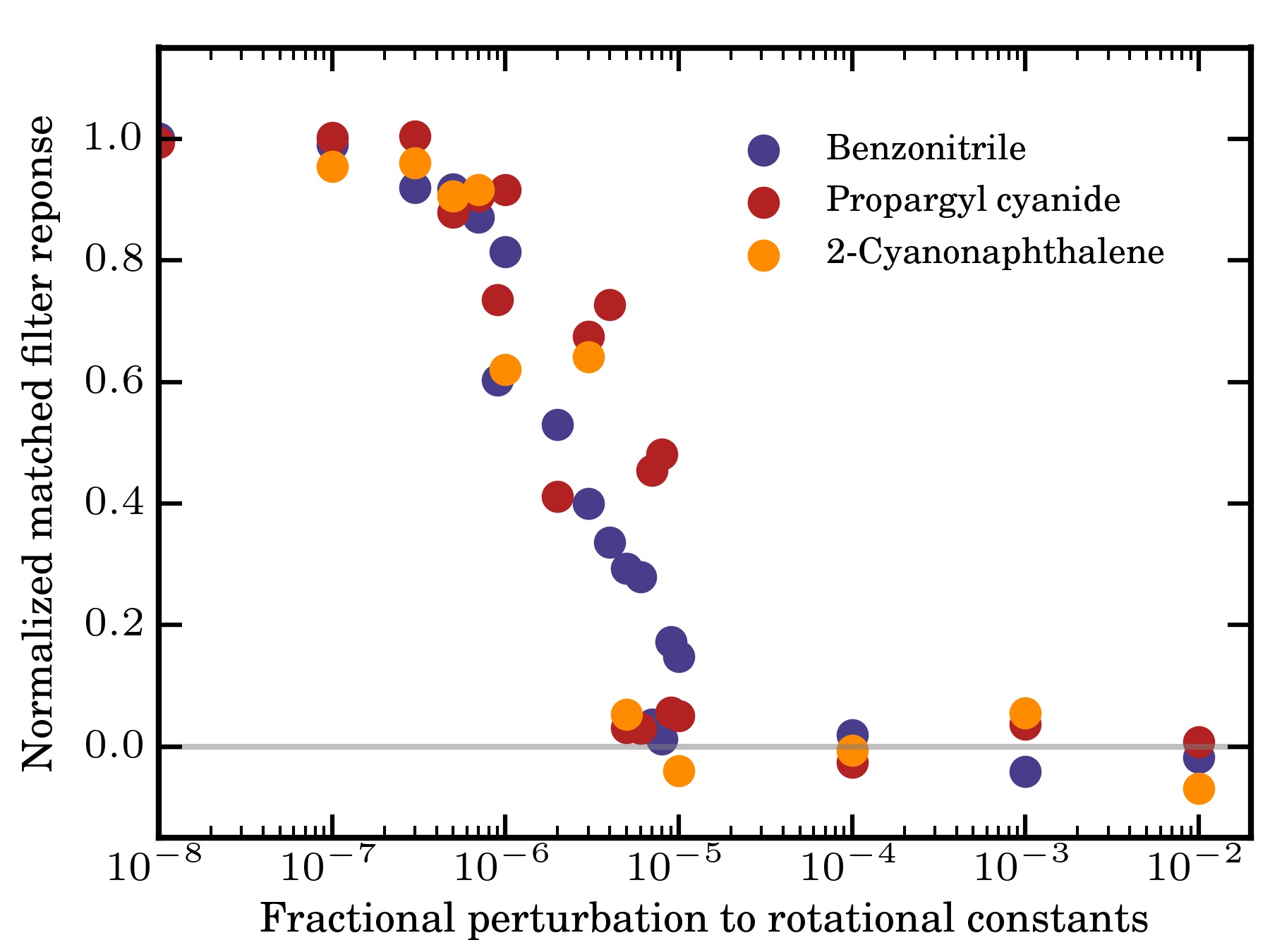}
    \caption{Fractional modification to rotational constants plotted versus normalized matched filter response for three species -- benzonitrile (blue), propargyl cyanide (red), and 2-cyanonaphthalene (orange).}
    \label{Fig_22}
\end{figure}

Finally, thus far we have made the assumption that the spectroscopic catalogs used in our spectral simulator are a fixed input, with no error. In reality, few large species of astronomical interest have precise laboratory constraints on their spectra, and several of our newly detected species in GOTHAM required significant refinement via new laboratory spectroscopic investigation \citep{McCarthy:2020aa, McGuire:2020bb, McGuire:2020aa}. To better understand the sensitivity of our stacking method to spectroscopic errors, we systematically introduced increasing amounts of Gaussian random noise to the rotational constants used to generate the catalogs for benzonitrile, propargyl cyanide \citep{McGuire:2020bb}, and 2-cyanonaphthalene. A plot of the fractional level of modification to the rotational constants versus the fractional peak filter response (normalized to the peak filter response for the nominal catalog) is shown in Figure \ref{Fig_22}. We find that for all three species, a modification of ten parts per million (ppm) is sufficient to effectively nullify the molecular detection. A relative precision of $\sim$100 parts per billion (ppb) is sufficient to recover most of the signal. This is roughly equivalent to the accuracy of a state-of-the-art high-resolution microwave spectrometer \citep{Crabtree:2016fj}, highlighting the necessity of modern laboratory constraints for the identification of large molecules in the interstellar medium.

This analysis also doubles as evidence that our stacking method is not likely to yield false positives given the line sparsity of TMC-1 -- a small change of a few ppm to rotational constants is sufficient to reduce the signal in stacked spectrum to nothing, making it unlikely that our stacking analysis would recover spurious signal. This point is discussed further in the Supplementary Materials, where we demonstrate the robustness of the method via jack-knifing the data.

\subsection{Detection of HC$_{11}$N}
\label{Observations}
\label{Observations_3}
\ce{HC11N} has a long and colorful history in radio astronomy.  Three radio lines were first reported toward IRC+10216 on the basis of a rotational constant derived by extrapolation from those measured experimentally for shorter members in this homologous series \citep{bell:389,oka:1982}. Any lingering doubt of the astronomical identification appeared to be put to rest with the observation of a fourth transition toward TMC-1 in 1985 \citep{bell:l63}. The subsequent laboratory detection of \ce{HC11N} \citep{travers:l65}, however, established that its rotational lines actually lie 0.13\% lower in frequency (a shift equivalent to 13 linewidths in IRC+10216 and nearly 800 linewidths in TMC-1) relative to those originally reported \citep{bell:389,bell:l63}.  The observed lines thus could not arise from \ce{HC11N}.  Subsequently, two new astronomical lines were detected in TMC-1 with the NRAO 43\,m radio telescope \citep{Bell_1997}, both in apparent agreement with the laboratory rest frequencies. Albeit based on slender astronomical data, the detection of \ce{HC11N} in space now appeared secure, or so it seemed.  In 2016, an attempt was made to verify the detection of \ce{HC11N} by analyzing archival observations towards TMC-1 with the 100\,m Green Bank Telescope (GBT; \citealt{Loomis_2016}). Even with substantially deeper integrations, no evidence was found for six consecutive transitions between 12.9 and 14.6\,GHz.  The non-detection of \ce{HC11N} toward TMC-1 was further supported by observations that were unable to detect two higher frequency transitions in a sensitive observation in K-band with the GBT \citep{Cordiner_2017}.

The apparent absence of \ce{HC11N} in TMC-1 and corresponding column density upper limit combined with a non-linear relationship of column density with chain length for shorter cyanopolyynes (HC$_x$N; \citealt{Bujarrabal_1981, Bell_1997, Ohishi_1998}) led \citet{Loomis_2016} to hypothesize that cyclization reactions may become important once a carbon chain reaches a critical size.  If correct, formation of ring isomers could then directly compete with linear isomers via `bottom-up' pathways. The detections of benzonitrile ($c$-\ce{C6H5CN}), the simplest aromatic nitrile \citep{McGuire_2018}, and now individual PAHs \citep{McGuire:2020aa}, in TMC-1 suggest that cyclic chemistry is far more widespread at these earliest stages of star formation than previously thought.

With confidence from the aforementioned tests that our method is able to rigorously detect not only species that show individual lines, but also those which sit below the visible noise, we turn back to the prior mysterious non-detection of HC$_{11}$N \citep{Loomis_2016}. A similar stacking and MCMC analysis was undertaken,\citep{Loomis_2016} but with significantly less data than was presently available in the GOTHAM observations.

\begin{figure*}[!t]
    \centering
    \includegraphics[width=0.9\textwidth]{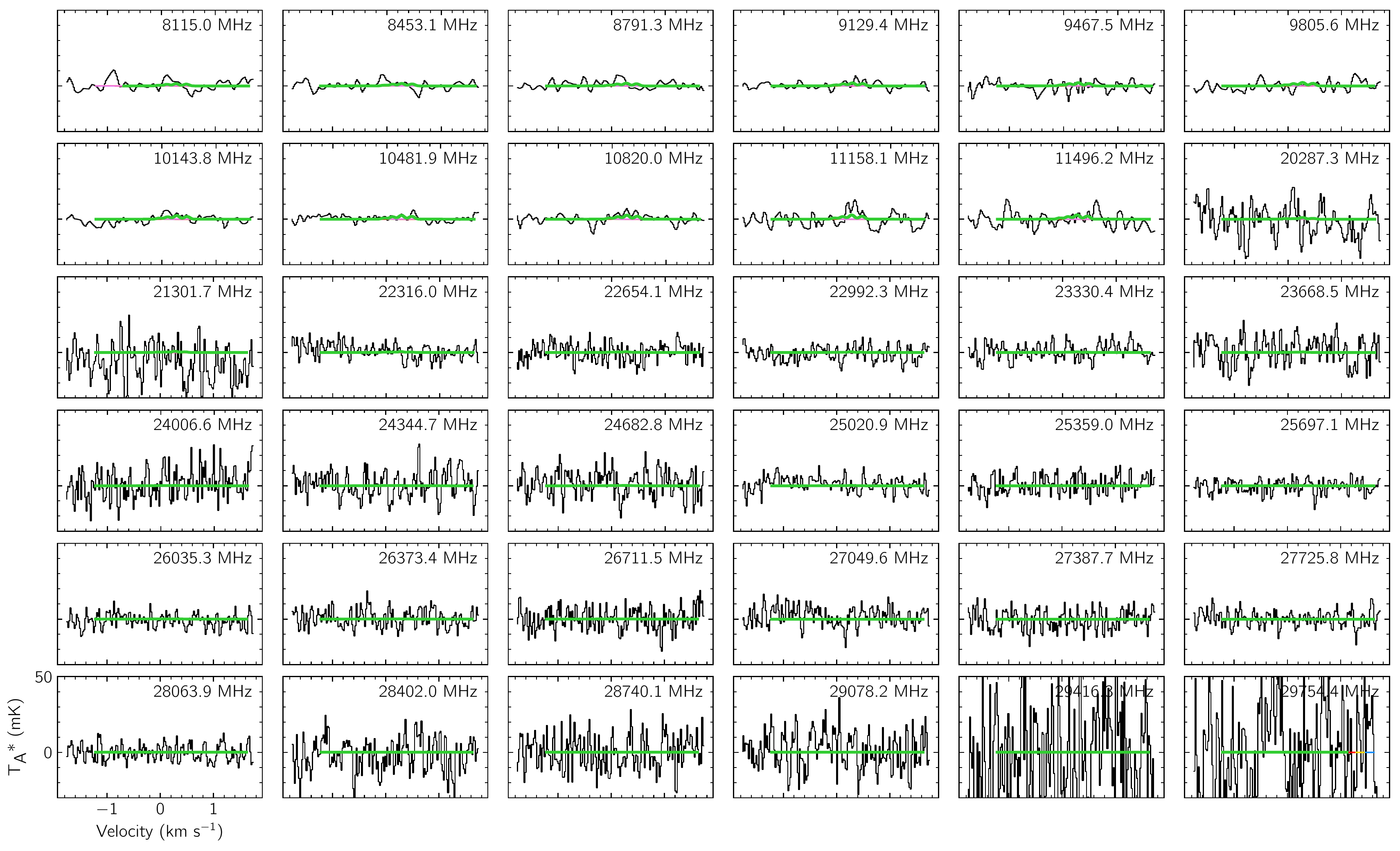}
    \caption{Individual line observations of HC$_{11}$N in the GOTHAM data.  The spectra (black) are displayed in velocity space relative to 5.8\,km\,s$^{-1}$, and using the rest frequency given in the top right of each panel. Quantum numbers are given in the top left of each panel, neglecting hyperfine splitting. The best-fit model to the data, including all velocity components, is overlaid in green.  Simulated spectra of the individual velocity components are shown in: blue (5.63\,km\,s$^{-1}$), gold (5.79\,km\,s$^{-1}$), red (5.91\,km\,s$^{-1}$), and violet (6.03\,km\,s$^{-1}$).}
    \label{Fig_16}
\end{figure*}

\begin{figure}
    \centering
    \includegraphics[width=0.45\textwidth]{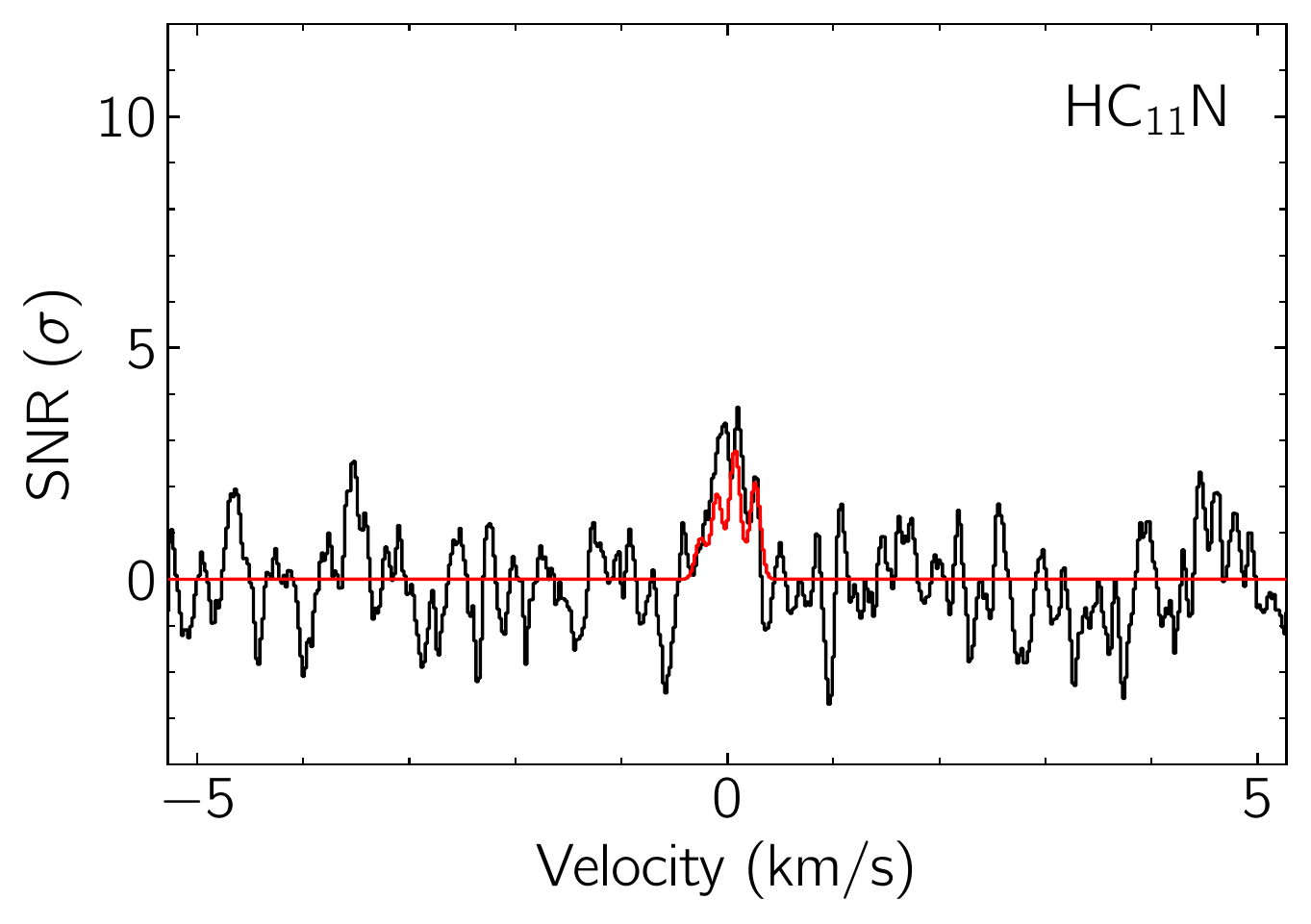}
    \includegraphics[width=0.45\textwidth]{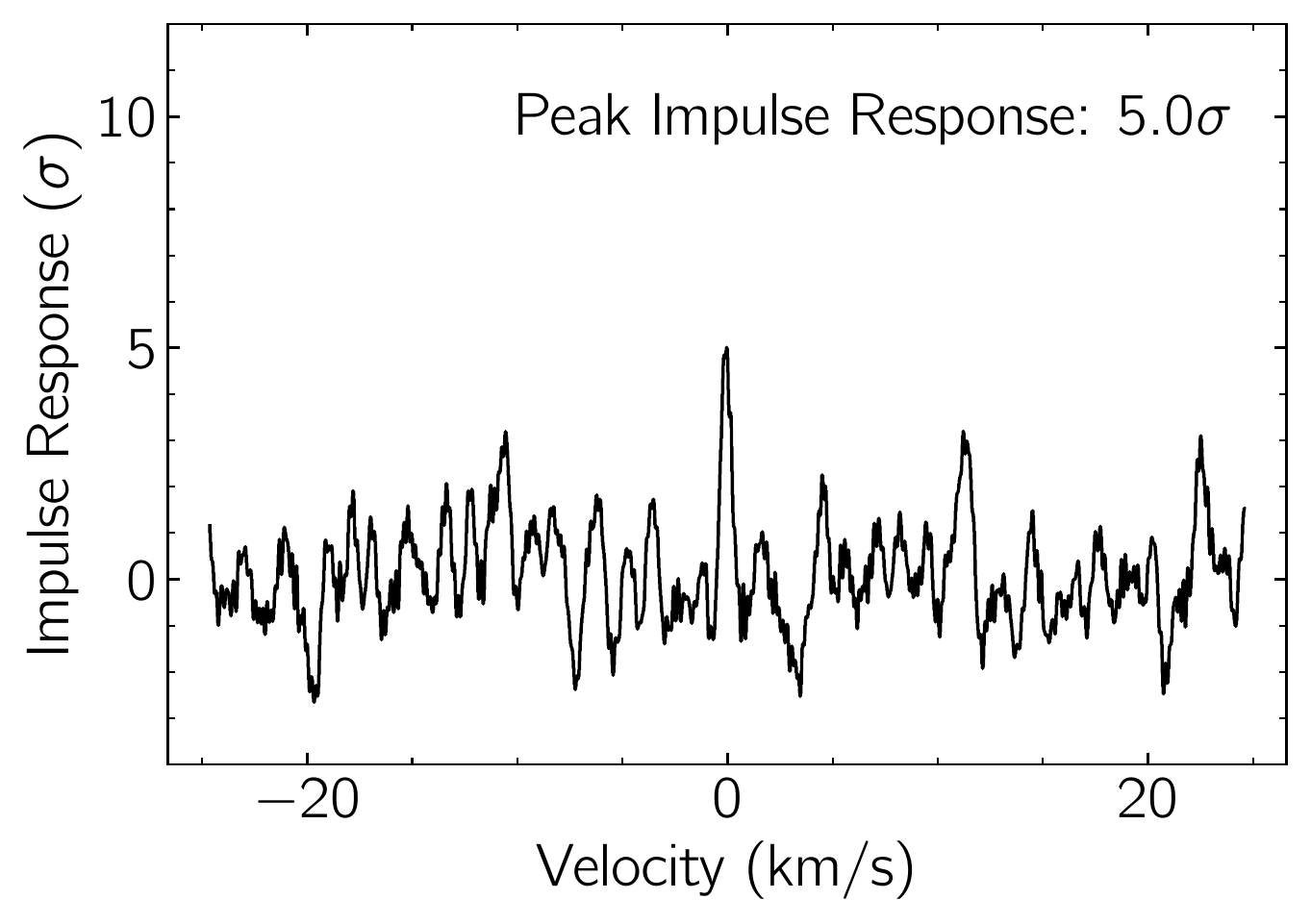}
    \caption{\textit{Left:} Velocity-stacked spectra of HC$_{11}$N in black, with the corresponding stack of the simulation using the best-fit parameters to the individual lines in red. The data have been uniformly sampled to a resolution of 0.02\,km\,s$^{-1}$. The intensity scale is the signal-to-noise ratio of the spectrum at any given velocity. \textit{Right:} Impulse response function of the stacked HC$_{11}$N spectrum using the simulated line profile as a matched filter.  The intensity scale is the signal-to-noise ratio of the response function when centered at a given velocity.  The peak of the impulse response function provides a minimum significance for the detection of 5.0$\sigma$.}
    \label{Fig_19}
\end{figure}

Unsurprisingly, we find that none of the brightest HC$_{11}$N lines are individually detected in our observations (Fig. \ref{Fig_16}). Fitting for HC$_{11}$N using priors from our HC$_{9}$N fit, however, we find column density posteriors that are consistent with a detection of HC$_{11}$N (Fig. \ref{Fig_17} and Table \ref{Table_3}). We visualize the significance of these posteriors through the same line stacking and matched filter analysis. The line stack shown in Fig. \ref{Fig_19} displays a tentative but encouraging 3.8$\sigma$ signal, and with a matched filter applied, the signal increases to a 5.0$\sigma$ detection (Fig. \ref{Fig_19}).

\begin{table*}
\centering
\caption{\ce{HC11N} best-fit parameters from MCMC analysis}
\begin{tabular}{c c c c c c}
\toprule
\multirow{2}{*}{Component}&	$v_{lsr}$					&	Size					&	\multicolumn{1}{c}{$N_T^\dagger$}					&	$T_{ex}$							&	$\Delta V$		\\
			&	(km s$^{-1}$)				&	($^{\prime\prime}$)		&	\multicolumn{1}{c}{(10$^{11}$ cm$^{-2}$)}		&	(K)								&	(km s$^{-1}$)	\\
\midrule
\hspace{0.1em}\vspace{-0.5em}\\
C1	&	$5.532^{+0.113}_{-0.022}$	&	$39^{+9}_{-8}$	&	$0.73^{+0.54}_{-0.32}$	&	\multirow{6}{*}{$6.6^{+0.3}_{-0.3}$}	&	\multirow{6}{*}{$0.117^{+0.012}_{-0.011}$}\\
\hspace{0.1em}\vspace{-0.5em}\\
C2	&	$5.722^{+0.043}_{-0.017}$	&	$21^{+7}_{-6}$	&	$2.60^{+3.73}_{-1.31}$	&		&	\\
\hspace{0.1em}\vspace{-0.5em}\\
C3	&	$5.887^{+0.027}_{-0.023}$	&	$56^{+18}_{-19}$	&	$0.36^{+0.32}_{-0.17}$	&		&	\\
\hspace{0.1em}\vspace{-0.5em}\\
C4	&	$6.034^{+0.052}_{-0.041}$	&	$9^{+10}_{-5}$	&	$4.12^{+16.68}_{-3.28}$	&		&	\\
\hspace{0.1em}\vspace{-0.5em}\\
\midrule
$N_T$ (Total)$^{\dagger\dagger}$	&	 \multicolumn{5}{c}{$7.8^{+21.27}_{-5.08}\times 10^{11}$~cm$^{-2}$}\\
\bottomrule
\end{tabular}

\begin{minipage}{0.75\textwidth}
	\footnotesize
	{Note} -- The quoted uncertainties represent the 16$^{th}$ and 84$^{th}$ percentile ($1\sigma$ for a Gaussian distribution) uncertainties.\\
	$^\dagger$Column density values are highly covariant with the derived source sizes.  The marginalized uncertainties on the column densities are therefore dominated by the largely unconstrained nature of the source sizes, and not by the signal-to-noise of the observations.
	$^{\dagger\dagger}$Uncertainties derived by adding the uncertainties of the individual components in quadrature.
\end{minipage}

\label{Table_3}

\end{table*}

The column density constraints from this analysis of HC$_{11}$N yield a total column density of $7.8^{+21.27}_{-5.08}\times 10^{11}$~cm$^{-2}$. {Three of the velocity components show well constrained column densities, while the fourth component column density can be best viewed as an upper limit. The total column density value is not directly comparable, however,} with the 2$\sigma$ upper limit of $9.4\times 10^{10}$~cm$^{-2}$ from \citet{Loomis_2016}, as that analysis did not constrain the HC$_{11}$N source size, instead assuming a much larger fixed source size of 6.0'$\times$1.3' which would fill the GBT beam (based on previous mapping observations of HC$_3$N). As seen in Fig. \ref{Fig_17}, column density is highly covariant with our derived source size, and the largest contribution to the total HC$_{11}$N column density comes from the fourth velocity component with a source size of $\sim$9\arcsec. With the brightest HC$_{11}$N lines originating in X-band, where the GBT beam size is $\sim$1.2', this source size would correspond to a beam dilution factor using Eq. 2 of $\sim$0.015. Thus under the same assumptions as \citet{Loomis_2016}, our newly measured total HC$_{11}$N column density would be roughly $1.2^{+3.2}_{-0.8}\times 10^{10}$~cm$^{-2}$, entirely consistent with the upper limit presented there of $9.4\times 10^{10}$~cm$^{-2}$.

\subsection{Discussion}
\label{Discussion}
\subsubsection{Cyanopolyyne column densities}
The prior analysis of relative cyanopolyyne column densities synthesized both GBT observations reported in that paper as well as previous literature values \citep{Loomis_2016}. In all cases, an assumption was made that emission filled the beam, and the individual velocity components were not considered.

These assumptions are reasonable for the smaller cyanopolyynes -- we find that `co-spatial' fits to HC$_3$N and HC$_5$N significantly better replicate the observed line profiles. Although the more optically thin larger cyanopolyyne species such as HC$_9$N and HC$_{11}$N are well fit by a `separate components' fit, the varying source sizes in these fits make it very difficult to compare column densities across the two different fitting methods. For the purposes of this comparison, we have therefore additionally fit all cyanopolyyne species with a `co-spatial' method, with results presented in detail in the Supplementary Materials. {The `separate components' and `co-spatial' results for the larger species are very similar, as would be expected for optically thin species.} Further discussion of the relative source sizes and distributions of the cyanopolyynes and the effect on their fits is presented below.

\begin{figure*}[t]
    \centering
    \includegraphics[width=0.45\textwidth]{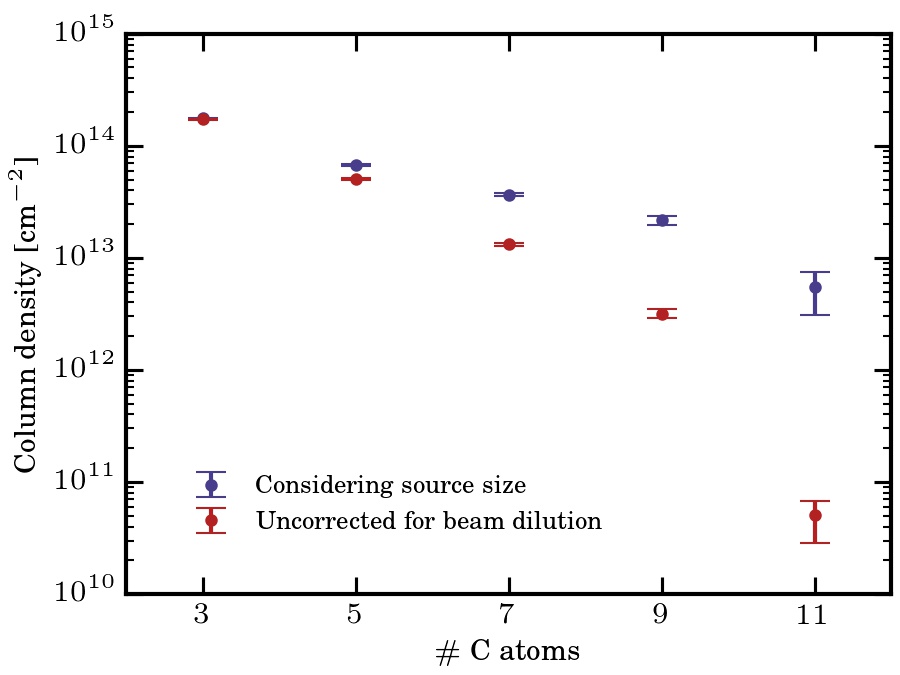}
    \includegraphics[width=0.45\textwidth]{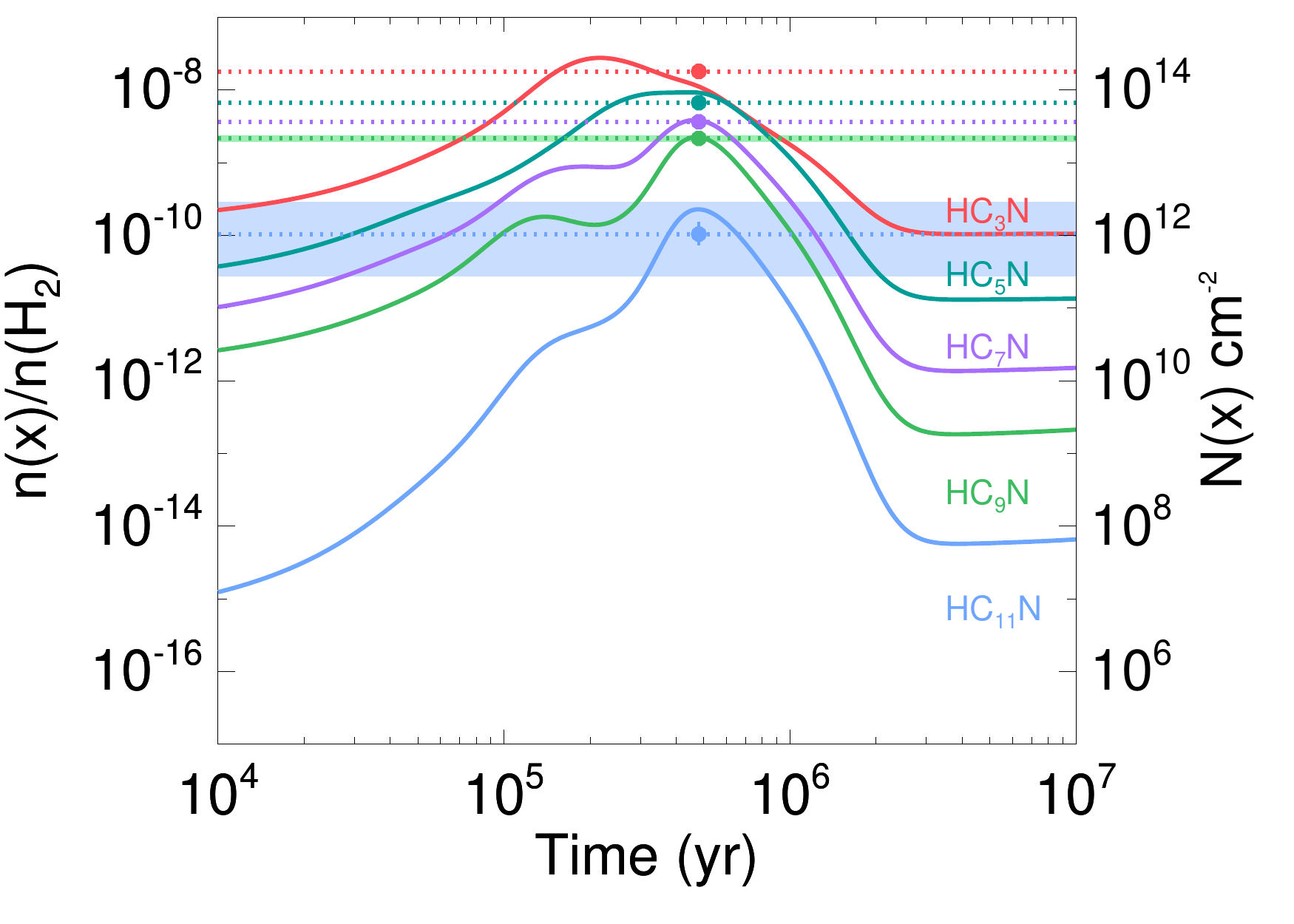}
    \caption{\textit{Left:} Cyanopolyyne total column densities from fits using the `co-spatial' approximation (see Supplementary Materials) are plotted against carbon-chain size, as in \citep{Loomis_2016}. Points in blue are taken directly from our fits, where the source size of each component was taken into account. Points in red have been adjusted back to the value which one would calculate under the assumption that the source fills the beam. These red points are directly comparable to those shown in Figure 5 in \citet{Loomis_2016}. \textit{Right:} Calculated abundances (solid lines), abundances from the co-spatial MCMC analysis (dotted lines), and best-fit times (dots) for the cyanopolyynes HC$_n$N, $n\in[3,5,7,9,11]$. Abundance ranges from the `separate components' MCMC analysis for \ce{HC9N} and \ce{HC11N} are shown by the green and blue bars, respectively. Equivalent column densities assuming $N(H_2)=10^{22}$ cm$^{-2}$ {\citep{Gratier_2016, Fuente_2019}} are shown on the right axis. Note: error bars for \ce{HC3N} - \ce{HC9N} are not visible at the scale used, but can be found in the tables in the Supplementary Materials.}
    \label{Fig_21}
\end{figure*}

Using the column densities derived from the fits presented in the Supplementary Materials, an updated version of Figure 5 from \citet{Loomis_2016} is shown in Fig. \ref{Fig_21}, along with a comparison to predictions from a chemical model, discussed in more detail in the Supplementary Materials. The general qualitative trend noted in that work is maintained, with a log-linear trend at smaller sizes, and a sharp decline at HC$_{11}$N.

\subsubsection{Spatial variations in cyanopolyyne chemistry}
\label{Discussion_1}
Previous spatially resolved observations of HC$_3$N, HC$_5$N, and HC$_7$N toward TMC-1 have shown them to be spatially extended on scales large enough to fill the GBT beam at the frequencies probed by GOTHAM \citep{Tolle_1981, Churchwell_1978, Olano_1988}. These observations were all taken at relatively coarse spatial resolution, however, and the detailed distribution of these species is unknown, as is the distribution of larger cyanopolyynes such as HC$_9$N. In particular, observations of cyanopolyynes at both high spectral and spatial resolution do not exist to date, making it difficult to spatially disentangle the four known velocity components in TMC-1.

Several pieces of evidence suggest that our two limiting sets of assumptions in the currently presented analysis of cyanopolyynes are insufficient, but also provide some hints at the true cyanopolyyne distribution. First, we note that `separate component' fits for HC$_3$N and HC$_5$N yield line profiles which poorly represent the data, while the `co-spatial' fits shown in the Supplementary Materials provide reasonable fits to the observational line profiles. This suggests that the velocity components are sufficiently co-spatial that when source sizes are large, they overlap significantly along the line of sight. Second, we find that for both the `co-spatial' and `separate component' fits, the source size(s) decrease with cyanopolyyne size as previously noted \citep{Bell_1998}, possibly suggesting spatially segregated chemical evolution within the source. Finally, for more optically thin species such as HC$_9$N, `separate component' fits yield widely varying source sizes for the components. This suggests that the source components are not purely co-spatial, and likely have some scatter within the beam.

{Our beam dilution and source-size fitting analysis is limited by both the sensitivity of our observations and the assumption that each source is centrally located within the beam. It is possible that the larger species have a broader distribution that is not well probed by our observations, due to sensitivity limitations.} If the spatio-kinematic structure of the cyanopolyynes is shared by other species, it may be possible to use a single set of interferometric observations as a template to unlock the GOTHAM observations, enabling more complicated fitting and thus better characterization of the true column density spatial distribution.

In conclusion, we have presented a new method for robustly characterizing and visualizing detections of new interstellar species in line sparse sources, even when individual lines of the species are not detected. These results of applying this method to the GOTHAM dataset have resulted in a total of six new interstellar species have been detected in TMC-1 \citep{McGuire:2020bb, McGuire:2020aa, McCarthy:2020aa, Burkhardt:2020aa, Xue:2020aa}. In particular, we have detected HC$_{11}$N in TMC-1 and derived a column density consistent with the previous upper limit presented in \citet{Loomis_2016}.

\section{Contributions}
R.A.L. wrote the manuscript and developed the MCMC and spectral stacking analysis code described here.
M.C.M. and K.L.K.L. performed the laboratory experiments and theoretical calculations for several of the catalogs used in this analysis, and helped revise the manuscript.
A.M.B and B.A.M. performed the astronomical observations and subsequent data reduction. E. H. determined and/or estimated rate coefficients and is the originator for many of the chemical simulations.
A.M.B. and C.N.S. contributed or undertook the astronomical modeling and simulations. E.R.W., M.A.C. and B.A.M. contributed to the design of the GOTHAM survey, and helped revise the manuscript. C.X. modified and contributed the chemical networks of the related species, and helped revise the manuscript. C.X. and A.J.R. performed the astronomical observations.

\section{Code statement}
All the codes used in the MCMC fitting and stacking analysis presented in this paper are open source and publicly available at \href{https://github.com/ryanaloomis/TMC1\_mcmc\_fitting}{https://github.com/ryanaloomis/TMC1\_mcmc\_fitting}.

\section{Competing interests statement}
The authors declare no competing interests.

\section{Data statement}
The datasets analyzed during the current study are available in the Green Bank Telescope archive (\href{https://archive.nrao.edu/archive/advquery.jsp}{https://archive.nrao.edu/archive/advquery.jsp}), PI: B. McGuire.  A user manual for their reduction and analysis is available as well (\href{https://greenbankobservatory.org/science/gbt-observers/visitor-facilities-policies/data-reduction-gbt-using-idl/}{https://greenbankobservatory.org/science/gbt-observers/visitor-facilities-policies/data-reduction-gbt-using-idl/}).  The complete, reduced survey data at X-band is available as Supplementary Information in \cite{McGuire:2020bb}.  The individual portions of reduced spectra used in the analysis of the individual species presented here is available in the Harvard Dataverse Archive \cite{GOTHAMDR1}.

\section{Acknowledgements}

The authors thank the anonymous reviewer for their helpful comments.  A.M.B. acknowledges support from the Smithsonian Institution as a Submillimeter Array (SMA) Fellow. M.C.M. and K.L.K. Lee acknowledge support from NSF grant AST-1615847 and NASA grant 80NSSC18K0396. Support for B.A.M. during the initial portions of this work was provided by NASA through Hubble Fellowship grant \#HST-HF2-51396 awarded by the Space Telescope Science Institute, which is operated by the Association of Universities for Research in Astronomy, Inc., for NASA, under contract NAS5-26555. C.N.S. thanks the Alexander von Humboldt Stiftung/Foundation for their generous support, as well as V. Wakelam for use of the \texttt{NAUTILUS} v1.1 code.  C.X. is a Grote Reber Fellow, and support for this work was provided by the NSF through the Grote Reber Fellowship Program administered by Associated Universities, Inc./National Radio Astronomy Observatory and the Virginia Space Grant Consortium.  E.H. thanks the National Science Foundation for support through grant AST 1906489.   S.B.C. and M.A.C. were supported by the NASA Astrobiology Institute through the Goddard Center for Astrobiology.  The National Radio Astronomy Observatory is a facility of the National Science Foundation operated under cooperative agreement by Associated Universities, Inc.  The Green Bank Observatory is a facility of the National Science Foundation operated under cooperative agreement by Associated Universities, Inc.


\section{Extended Data}

\begin{figure*}[!hb]
\centering
\includegraphics[width=\textwidth]{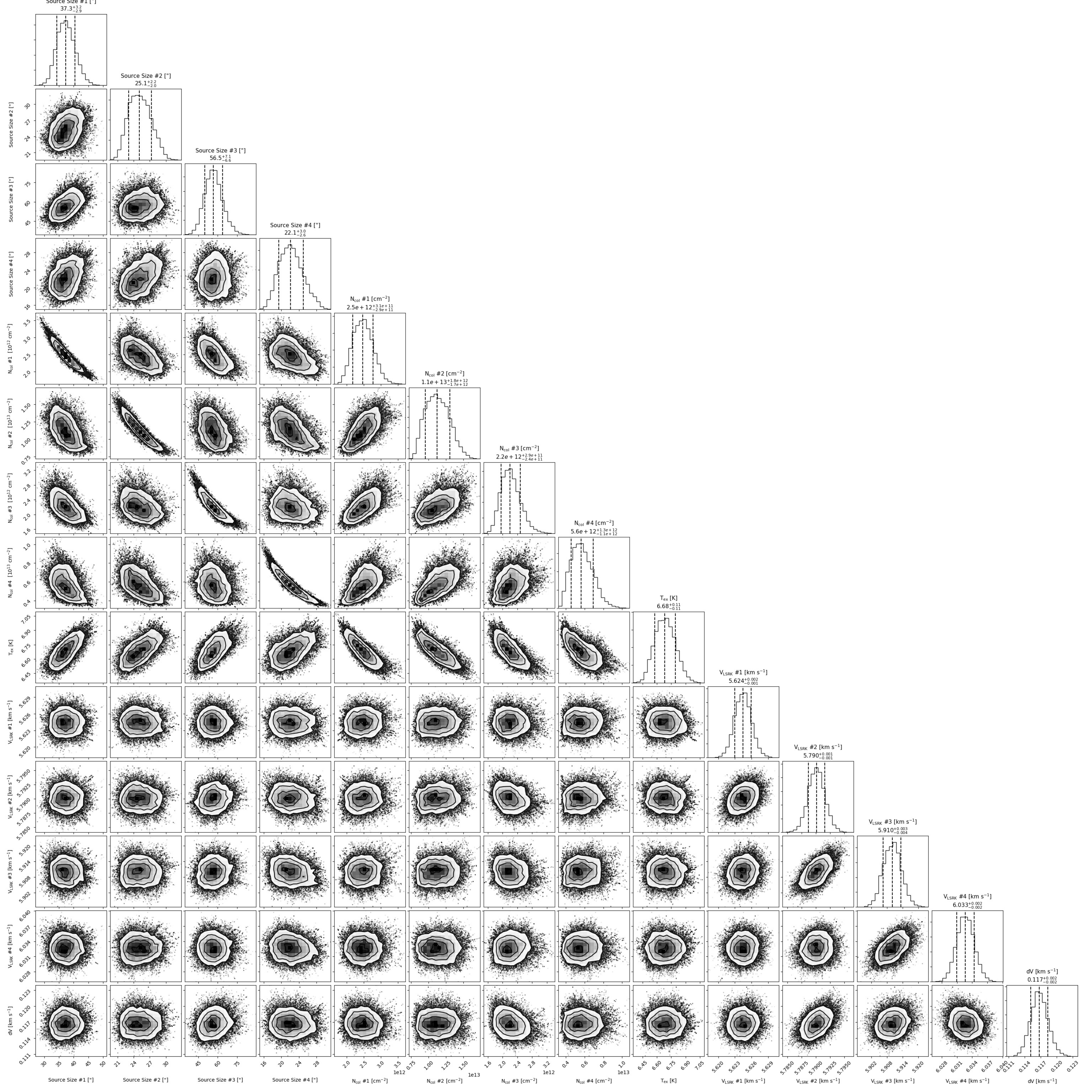}
\caption{Parameter covariances and marginalized posterior distributions for the \ce{HC9N} MCMC fit. 16$^{th}$, 50$^{th}$, and 84$^{th}$ confidence intervals (corresponding to $\pm$1 sigma for a Gaussian posterior distribution) are shown as vertical lines.}
\label{Fig_3}
\end{figure*}
\clearpage

\begin{table*}
\centering
\caption{\ce{HC9N} best-fit parameters from MCMC analysis}
\begin{tabular}{c c c c c c}
\toprule
\multirow{2}{*}{Component}&	$v_{lsr}$					&	Size					&	\multicolumn{1}{c}{$N_T^\dagger$}					&	$T_{ex}$							&	$\Delta V$		\\
			&	(km s$^{-1}$)				&	($^{\prime\prime}$)		&	\multicolumn{1}{c}{(10$^{12}$ cm$^{-2}$)}		&	(K)								&	(km s$^{-1}$)	\\
\midrule
\hspace{0.1em}\vspace{-0.5em}\\
C1	&	$5.624^{+0.002}_{-0.001}$	&	$37^{+3}_{-2}$	&	$2.47^{+0.31}_{-0.29}$	&	\multirow{6}{*}{$6.7^{+0.1}_{-0.1}$}	&	\multirow{6}{*}{$0.117^{+0.002}_{-0.002}$}\\
\hspace{0.1em}\vspace{-0.5em}\\
C2	&	$5.790^{+0.001}_{-0.001}$	&	$25^{+2}_{-2}$	&	$11.19^{+1.83}_{-1.67}$	&		&	\\
\hspace{0.1em}\vspace{-0.5em}\\
C3	&	$5.910^{+0.003}_{-0.004}$	&	$56^{+7}_{-6}$	&	$2.20^{+0.29}_{-0.24}$	&		&	\\
\hspace{0.1em}\vspace{-0.5em}\\
C4	&	$6.033^{+0.002}_{-0.002}$	&	$22^{+2}_{-2}$	&	$5.64^{+1.30}_{-1.07}$	&		&	\\
\hspace{0.1em}\vspace{-0.5em}\\
\midrule
$N_T$ (Total)$^{\dagger\dagger}$	&	 \multicolumn{5}{c}{$2.15^{+0.23}_{-0.20}\times 10^{13}$~cm$^{-2}$}\\
\bottomrule
\end{tabular}

\begin{minipage}{0.75\textwidth}
	\footnotesize
	{Note} -- The quoted uncertainties represent the 16$^{th}$ and 84$^{th}$ percentile ($1\sigma$ for a Gaussian distribution) uncertainties.\\
	$^\dagger$Column density values are highly covariant with the derived source sizes.  The marginalized uncertainties on the column densities are therefore dominated by the largely unconstrained nature of the source sizes, and not by the signal-to-noise of the observations.
	$^{\dagger\dagger}$Uncertainties derived by adding the uncertainties of the individual components in quadrature.
\end{minipage}
\label{Table_1}
\end{table*}

\begin{figure*}
\centering
\includegraphics[width=0.5\textwidth]{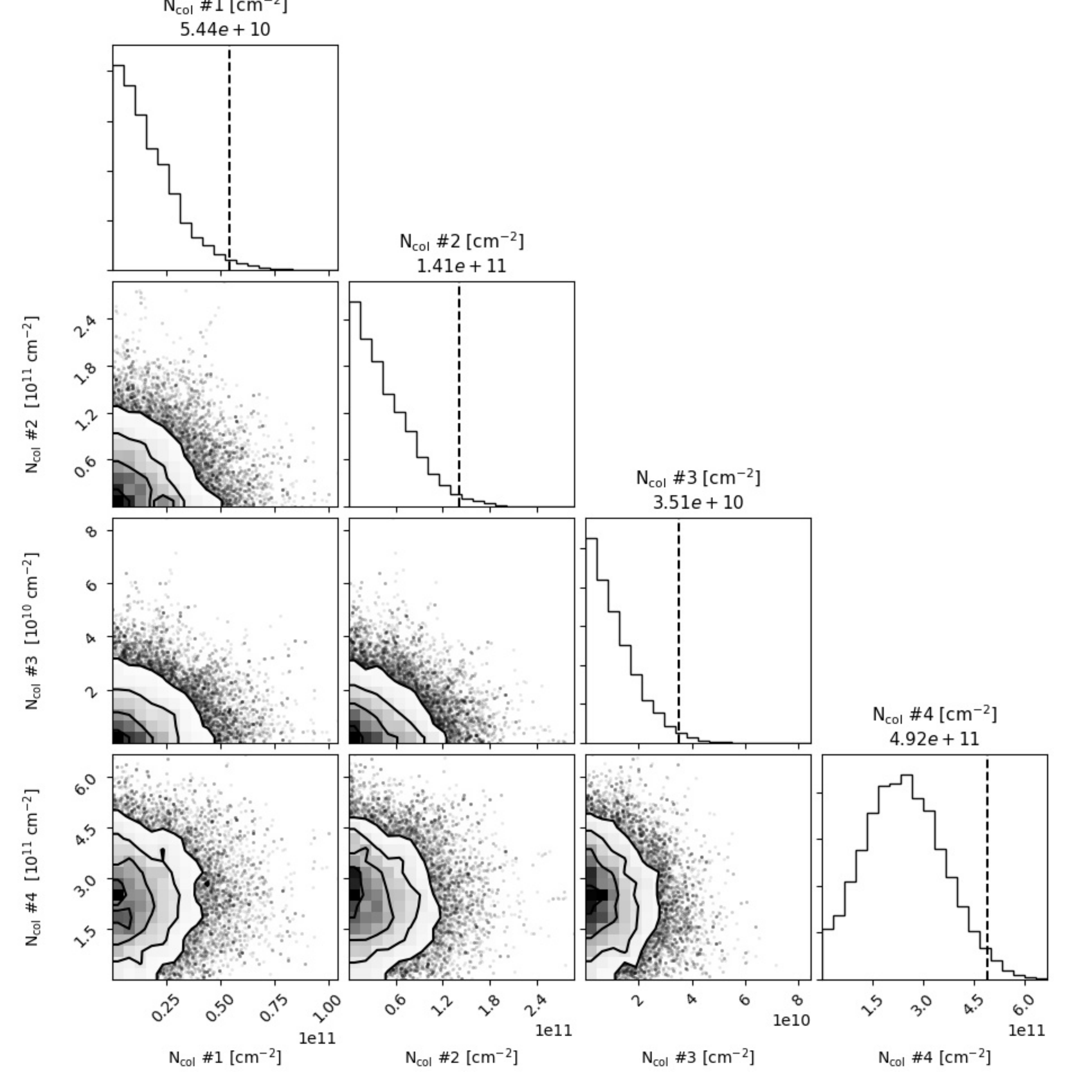}
\caption{Parameter covariances and marginalized posterior distributions for the HC$_{13}$N MCMC fit. The 97.8$^{th}$ confidence interval (corresponding to 2 sigmas for a Gaussian posterior distribution) is shown as a vertical line.}
\label{Fig_7}
\end{figure*}

\begin{figure*}
\centering
\includegraphics[width=\textwidth]{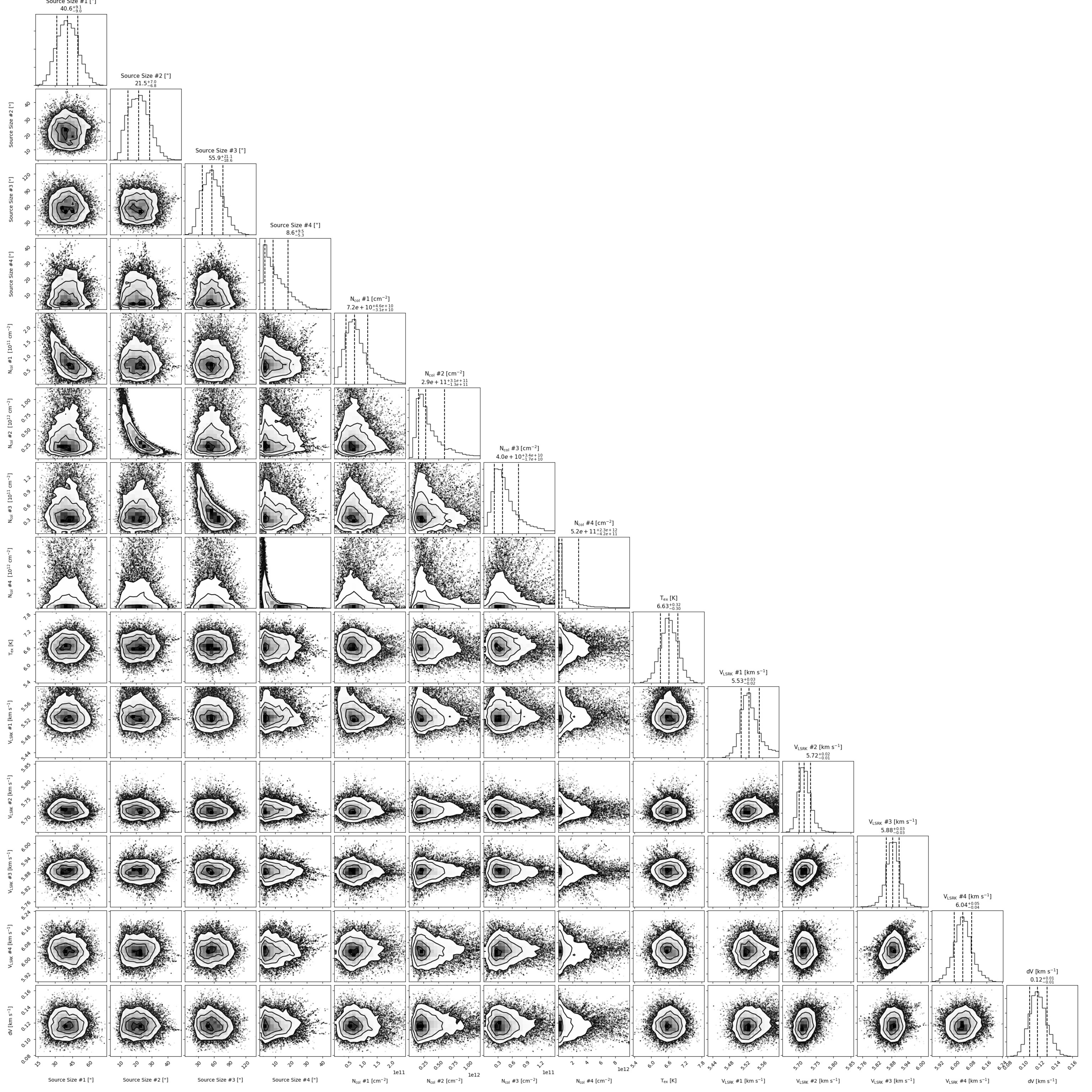}
\caption{Parameter covariances and marginalized posterior distributions for the HC$_{11}$N MCMC fit. 16$^{th}$, 50$^{th}$, and 84$^{th}$ confidence intervals (corresponding to $\pm$1 sigma for a Gaussian posterior distribution) are shown as vertical lines.}
\label{Fig_17}
\end{figure*}

\clearpage
\section{Supplementary Information}
\subsection{Demonstration of robustness}
When stacking  many lines across a wide bandwidth, there is a clear risk that signal from an interloping line of a different species may bleed into the analysis. Here, we demonstrate that in applying our technique to GOTHAM observations of TMC-1, this risk can be mitigated, and our detections are robust.

\subsubsection{Application to benzonitrile}
\begin{figure*}
\centering
\includegraphics[width=\textwidth]{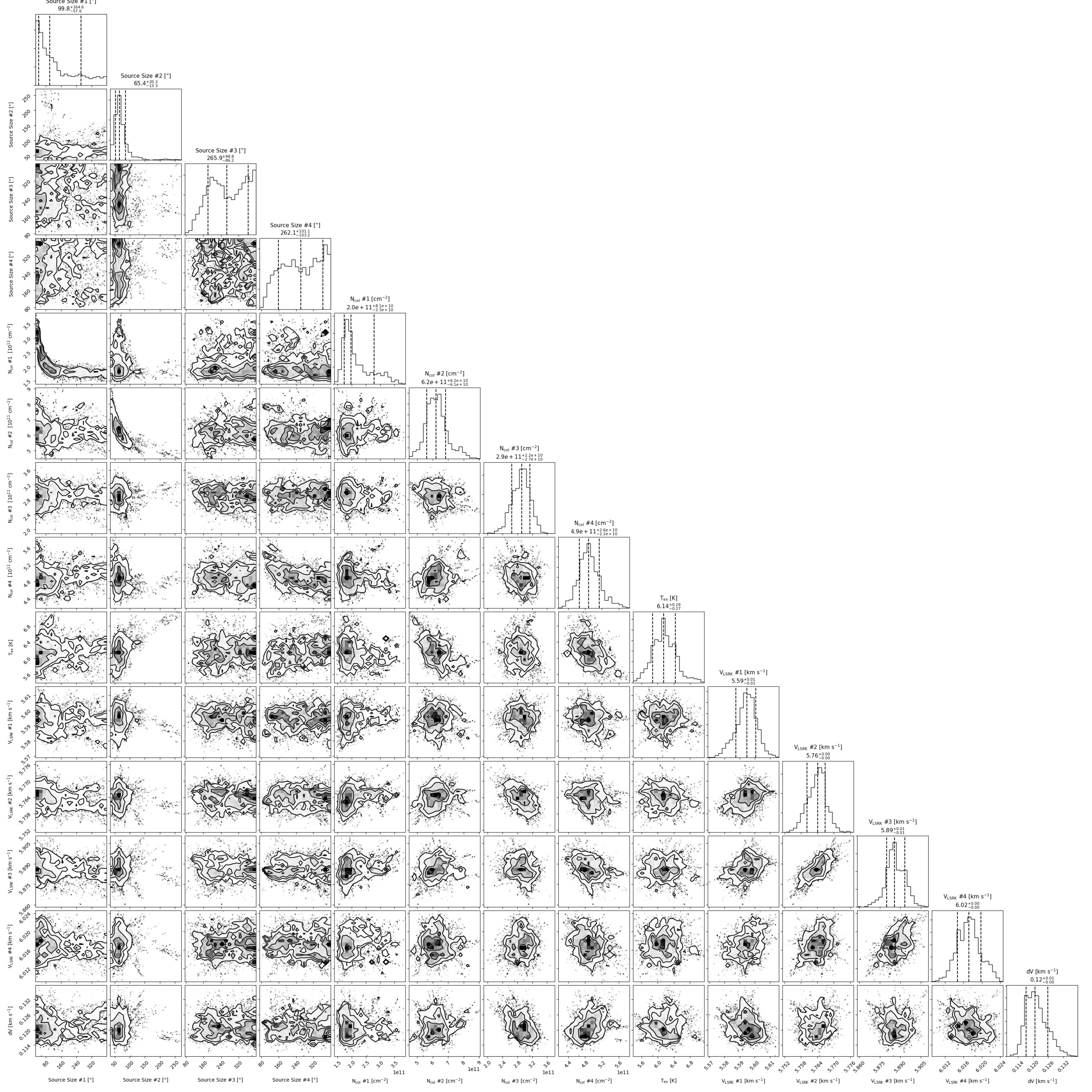}
\caption{Parameter covariances and marginalized posterior distributions for the benzonitrile MCMC fit. 16$^{th}$, 50$^{th}$, and 84$^{th}$ confidence intervals (corresponding to $\pm$1 sigma for a Gaussian posterior distribution) are shown as vertical lines. Source sizes for components 3 and 4 are consistent with a source that fills the GBT beam (i.e. they are unconstrained on the upper bound).}
\label{Fig_8}
\end{figure*}

\begin{figure*}[!t]
    \centering
    \includegraphics[width=\textwidth]{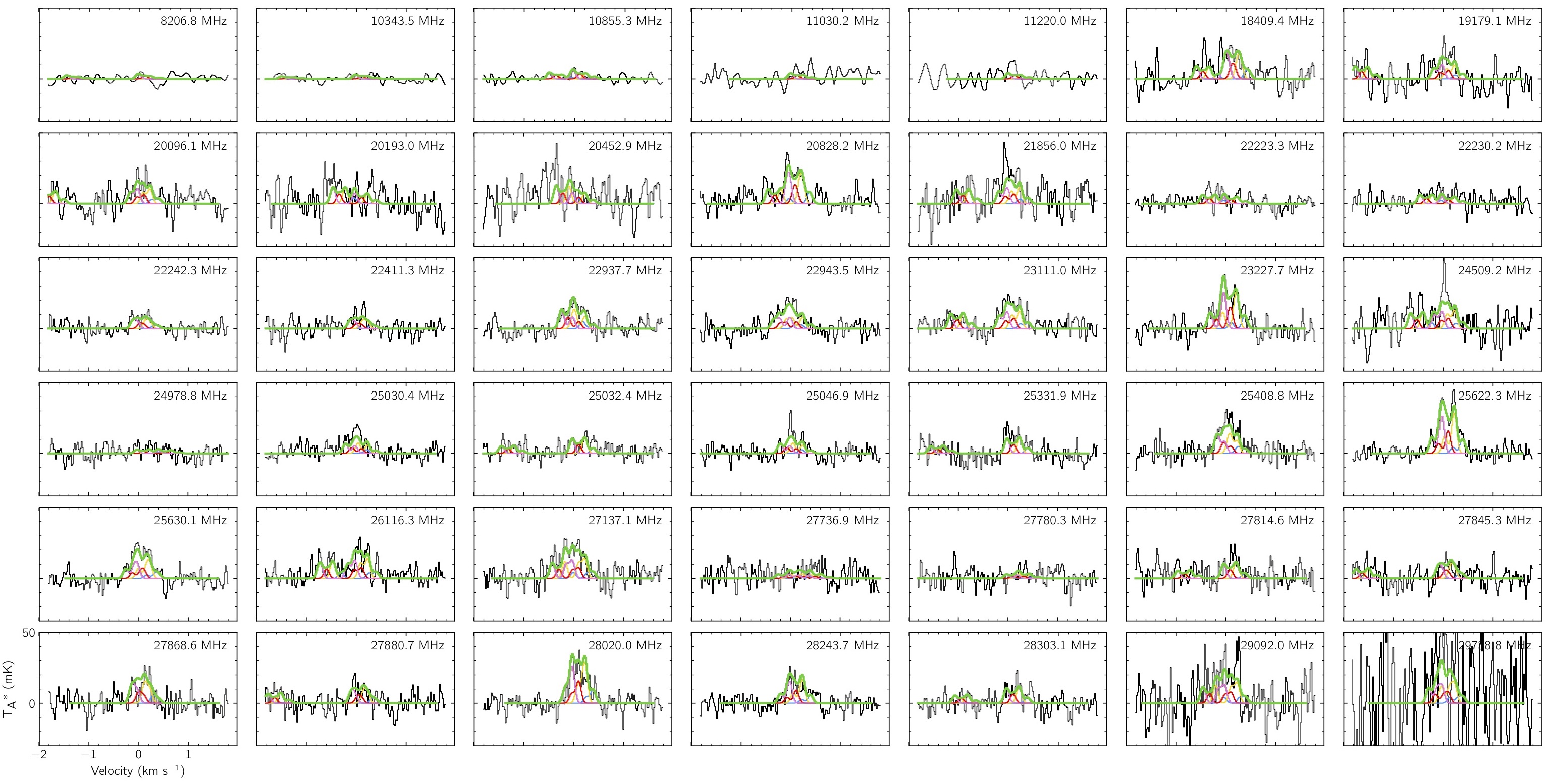}
    \caption{Individual line detections of benzonitrile in the GOTHAM data.  The spectra (black) are displayed in velocity space relative to 5.8\,km\,s$^{-1}$, and using the rest frequency given in the top right of each panel. Quantum numbers are given in the top left of each panel, neglecting hyperfine splitting. The best-fit model to the data, including all velocity components, is overlaid in green.  Simulated spectra of the individual velocity components are shown in: blue (5.63\,km\,s$^{-1}$), gold (5.79\,km\,s$^{-1}$), red (5.91\,km\,s$^{-1}$), and violet (6.03\,km\,s$^{-1}$).  See Table~\ref{Table_2}.}
    \label{Fig_9}
\end{figure*}

The presence of benzonitrile in TMC-1 has been further verified by our GOTHAM observations since its initial detection \citep{McGuire_2018}. Multiple individual lines are detected in our data, and the parameters from our MCMC fit (Table \ref{Table_2} and Fig. \ref{Fig_8}) can be seen to well-replicate the observations (Fig. \ref{Fig_9}), motivating our use of benzonitrile as a template for analysis of all aromatic species in the GOTHAM data.

\begin{table*}[!hb]
\centering
\caption{Benzonitrile best-fit parameters from MCMC analysis}
\begin{tabular}{c c c c c c}
\toprule
\multirow{2}{*}{Component}&	$v_{lsr}$					&	Size					&	\multicolumn{1}{c}{$N_T^\dagger$}					&	$T_{ex}$							&	$\Delta V$		\\
			&	(km s$^{-1}$)				&	($^{\prime\prime}$)		&	\multicolumn{1}{c}{(10$^{11}$ cm$^{-2}$)}		&	(K)								&	(km s$^{-1}$)	\\
\midrule
\hspace{0.1em}\vspace{-0.5em}\\
C1	&	$5.595^{+0.006}_{-0.007}$	&	$99^{+164}_{-57}$	&	$1.98^{+0.81}_{-0.23}$	&	\multirow{6}{*}{$6.1^{+0.3}_{-0.3}$}	&	\multirow{6}{*}{$0.121^{+0.005}_{-0.004}$}\\
\hspace{0.1em}\vspace{-0.5em}\\
C2	&	$5.764^{+0.003}_{-0.004}$	&	$65^{+20}_{-13}$	&	$6.22^{+0.62}_{-0.61}$	&		&	\\
\hspace{0.1em}\vspace{-0.5em}\\
C3	&	$5.886^{+0.007}_{-0.006}$	&	$265^{+98}_{-86}$	&	$2.92^{+0.22}_{-0.27}$	&	    &	\\
\hspace{0.1em}\vspace{-0.5em}\\
C4	&	$6.017^{+0.003}_{-0.002}$	&	$262^{+101}_{-103}$	&	$4.88^{+0.26}_{-0.22}$	&		&	\\
\hspace{0.1em}\vspace{-0.5em}\\
\midrule
$N_T$ (Total)$^{\dagger\dagger}$	&	 \multicolumn{5}{c}{$1.6^{+0.19}_{-0.13}\times 10^{12}$~cm$^{-2}$}\\
\bottomrule
\end{tabular}

\begin{minipage}{0.75\textwidth}
	\footnotesize
	{Note} -- The quoted uncertainties represent the 16$^{th}$ and 84$^{th}$ percentile ($1\sigma$ for a Gaussian distribution) uncertainties.\\
	$^\dagger$Column density values are highly covariant with the derived source sizes.  The marginalized uncertainties on the column densities are therefore dominated by the largely unconstrained nature of the source sizes, and not by the signal-to-noise of the observations.
	$^{\dagger\dagger}$Uncertainties derived by adding the uncertainties of the individual components in quadrature.
\end{minipage}

\label{Table_2}

\end{table*}

\begin{figure}
    \centering
    \includegraphics[width=0.5\columnwidth]{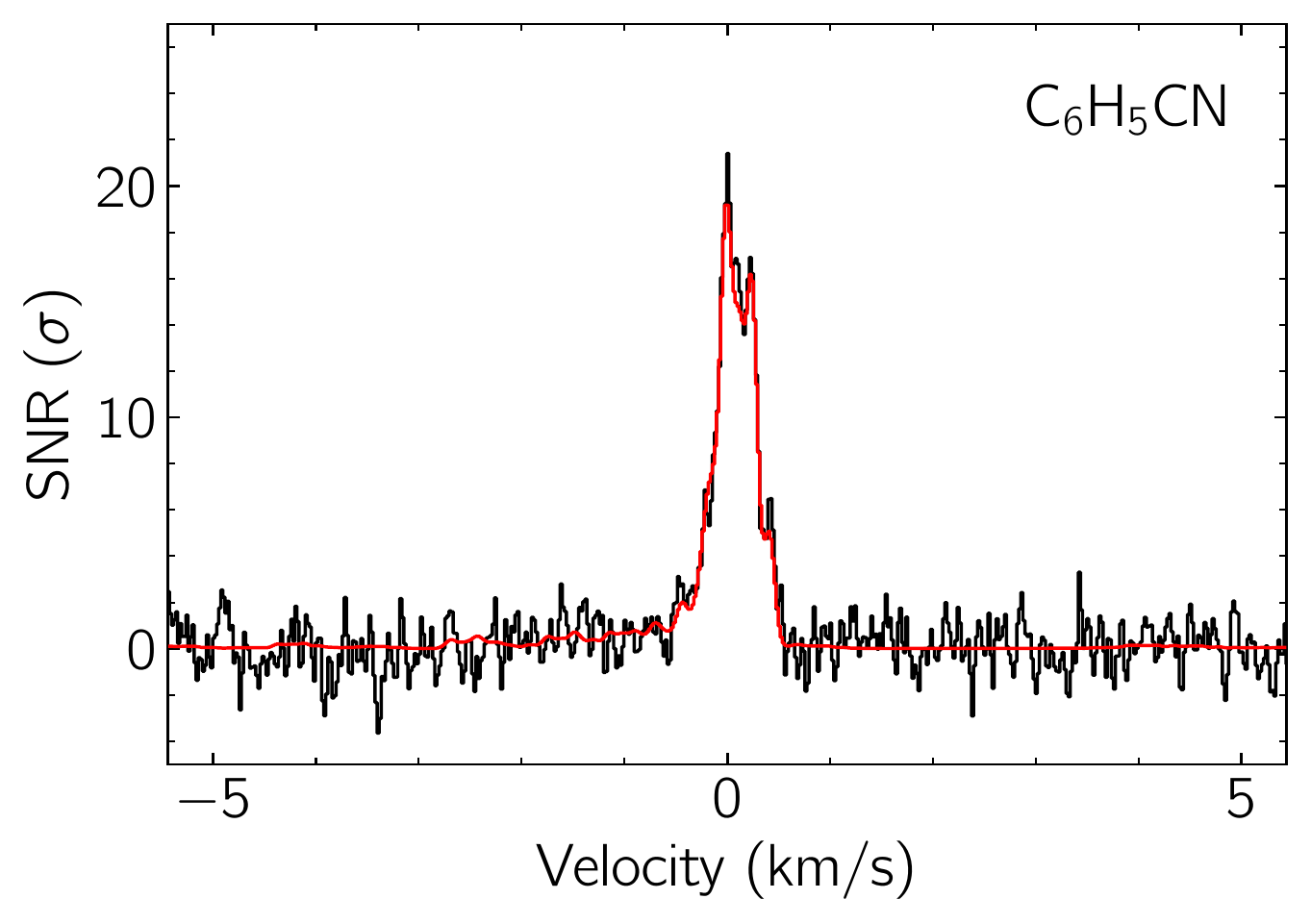}
    \caption{Velocity-stacked spectra of benzonitrile in black, with the corresponding stack of the simulation using the best-fit parameters to the individual lines in red. The data have been uniformly sampled to a resolution of 0.02\,km\,s$^{-1}$. The intensity scale is the signal-to-noise ratio of the spectrum at any given velocity.}
    \label{Fig_10}
\end{figure}

\begin{figure}
    \centering
    \includegraphics[width=0.5\columnwidth]{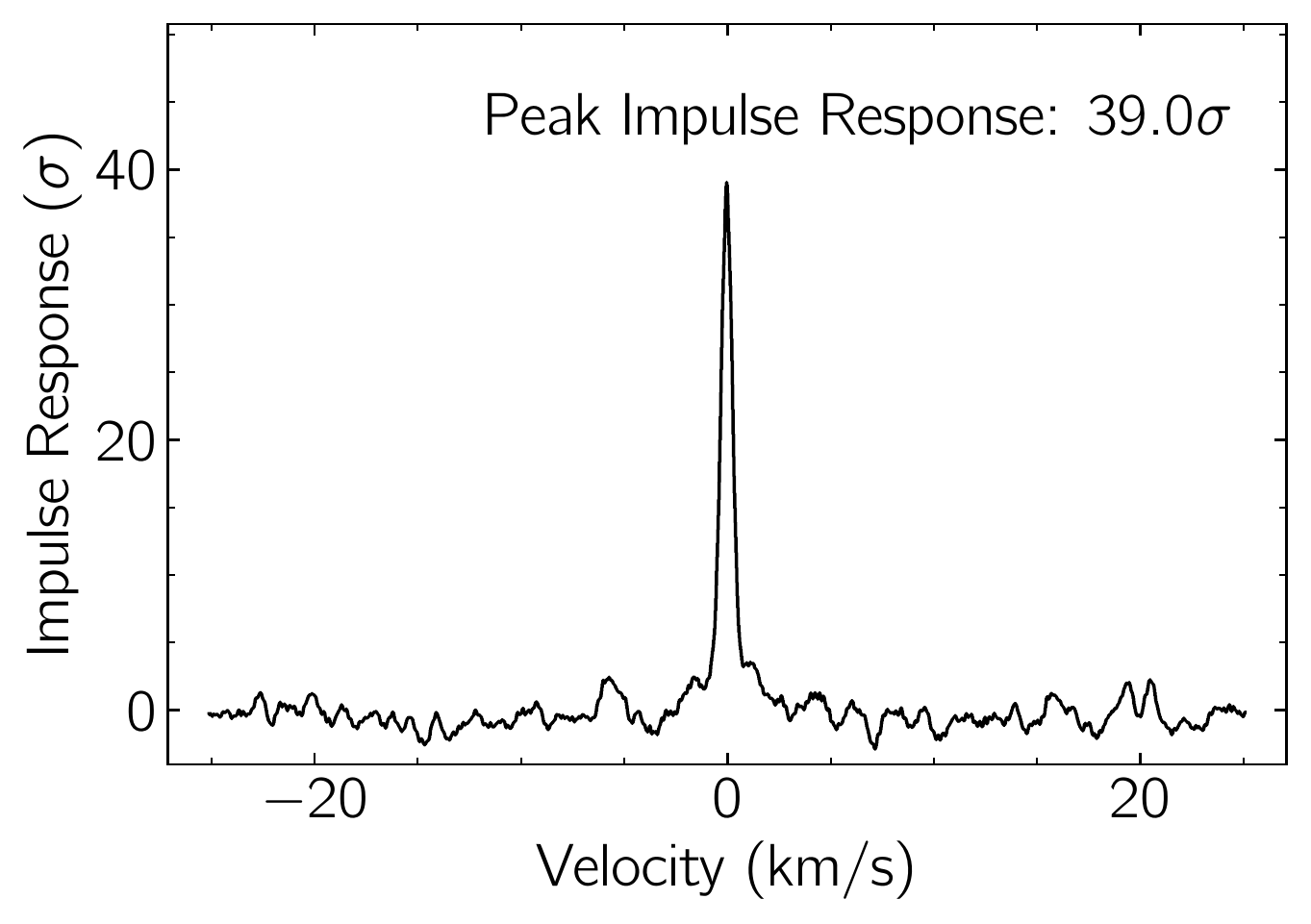}
    \caption{Impulse response function of the stacked benzonitrile spectrum using the simulated line profile as a matched filter.  The intensity scale is the signal-to-noise ratio of the response function when centered at a given velocity.  The peak of the impulse response function provides a minimum significance for the detection of 39.0$\sigma$.}
    \label{Fig_11}
\end{figure}

These lines were then stacked and the model stack was applied to the data as a matched filter. The resultant stacks and matched filter response are shown in Fig. \ref{Fig_10} and \ref{Fig_11}. The line detection significance is dramatically improved, from $\sim$1-12$\sigma$ in the individual lines, to 21.8$\sigma$ in the line stack, and then 39.0$\sigma$ in the matched filter response. The velocity structure shown in the individual lines in Fig. \ref{Fig_9} is also more clearly seen in the stack in Fig. \ref{Fig_10}.

\subsubsection{Application to 2-cyanonaphthalene}
\begin{figure*}[!t]
    \centering
    \includegraphics[width=\textwidth]{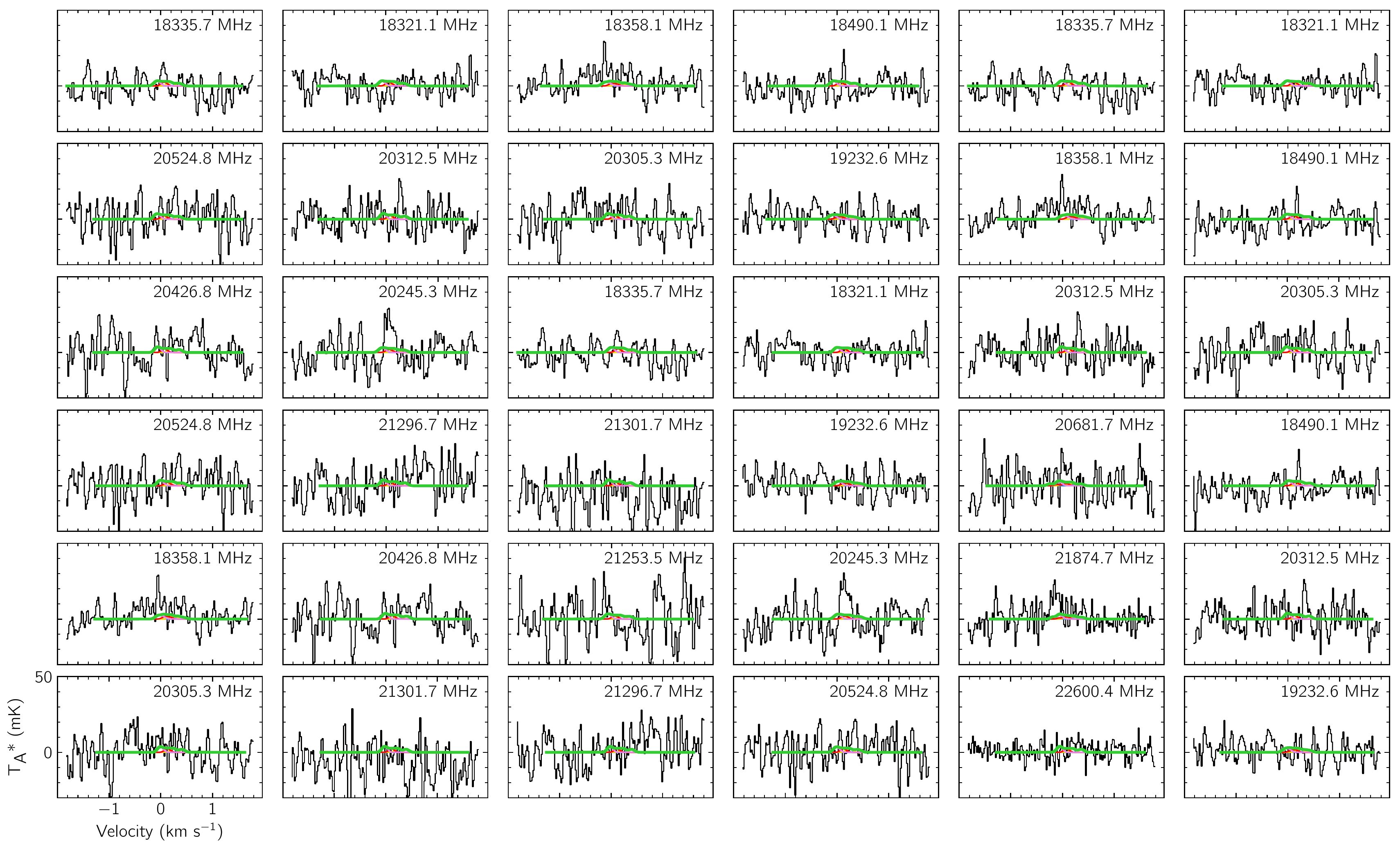}
    \caption{Individual line observations of 2-cyanonaphthalene in the GOTHAM data.  The spectra (black) are displayed in velocity space relative to 5.8\,km\,s$^{-1}$, and using the rest frequency given in the top right of each panel. Quantum numbers are given in the top left of each panel, neglecting hyperfine splitting. The best-fit model to the data, including all velocity components, is overlaid in green.  Simulated spectra of the individual velocity components are shown in: blue (5.63\,km\,s$^{-1}$), gold (5.79\,km\,s$^{-1}$), red (5.91\,km\,s$^{-1}$), and violet (6.03\,km\,s$^{-1}$).}
    \label{Fig_12}
\end{figure*}

\begin{figure*}
\centering
\includegraphics[width=\textwidth]{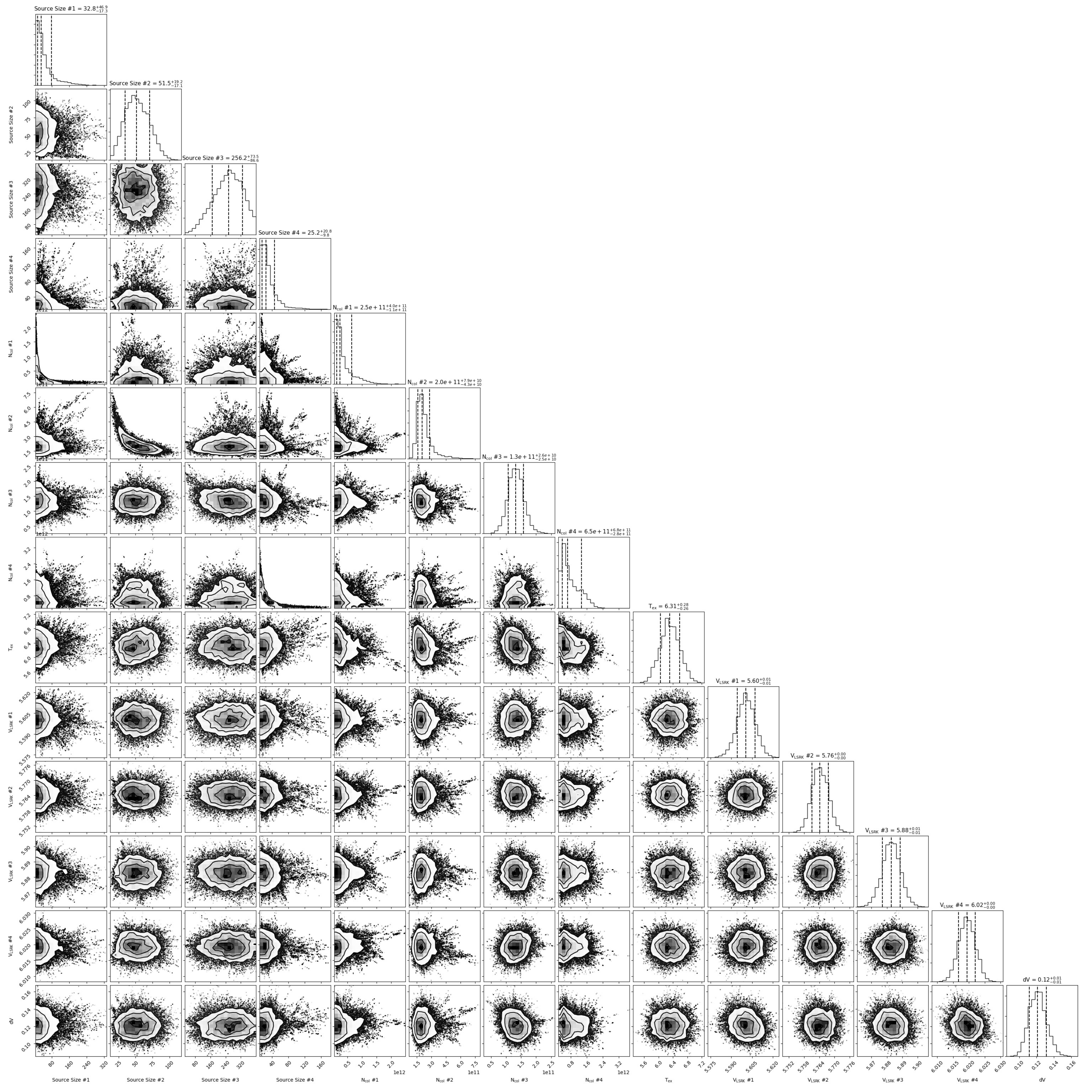}
\caption{Parameter covariances and marginalized posterior distributions for the 2-cyanonaphthalene MCMC fit. 16$^{th}$, 50$^{th}$, and 84$^{th}$ confidence intervals (corresponding to $\pm$1 sigma for a Gaussian posterior distribution) are shown as vertical lines.}
\label{Fig_13}
\end{figure*}

\begin{figure}
    \centering
    \includegraphics[width=0.5\columnwidth]{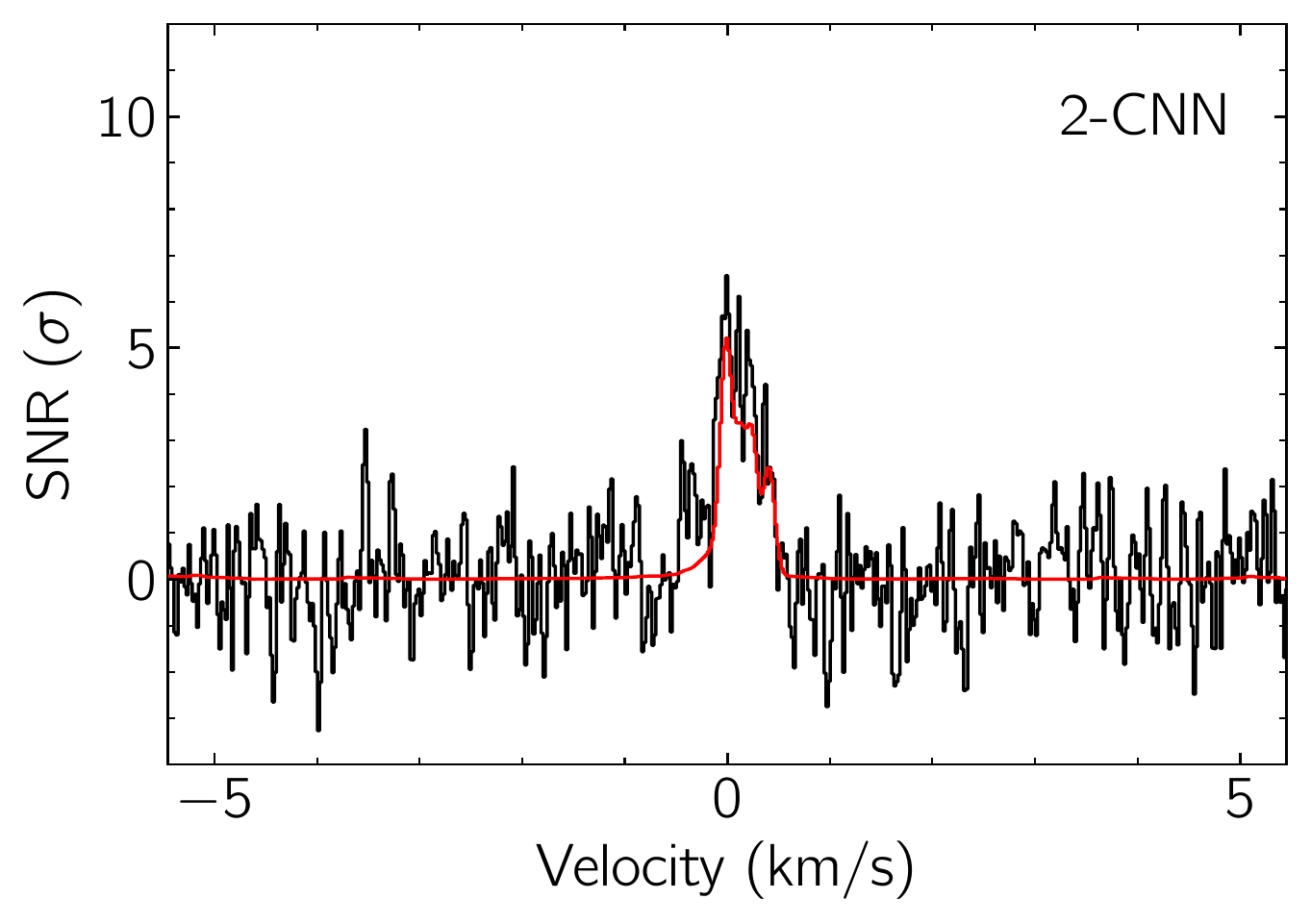}
    \caption{Velocity-stacked spectra of 2-cyanonaphthalene in black, with the corresponding stack of the simulation using the best-fit parameters to the individual lines in red. The data have been uniformly sampled to a resolution of 0.02\,km\,s$^{-1}$. The intensity scale is the signal-to-noise ratio of the spectrum at any given velocity.}
    \label{Fig_14}
\end{figure}

\begin{figure}
    \centering
    \includegraphics[width=0.5\columnwidth]{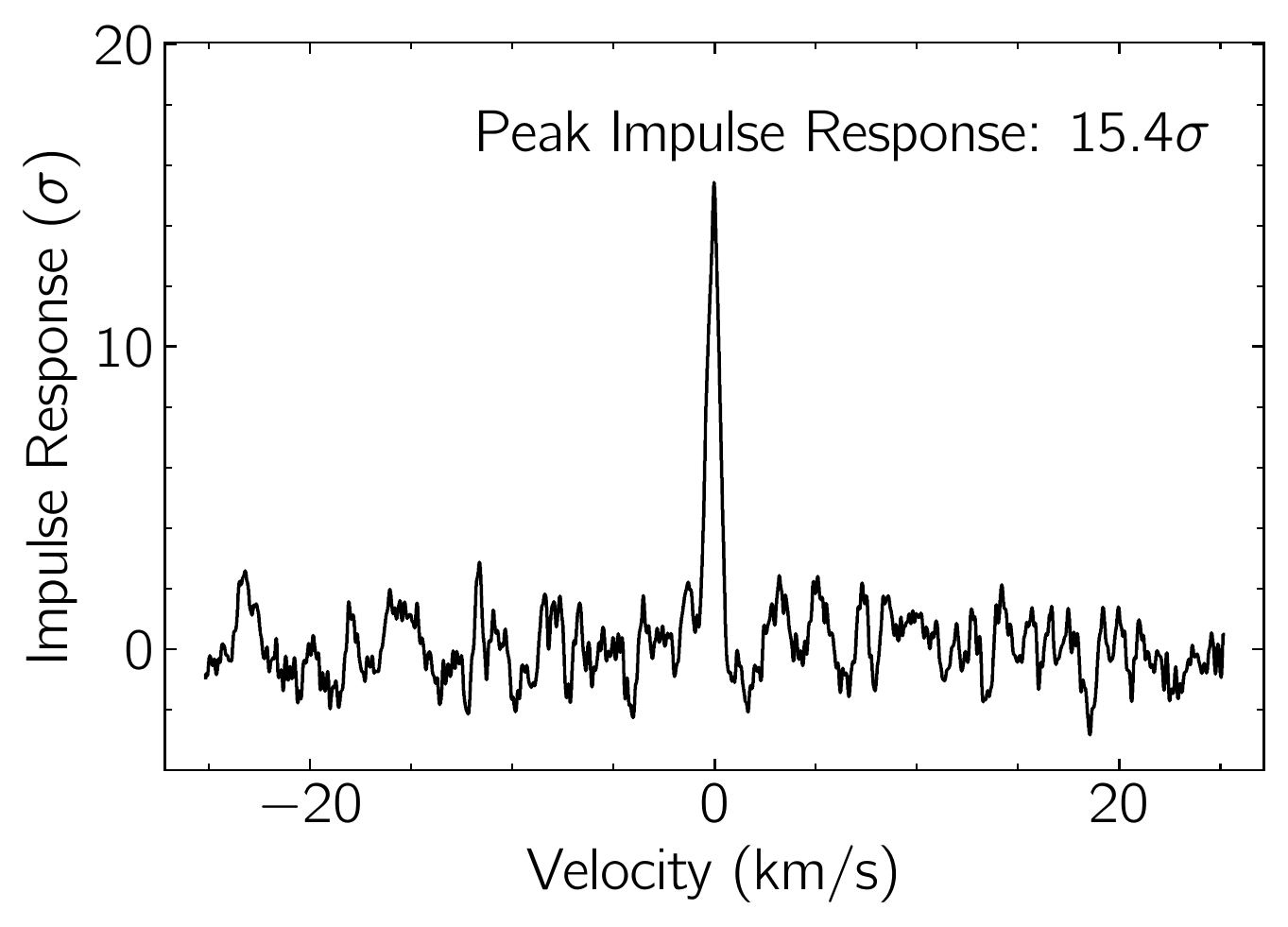}
    \caption{Impulse response function of the stacked 2-cyanonaphthalene spectrum using the simulated line profile as a matched filter.  The intensity scale is the signal-to-noise ratio of the response function when centered at a given velocity.  The peak of the impulse response function provides a minimum significance for the detection of 15.4$\sigma$.}
    \label{Fig_15}
\end{figure}

Beyond the example shown for benzonitrile, where individual lines are detected above the noise, it is also possible to apply this method to species where individual lines are not detected. The partition function of 2-cyanonaphthalene is such that, even at the low temperatures of TMC-1 ($\sim$10~K), thousands of lines within the GOTHAM frequency coverage have similar emission intensities, and are all quite weak. As shown in Fig. \ref{Fig_12}, none of the 36 strongest lines (based on the initial simulation parameters described in the main text) are detected in the GOTHAM data. Despite this, a global MCMC analysis using priors from our benzonitrile fit (Table~\ref{Table_2}) is able to provide useful constraints, and posterior distributions are shown in Fig. \ref{Fig_13}. The weighted stack of all observed lines is shown in Fig. \ref{Fig_14}, with a clear detection now present at 6.9$\sigma$, including visible velocity structure. Applying the model stack as a matched filter yields a 15.4$\sigma$ detection significance (Fig. \ref{Fig_15}). Further discussion and analysis of this molecular detection is presented in \citet{McGuire:2020aa}.

\subsubsection{Robustness to false positives}
\label{Appendix_B}
A molecular detection method must not only detect signal that is present in data, but it must also not yield false positives. Stacking techniques in particular are susceptible to `bleed-in' from interloping lines when line density is too high. In our technique, we first take a thresholding step when searching for a weak species (where no individual lines are expected to be present above the noise), discarding any windows that contain signal above 6$\sigma$, as these windows likely contain an interloping line.

Although this method is likely to remove the most egregious interloping lines, there very well may be many lines sitting below the noise that could cause false positives. Indeed, we know that there is signal below the visible noise level, as there are thousands of weak lines that we intend to stack for species such as 2-cyanonaphthalene. Here, it is the extremely narrow linewidth of TMC-1 and the corresponding line sparsity discussed earlier that reduce the risk of collisions. This is demonstrated by how sensitive our filter responses are to even minuscule ($\sim$ppm) changes to rotational constants.

A common method to further illustrate that we are uncovering emission from the species of interested, and not from an unexpected contaminating line or set of lines, is to jack-knife the data. For a species such as 2-cyanonaphthalene, there are thousands of lines of similar strength, and by splitting the set of analyzed lines in half we can examine whether an outsized portion of the signal is coming from one half of the data. If the signal is real and our noise is mainly white, we would expect each half of the data to have roughly $\frac{1}{\sqrt{2}} \times$ the filter response of the full dataset. In contrast, if an interloping line was dominating the response, we would expect the two halves to have very unbalanced responses, with one consistent with pure noise and the other close to the original filter response.

\begin{figure*}[!h]
    \centering
    \includegraphics[width=0.35\textwidth]{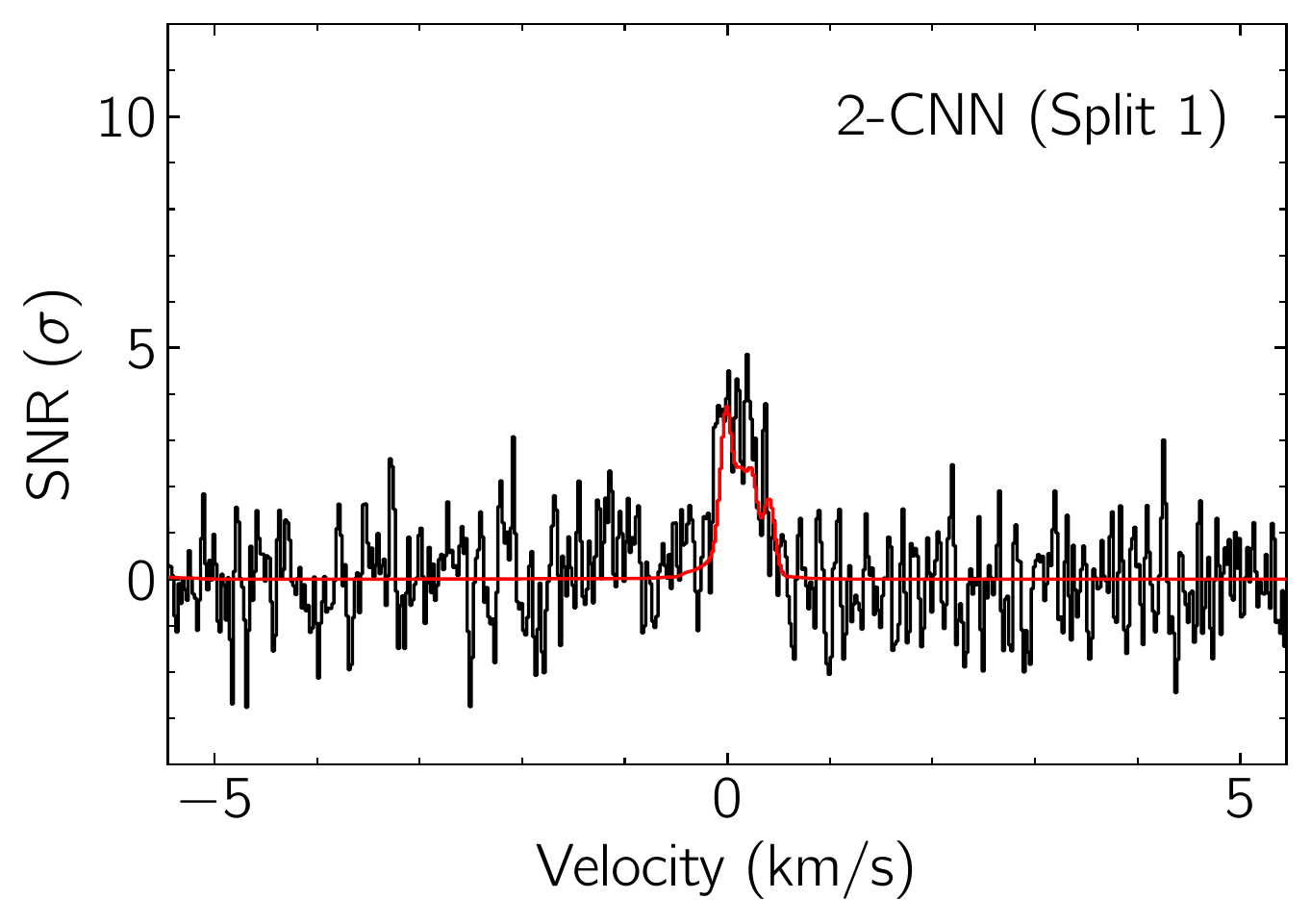}
    \includegraphics[width=0.35\textwidth]{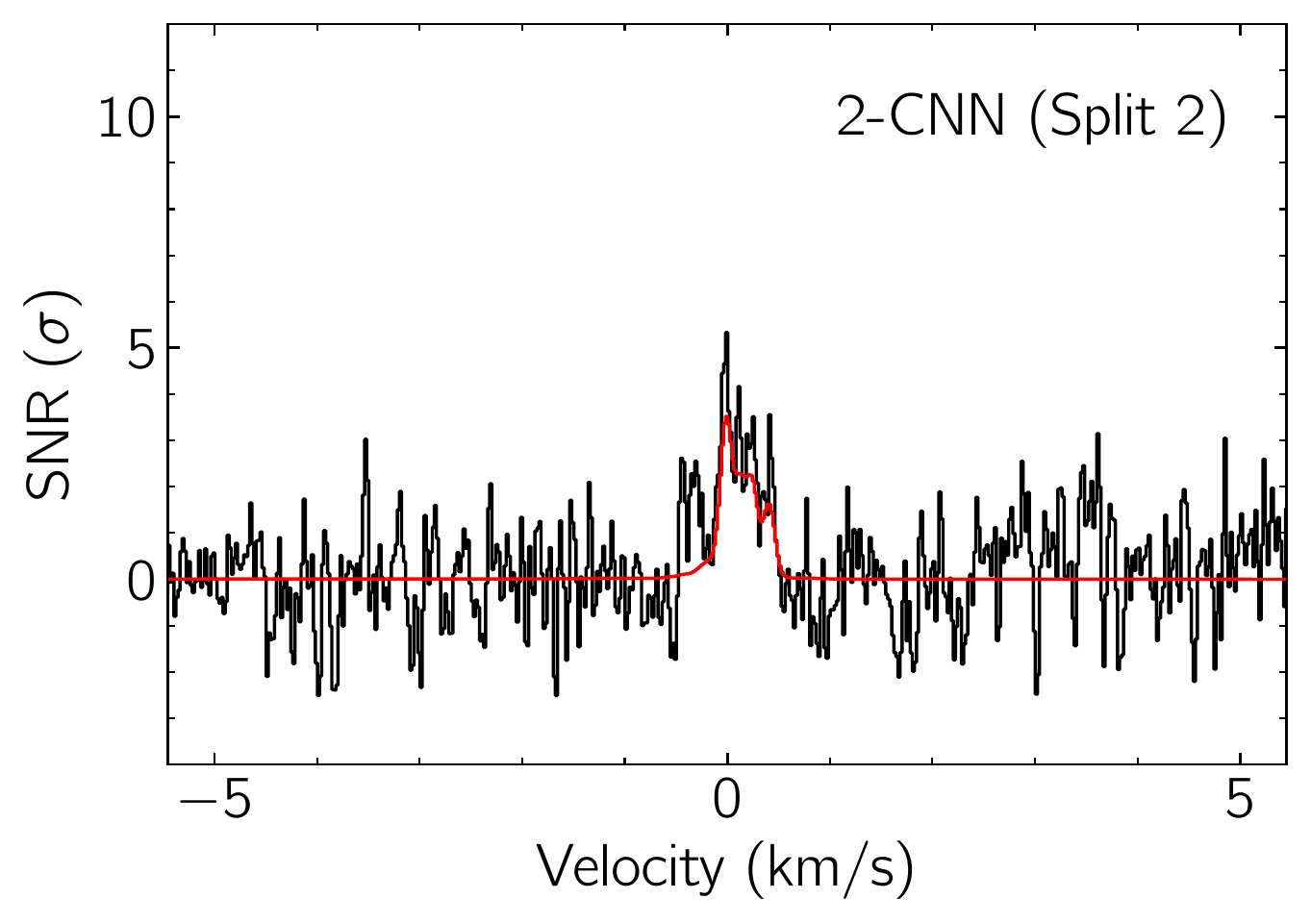}
        \includegraphics[width=0.35\textwidth]{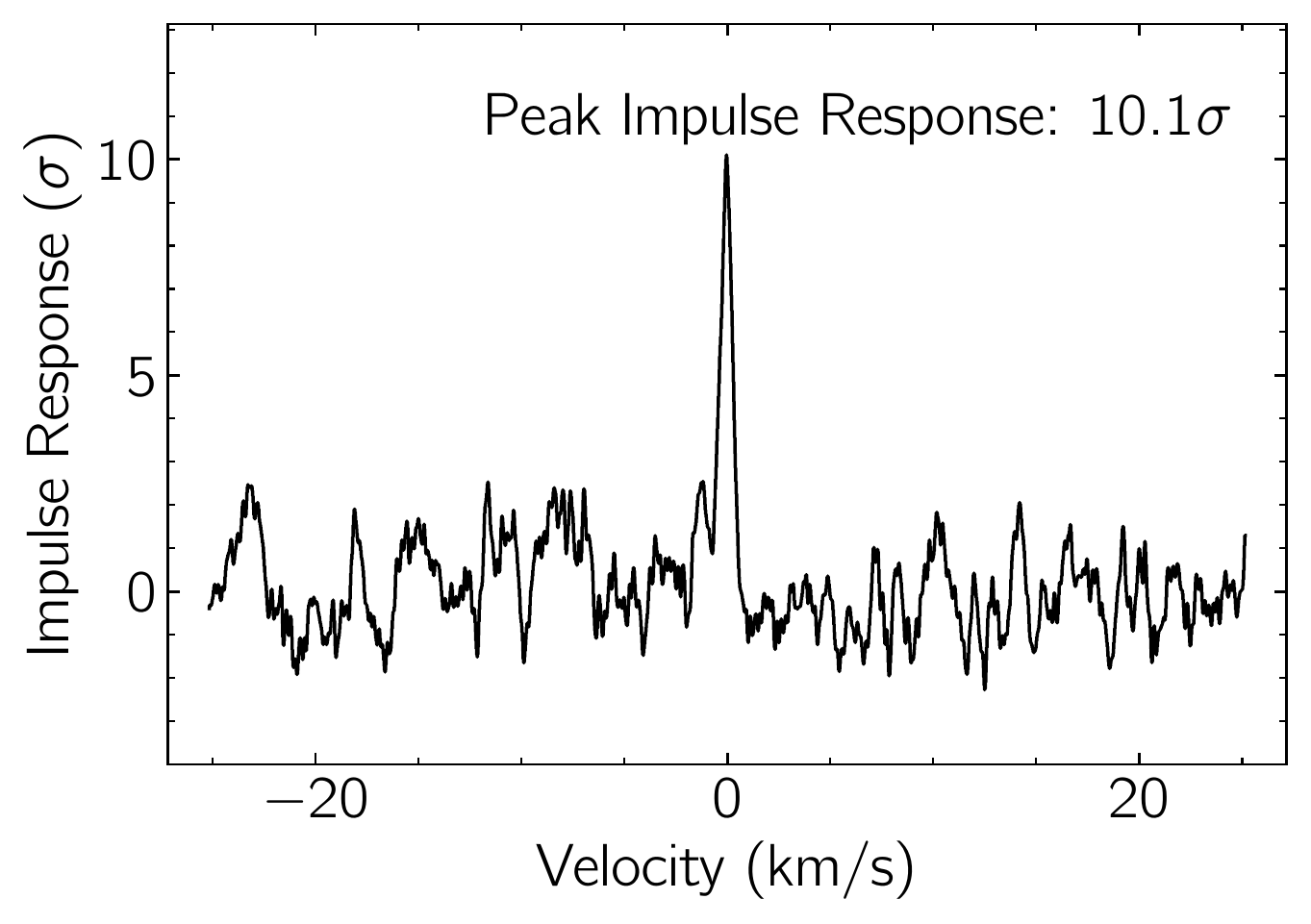}
    \includegraphics[width=0.35\textwidth]{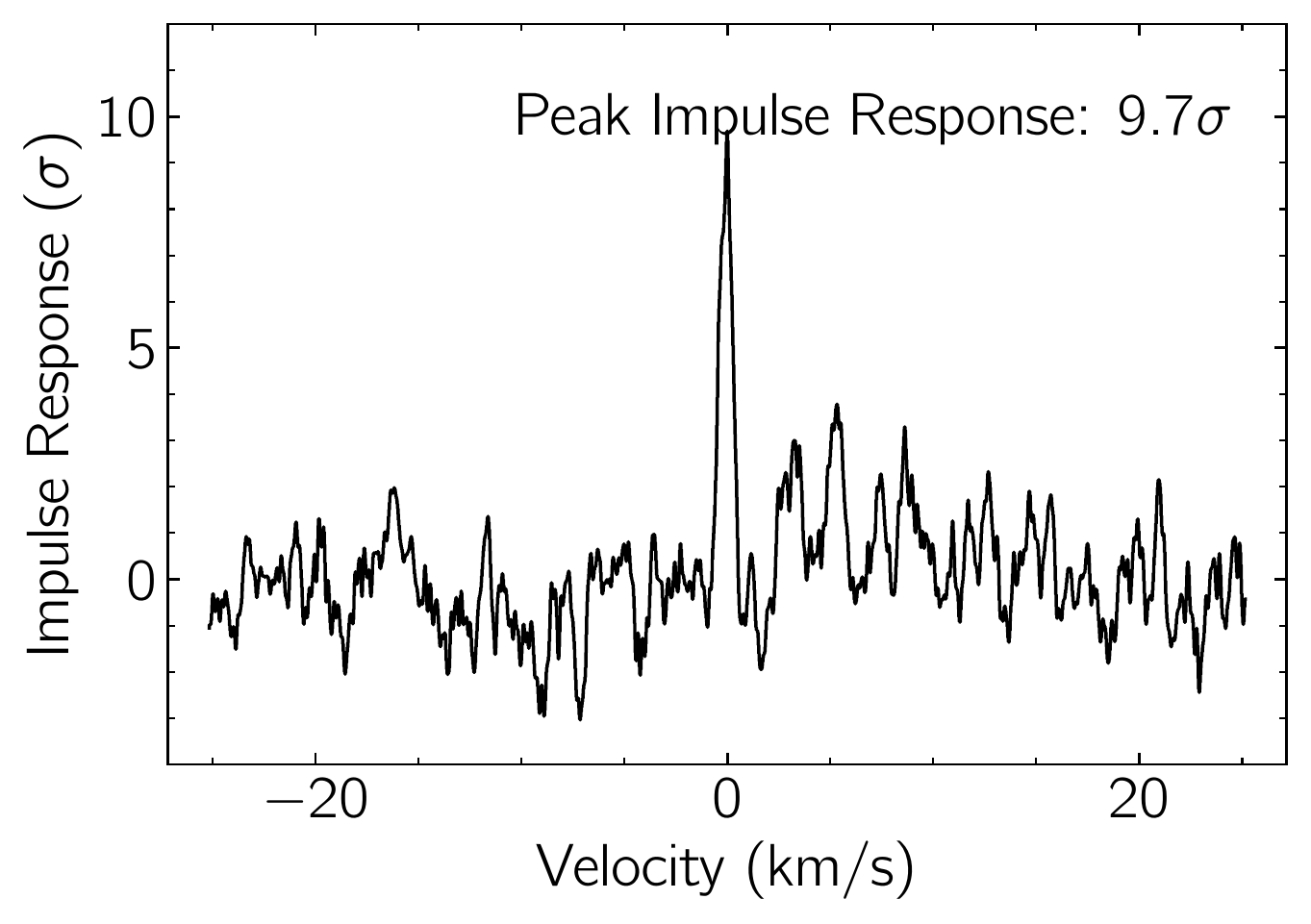}
    \caption{\emph{Upper left:} Velocity-stacked spectra of 2-cyanonaphthalene in black, with the corresponding stack of the simulation using the best-fit parameters to the individual lines in red. Every other 2-cyanonaphthalene was used for this stack (split 1), and the data have been uniformly sampled to a resolution of 0.02\,km\,s$^{-1}$.  The intensity scale is the signal-to-noise ratio of the spectrum at any given velocity. \emph{Upper right:} Same as left, but for the other half of lines (split 2). \emph{Lower left:} Impulse response function of the stacked spectrum (split 1) using the simulated line profile as a matched filter.  The intensity scale is the signal-to-noise ratio of the response function when centered at a given velocity. The peak of the impulse response function provides a minimum significance for the detection of 10.3$\sigma$. \emph{Lower right:} same as lower left, but for the other half of the lines (split 2). The peak of the impulse response function provides a minimum significance for the detection of 9.8$\sigma$.}
    \label{Fig_B1}
\end{figure*}

A jack-knifing test on 2-cyanonaphthalene is shown in Figure \ref{Fig_B1}. The set of windowed 2-cyanonaphthalene lines were split by taking every other line (treating sets of hyperfine components as a single line for this purpose). We find that the signal is quite evenly split over the two datasets, illustrating that the filter response is not a false positive, and does not arise from interloping lines.


\subsection{Co-spatial cyanopolyyne fitting results}
\label{Appendix_A}
\subsubsection{HC$_3$N}
\begin{figure*}[!b]
    \centering
    \includegraphics[width=0.8\textwidth]{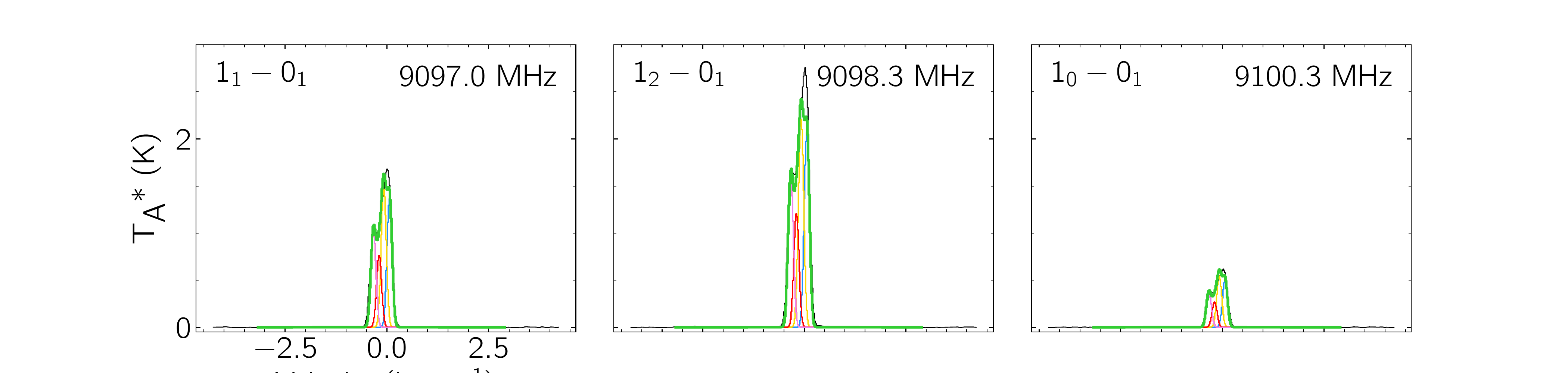}
    \includegraphics[width=0.8\textwidth]{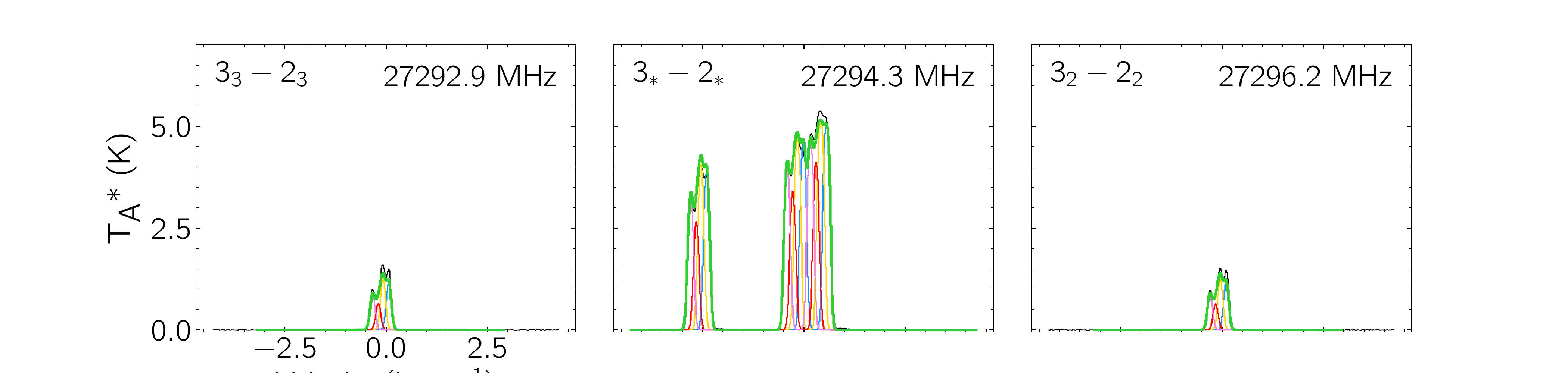}
    \caption{Individual line detections of \ce{HC3N} in the GOTHAM data.  The spectra (black) are displayed in velocity space relative to 5.8\,km\,s$^{-1}$, and using the rest frequency given in the top right of each panel. Quantum numbers are given in the top left of each panel, neglecting hyperfine splitting. The best-fit model to the data, including all velocity components, is overlaid in green.  Simulated spectra of the individual velocity components are shown in: blue (5.63\,km\,s$^{-1}$), gold (5.79\,km\,s$^{-1}$), red (5.91\,km\,s$^{-1}$), and violet (6.03\,km\,s$^{-1}$).  See Table~\ref{Table_A1}.}
    \label{Fig_A1}
\end{figure*}

\begin{figure*}[!b]
    \centering
    \includegraphics[width=0.4\textwidth]{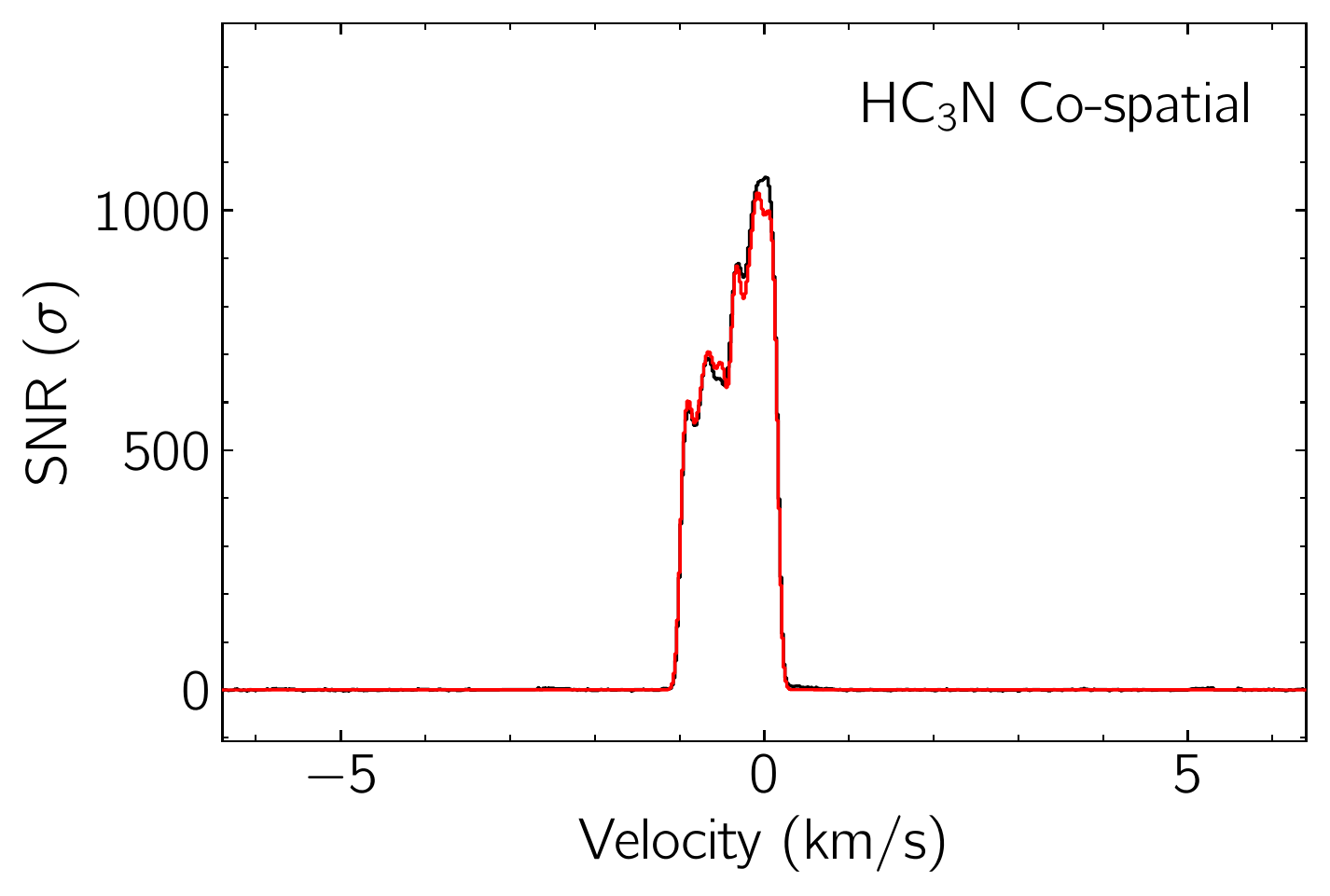}
    \includegraphics[width=0.4\textwidth]{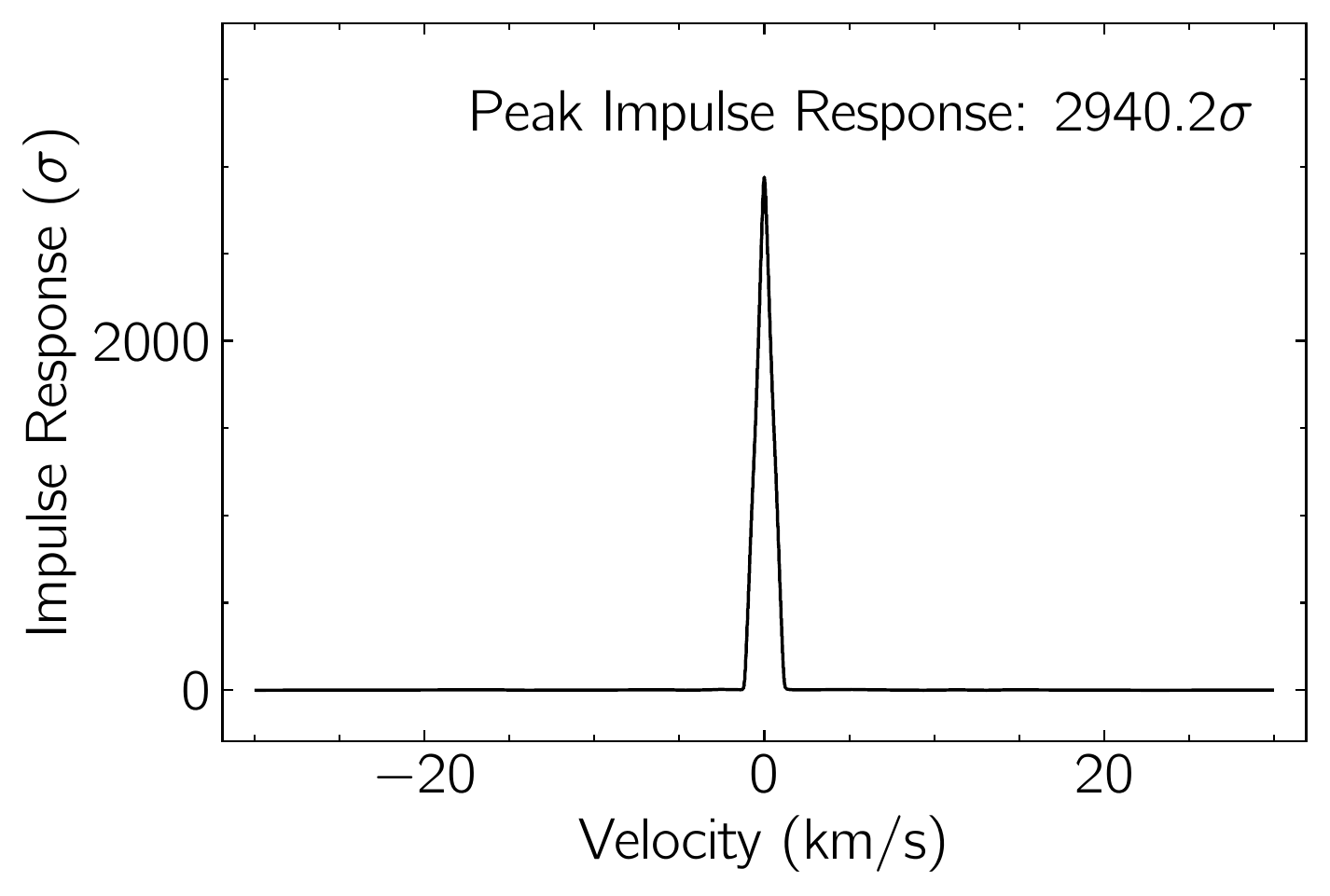}
    \caption{\emph{Left:} Velocity-stacked spectra of \ce{HC3N} in black, with the corresponding stack of the simulation using the best-fit parameters to the individual lines in red.  The data have been uniformly sampled to a resolution of 0.02\,km\,s$^{-1}$.  The intensity scale is the signal-to-noise ratio of the spectrum at any given velocity. \emph{Right:} Impulse response function of the stacked spectrum using the simulated line profile as a matched filter.  The intensity scale is the signal-to-noise ratio of the response function when centered at a given velocity.  The peak of the impulse response function provides a minimum significance for the detection of 2940.2$\sigma$.}
    \label{Fig_A2}
\end{figure*}

\begin{figure*}
    \centering
    \includegraphics[width=\textwidth]{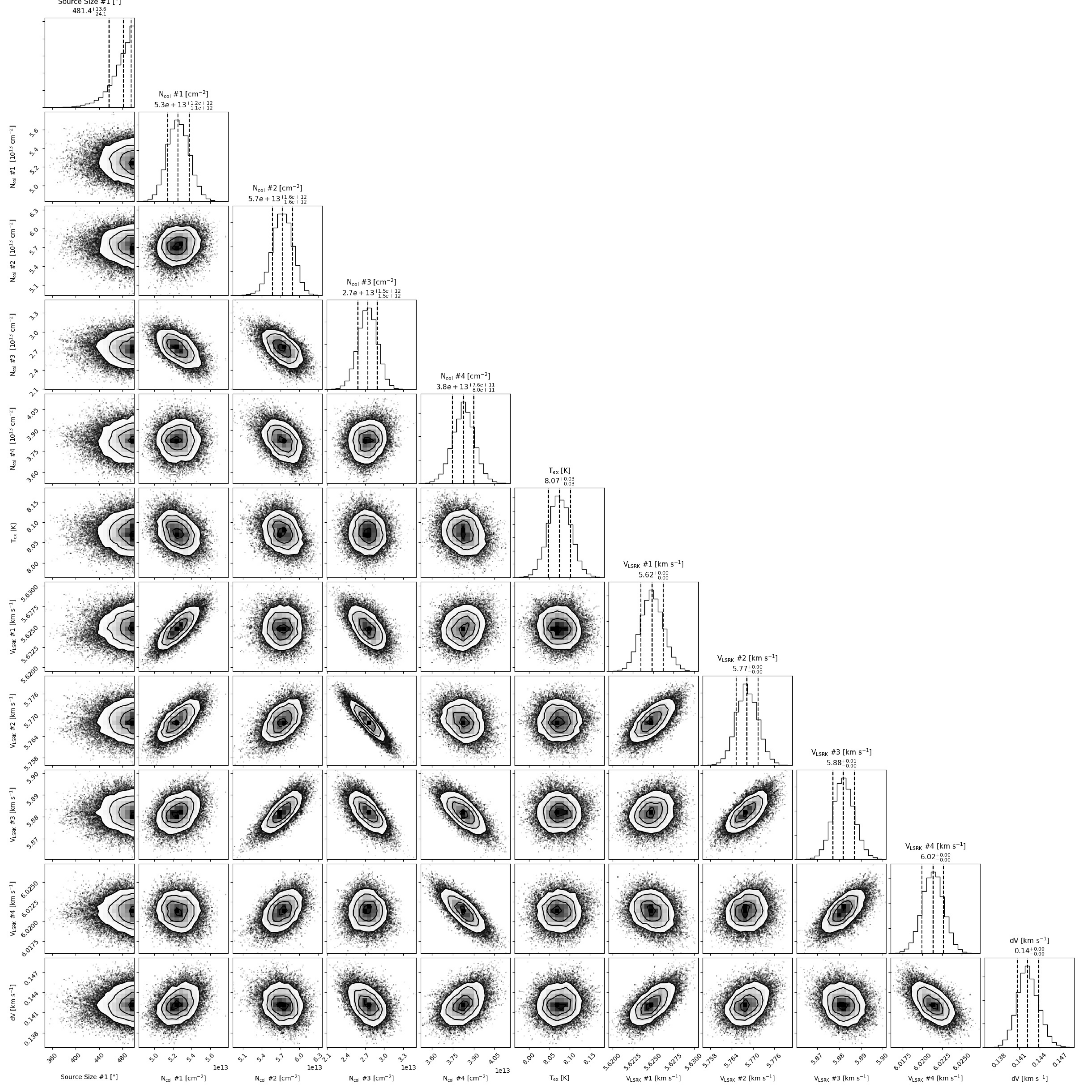}
    \caption{Parameter covariances and marginalized posterior distributions for the `co-spatial' HC$_{3}$N MCMC fit. 16$^{th}$, 50$^{th}$, and 84$^{th}$ confidence intervals (corresponding to $\pm$1 sigma for a Gaussian posterior distribution) are shown as vertical lines.}
    \label{Fig_A3}
\end{figure*}

The best-fit parameters for the MCMC analysis of \ce{HC3N}, under the `co-spatial' approximation, are given in Table~\ref{Table_A1}.  The individual detected lines are shown in Figure~\ref{Fig_A1}, while the stacked spectrum and matched filter results are shown in Figure~\ref{Fig_A2}. A corner plot of the parameter covariances for the \ce{HC3N} MCMC fit is shown in Figure~\ref{Fig_A3}.

\begin{table*}[!h]
\centering
\caption{HC$_{3}$N `co-spatial' best-fit parameters from MCMC analysis}
\begin{tabular}{c c c c c c}
\toprule
\multirow{2}{*}{Component}&	$v_{lsr}$					&	Size					&	\multicolumn{1}{c}{$N_T^\dagger$}					&	$T_{ex}$							&	$\Delta V$		\\
			&	(km s$^{-1}$)				&	($^{\prime\prime}$)		&	\multicolumn{1}{c}{(10$^{13}$ cm$^{-2}$)}		&	(K)								&	(km s$^{-1}$)	\\
\midrule
\hspace{0.1em}\vspace{-0.5em}\\
C1	&	$5.625^{+0.001}_{-0.001}$	&	\multirow{6}{*}{$481^{+13}_{-24}$}	&	$5.25^{+0.12}_{-0.11}$	&	\multirow{6}{*}{$8.1^{+0.0}_{-0.0}$}	&	\multirow{6}{*}{$0.142^{+0.002}_{-0.001}$}\\
\hspace{0.1em}\vspace{-0.5em}\\
C2	&	$5.768^{+0.003}_{-0.003}$	&		&	$5.73^{+0.16}_{-0.16}$	&		&	\\
\hspace{0.1em}\vspace{-0.5em}\\
C3	&	$5.882^{+0.005}_{-0.005}$	&		&	$2.74^{+0.15}_{-0.15}$	&		&	\\
\hspace{0.1em}\vspace{-0.5em}\\
C4	&	$6.021^{+0.001}_{-0.001}$	&		&	$3.83^{+0.08}_{-0.08}$	&		&	\\
\hspace{0.1em}\vspace{-0.5em}\\
\midrule
$N_T$ (Total)$^{\dagger\dagger}$	&	 \multicolumn{5}{c}{$1.8^{+0.05}_{-0.05}\times 10^{14}$~cm$^{-2}$}\\
\bottomrule
\end{tabular}

\begin{minipage}{0.75\textwidth}
	\footnotesize
	{Note} -- The quoted uncertainties represent the 16$^{th}$ and 84$^{th}$ percentile ($1\sigma$ for a Gaussian distribution) uncertainties.\\
	$^\dagger$Column density values are highly covariant with the derived source sizes.  The marginalized uncertainties on the column densities are therefore dominated by the largely unconstrained nature of the source sizes, and not by the signal-to-noise of the observations.\\
	$^{\dagger\dagger}$Uncertainties derived by adding the uncertainties of the individual components in quadrature.
\end{minipage}

\label{Table_A1}
\end{table*}

\clearpage
\subsubsection{HC$_5$N}
\begin{figure*}[!b]
    \centering
    \includegraphics[width=0.2\textwidth]{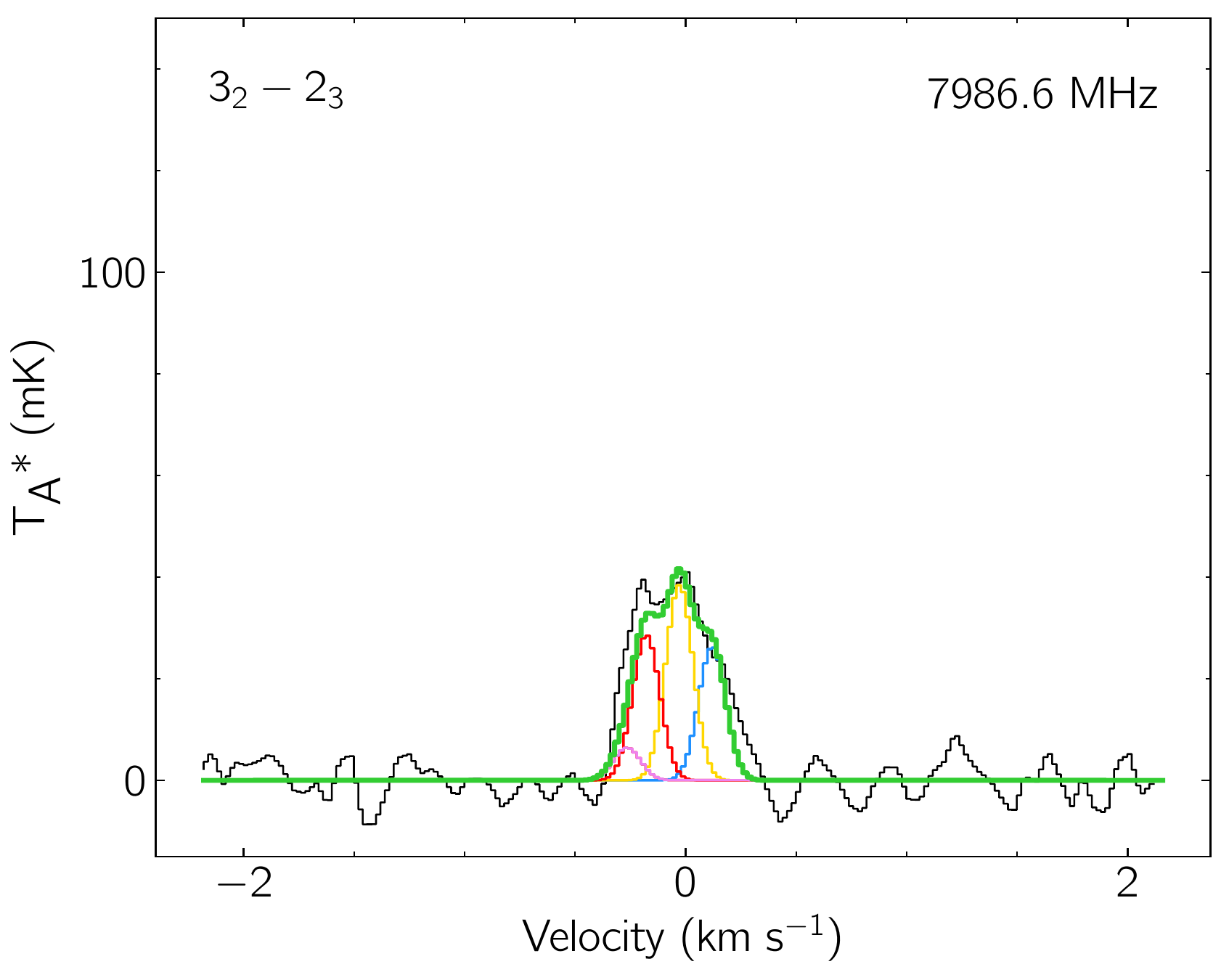}
    \includegraphics[width=0.2\textwidth]{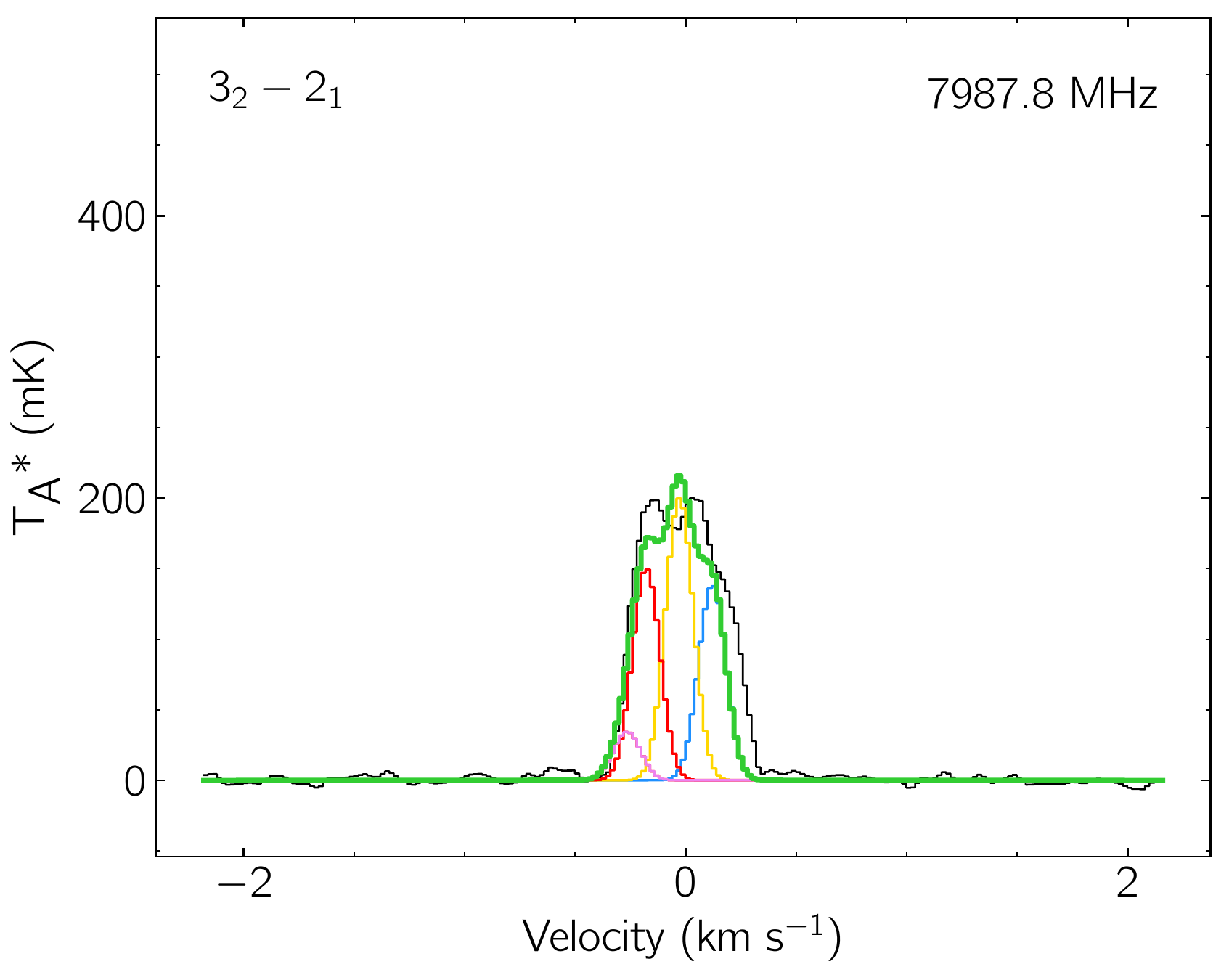}
    \includegraphics[width=0.2\textwidth]{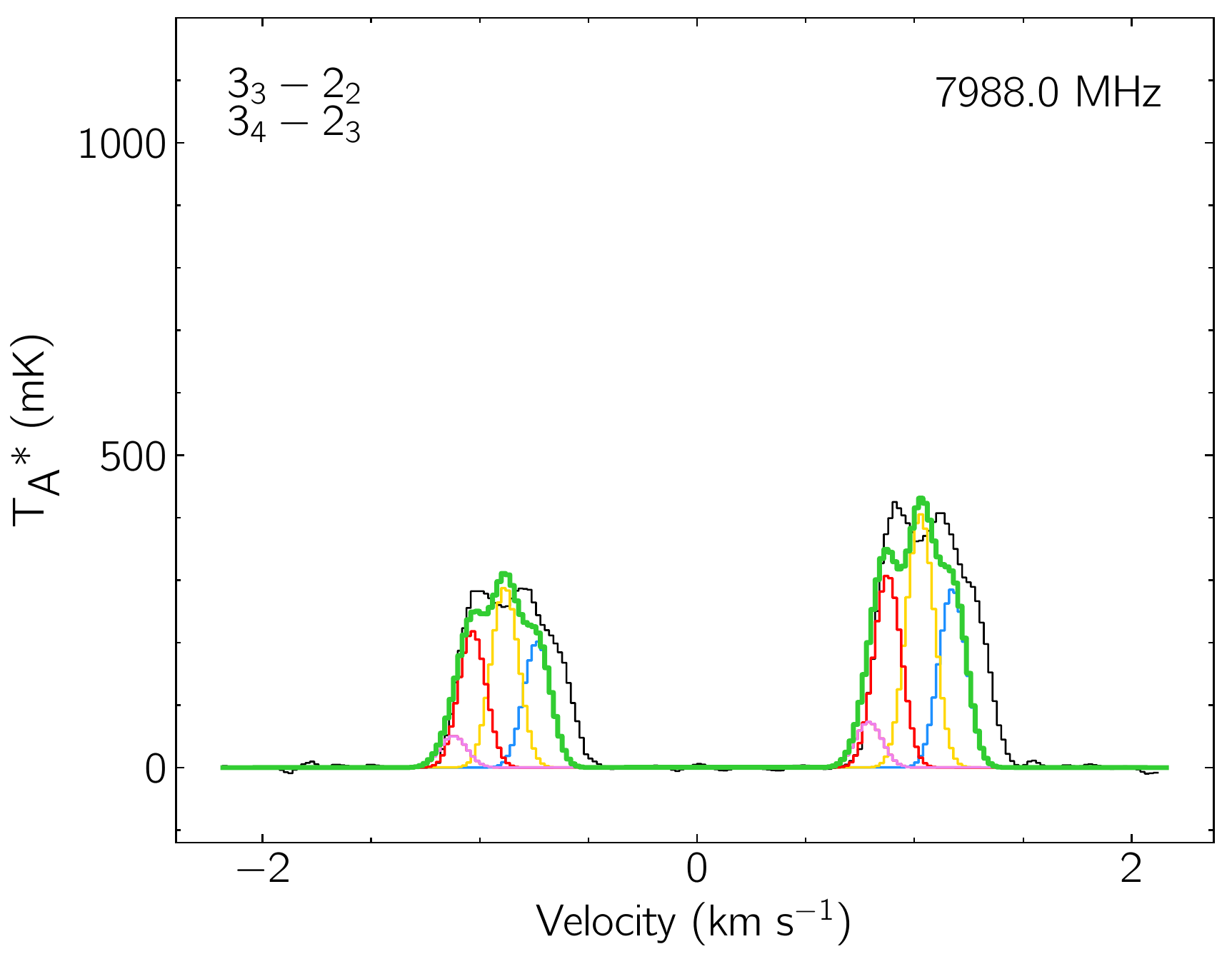}
    \includegraphics[width=0.2\textwidth]{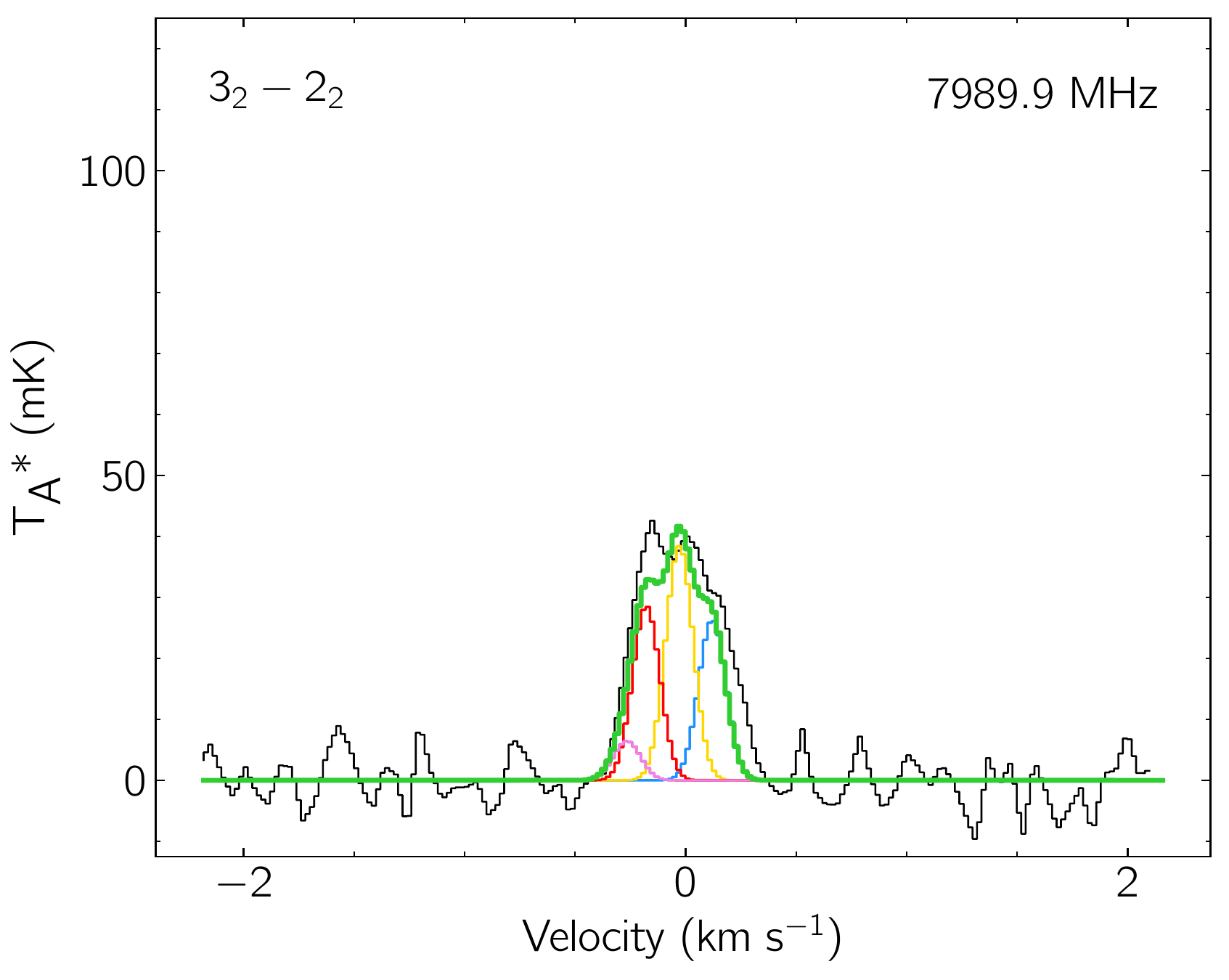}
    \includegraphics[width=0.2\textwidth]{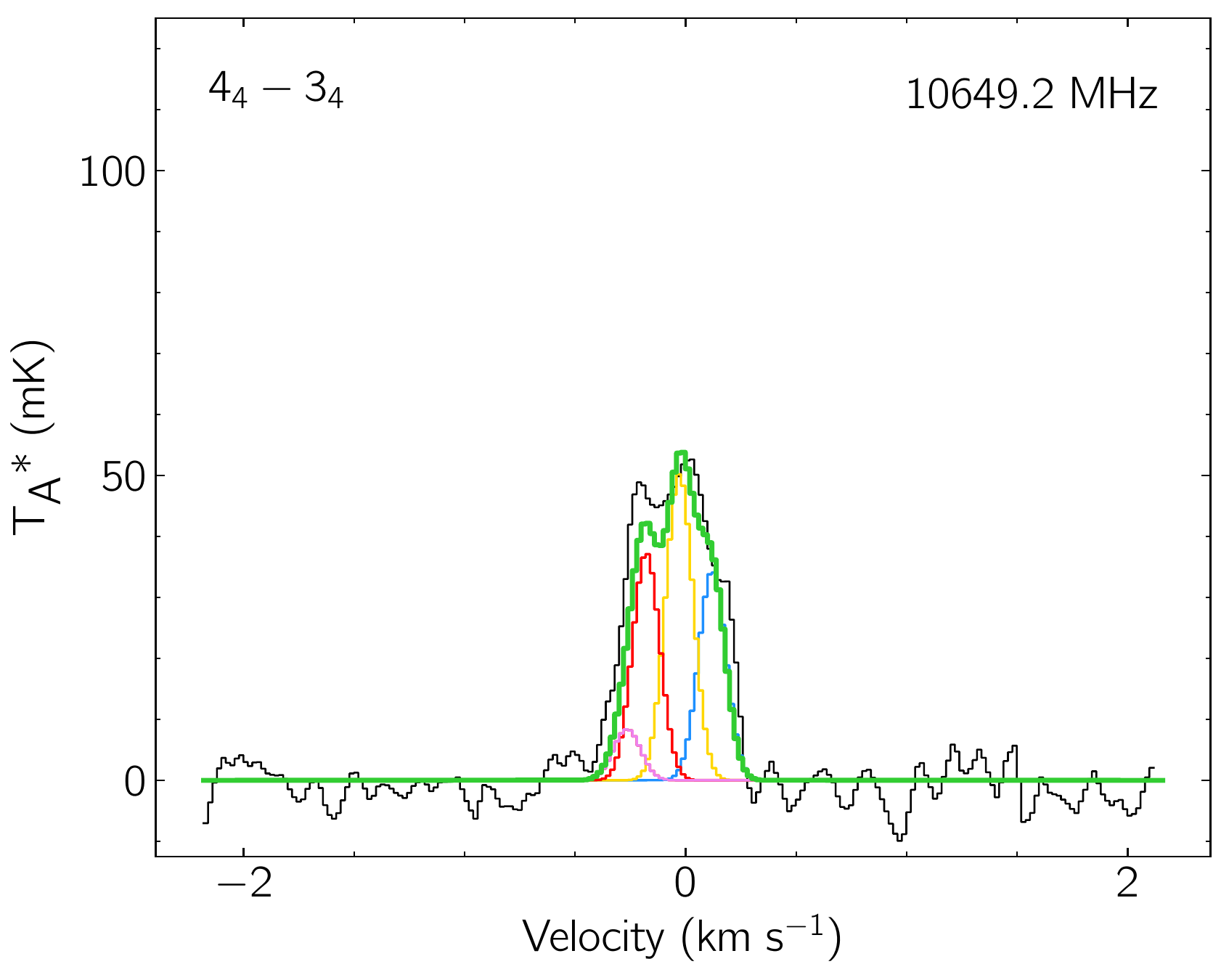}
    \includegraphics[width=0.2\textwidth]{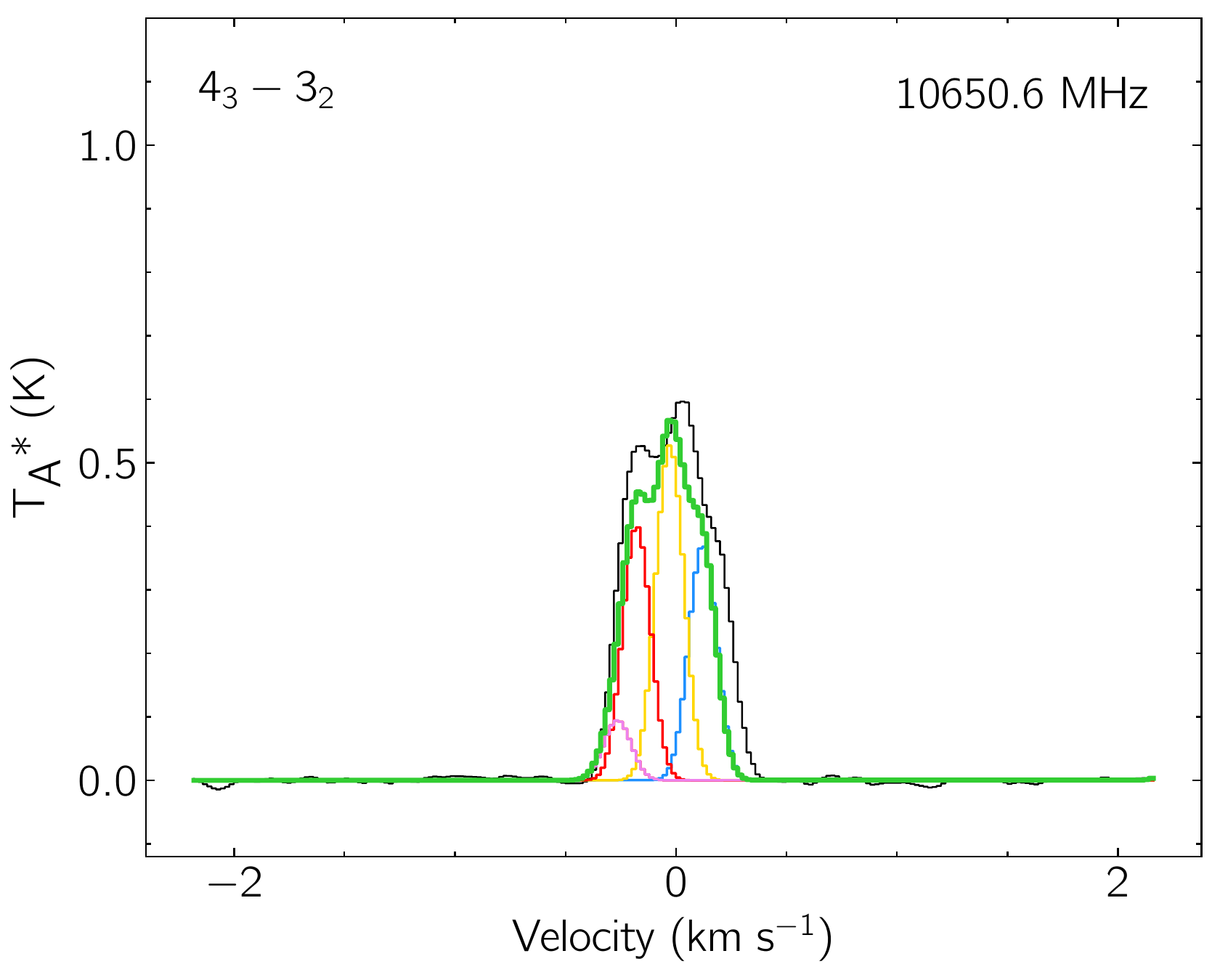}
    \includegraphics[width=0.2\textwidth]{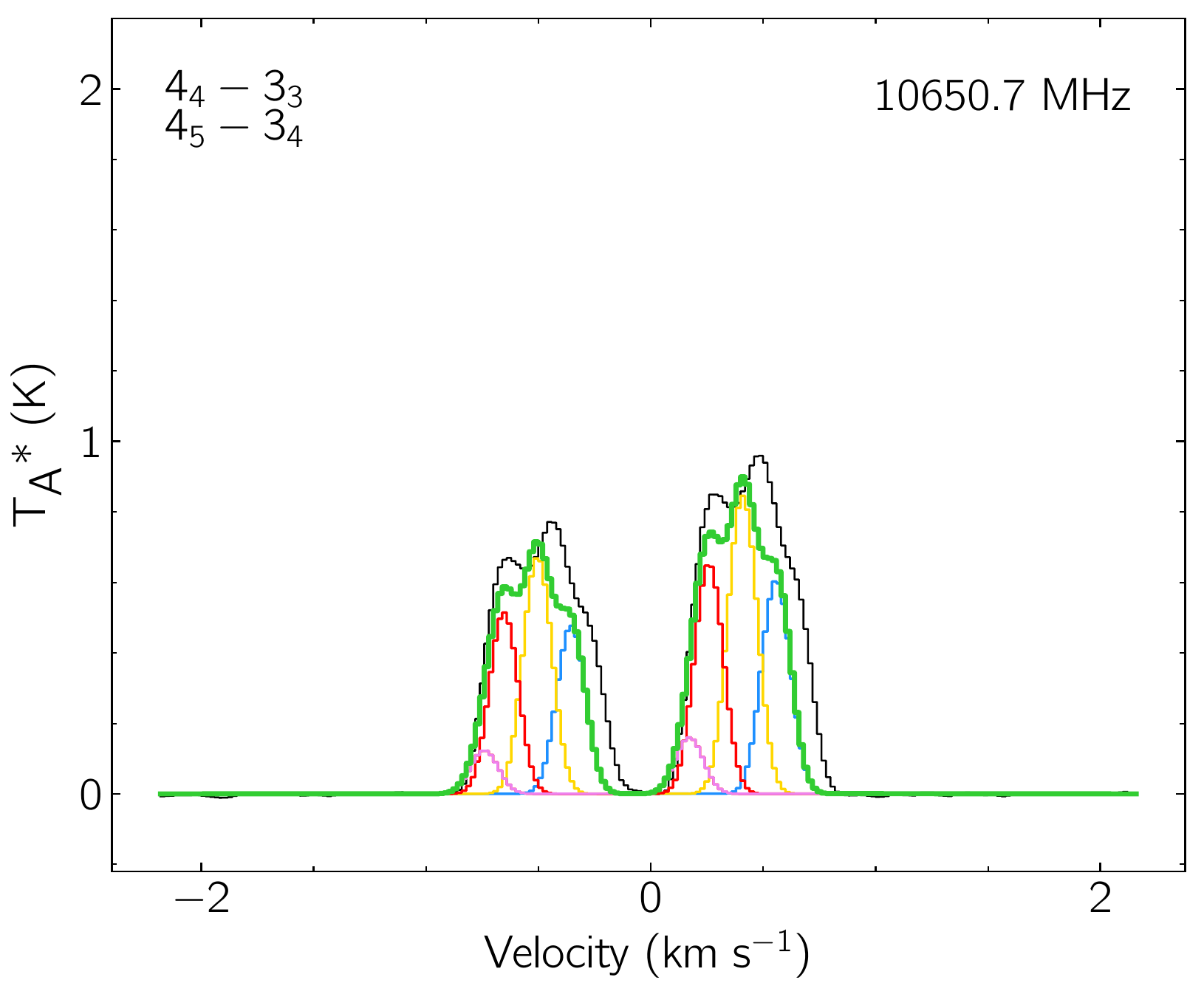}
    \includegraphics[width=0.2\textwidth]{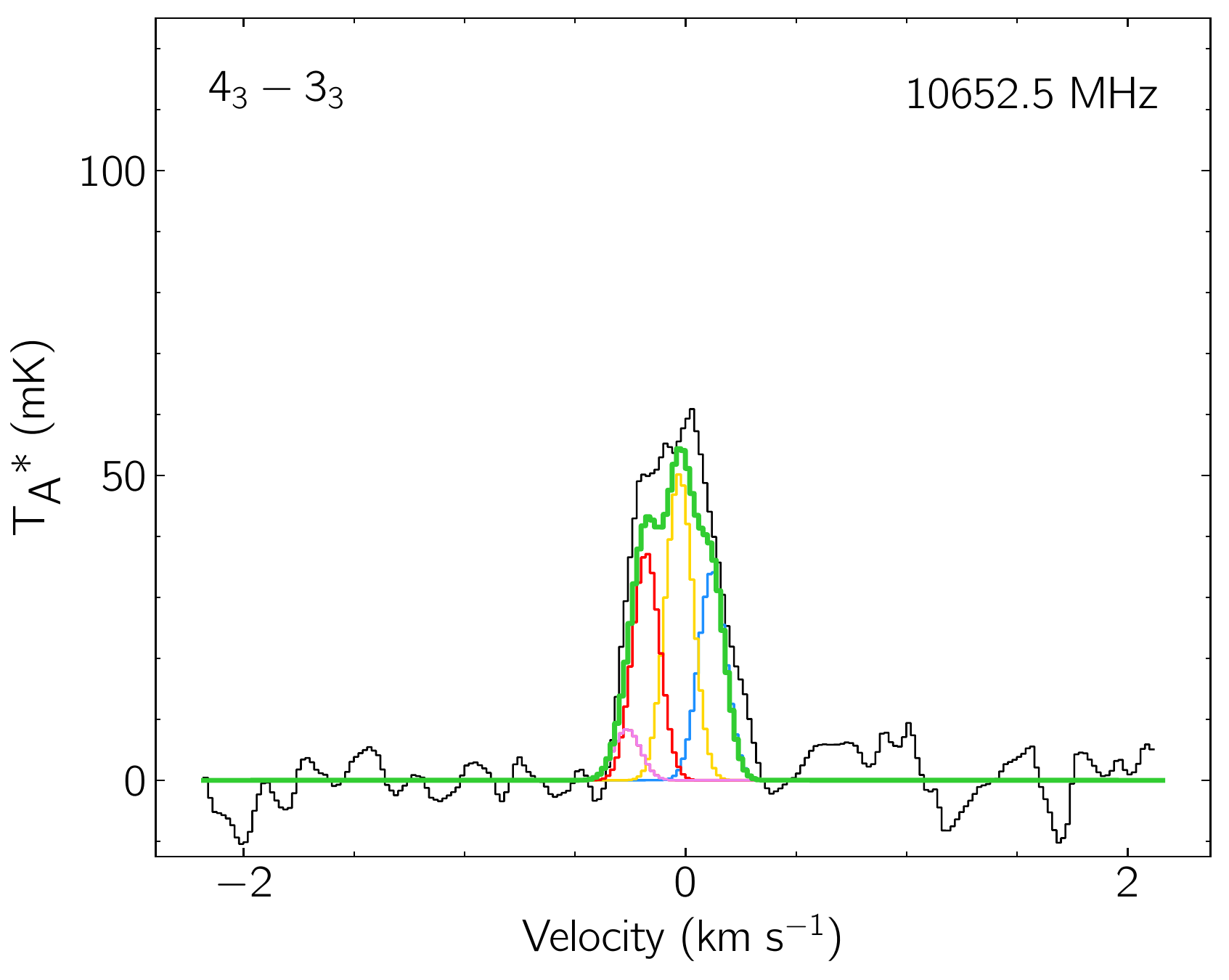}
    \includegraphics[width=0.2\textwidth]{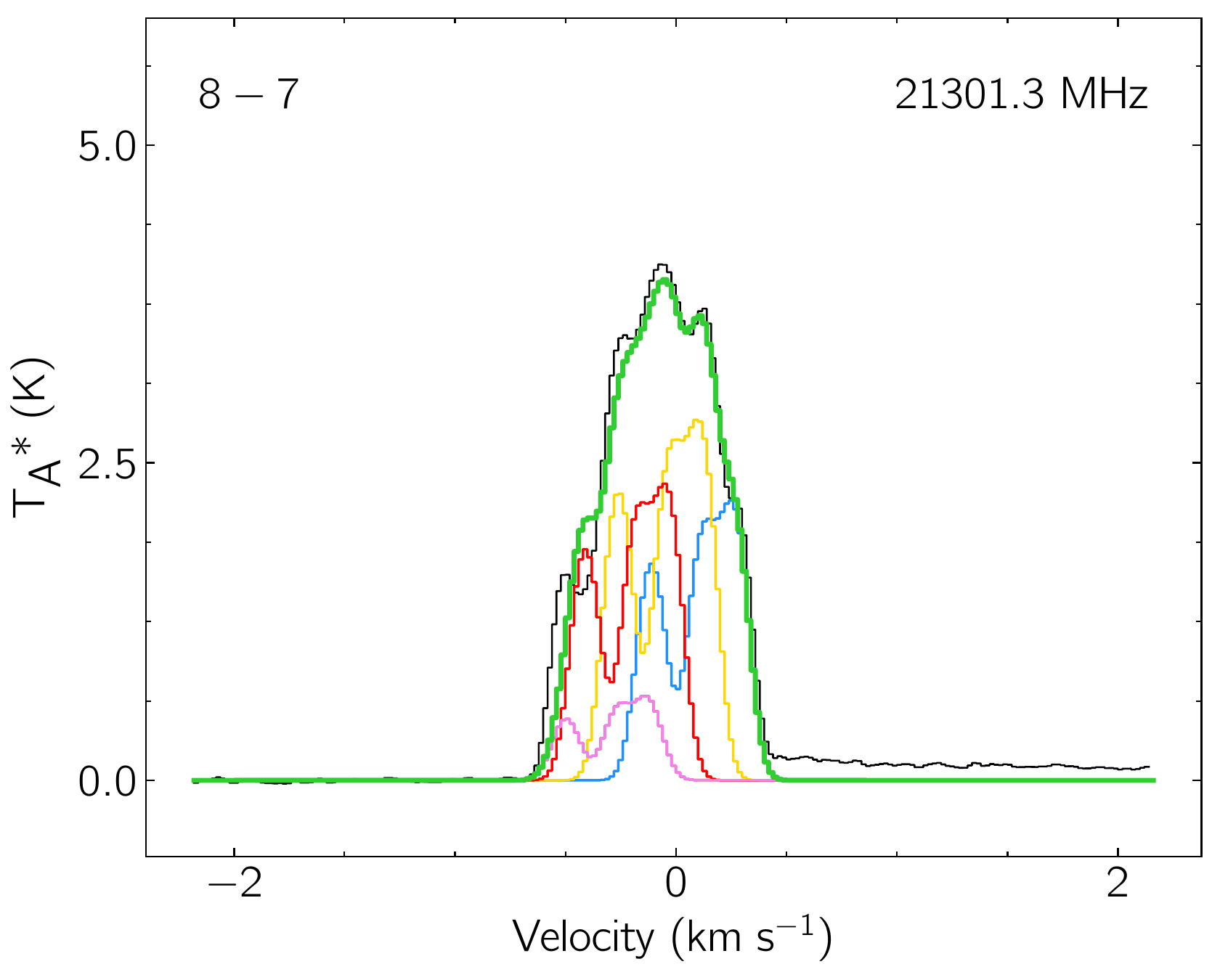}
    \includegraphics[width=0.2\textwidth]{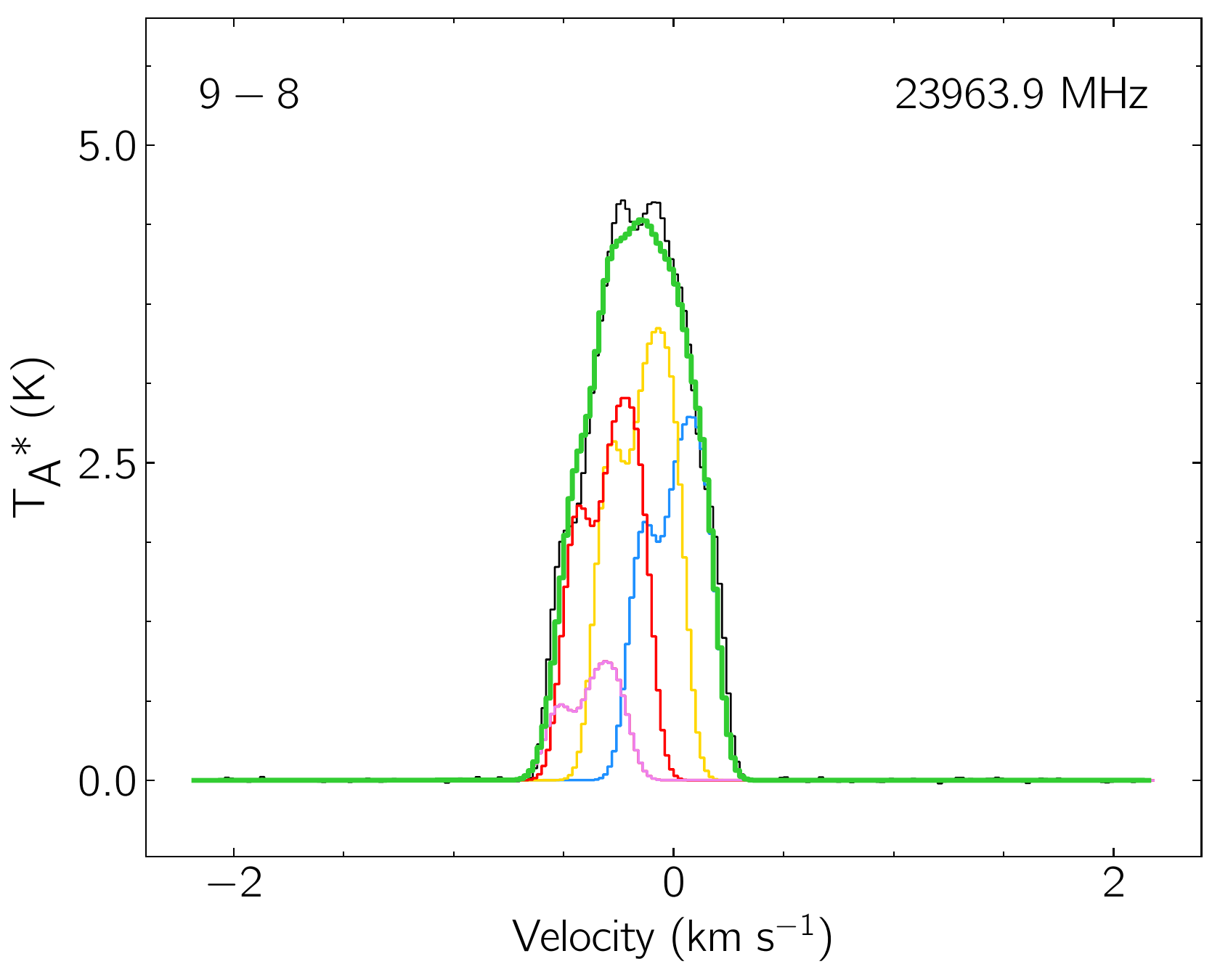}
    \includegraphics[width=0.2\textwidth]{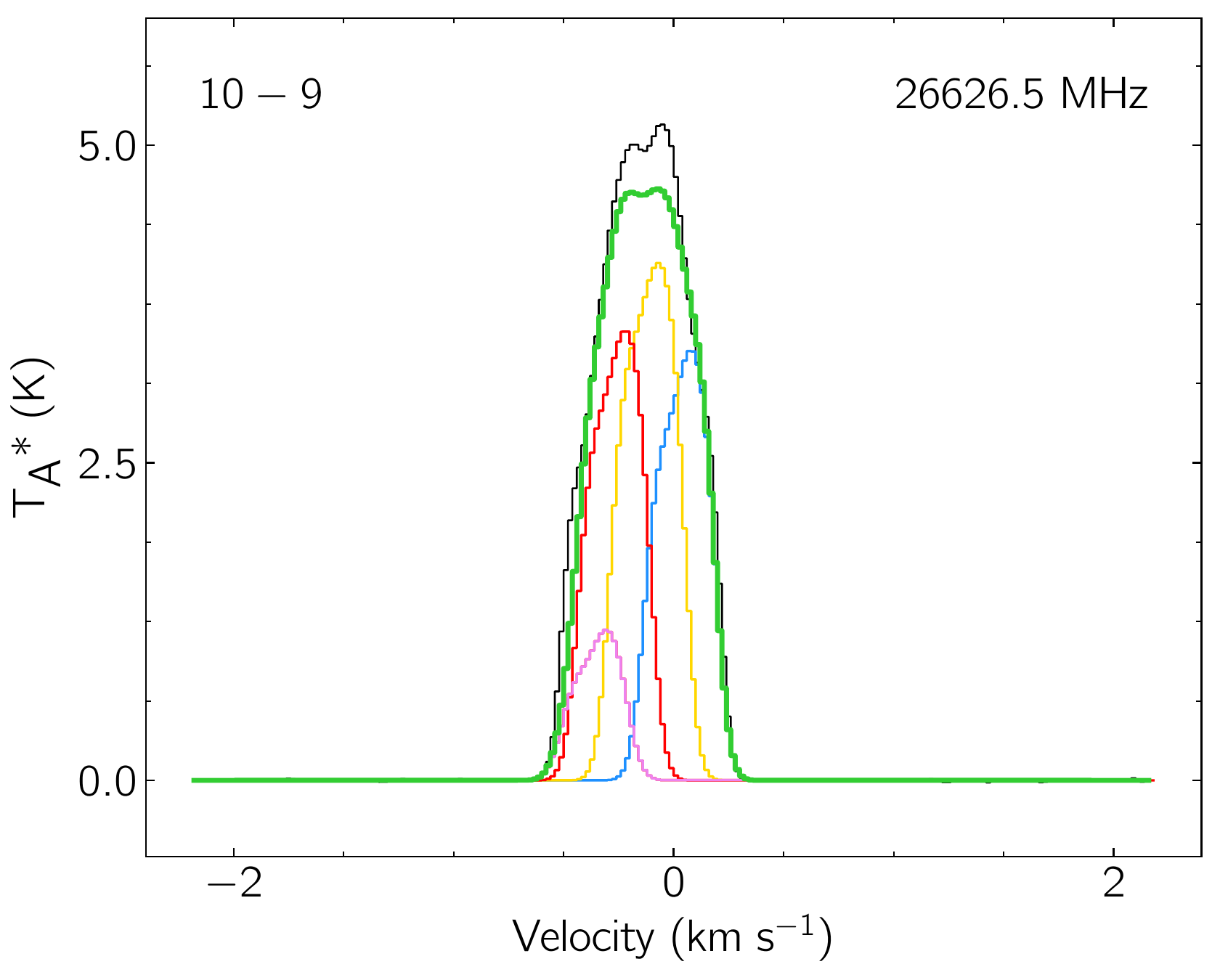}
    \includegraphics[width=0.2\textwidth]{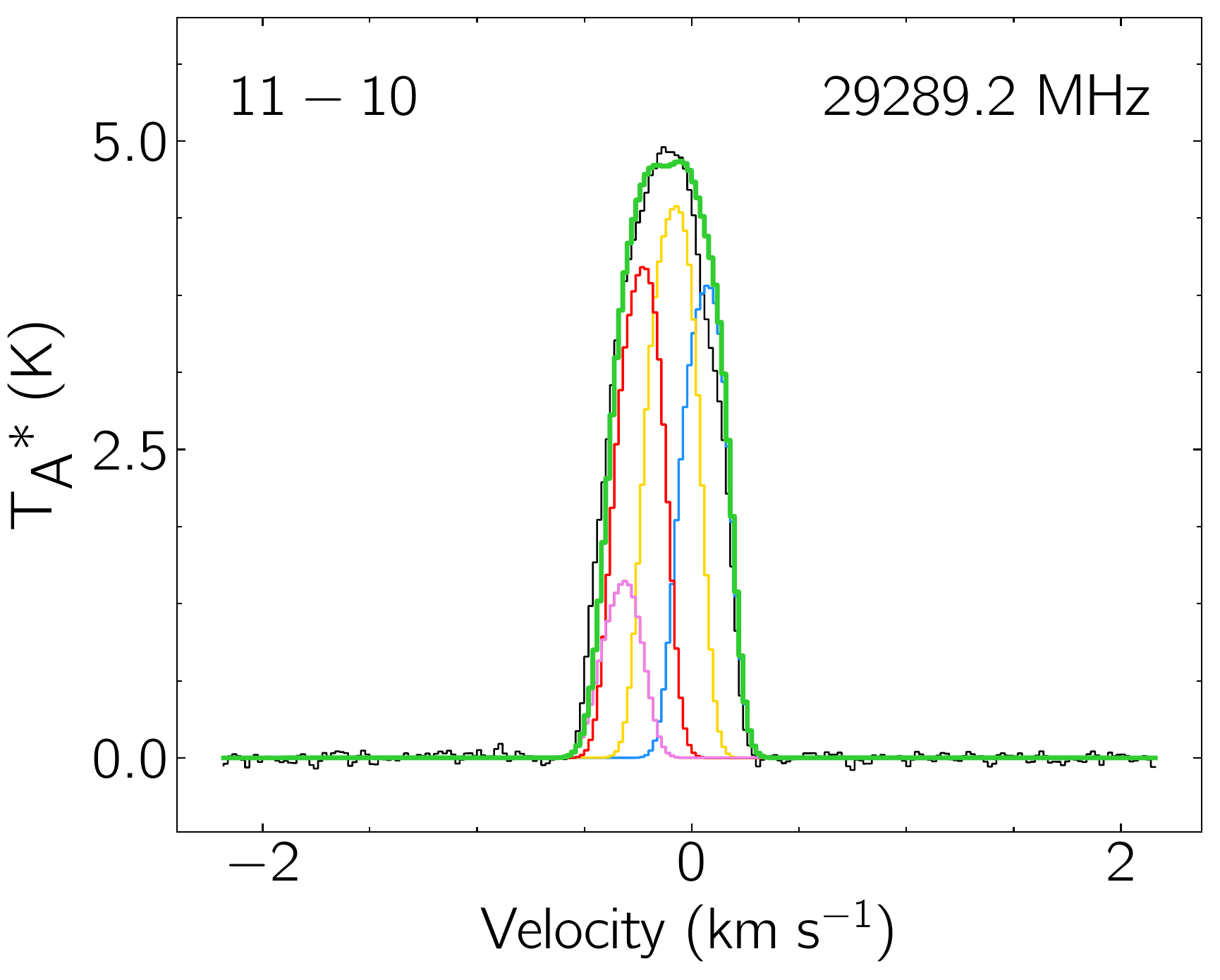}
    \caption{Individual line detections of \ce{HC5N} in the GOTHAM data.  The spectra (black) are displayed in velocity space relative to 5.8\,km\,s$^{-1}$, and using the rest frequency given in the top right of each panel. Quantum numbers are given in the top left of each panel, neglecting hyperfine splitting. The best-fit model to the data, including all velocity components, is overlaid in green.  Simulated spectra of the individual velocity components are shown in: blue (5.63\,km\,s$^{-1}$), gold (5.79\,km\,s$^{-1}$), red (5.91\,km\,s$^{-1}$), and violet (6.03\,km\,s$^{-1}$).  See Table~\ref{Table_A2}.}
    \label{Fig_A4}
\end{figure*}

\begin{figure*}[!b]
    \centering
    \includegraphics[width=0.4\textwidth]{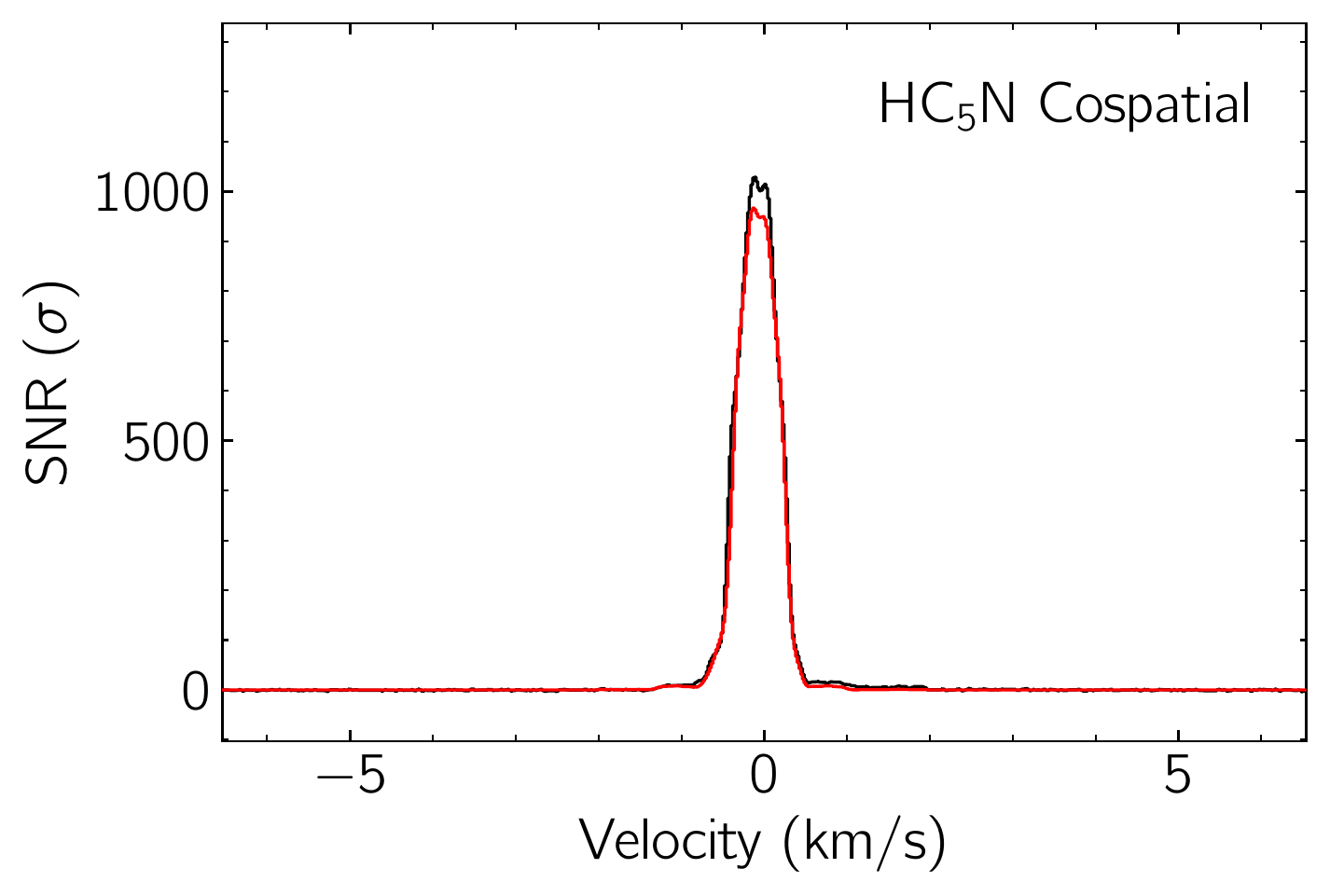}
    \includegraphics[width=0.4\textwidth]{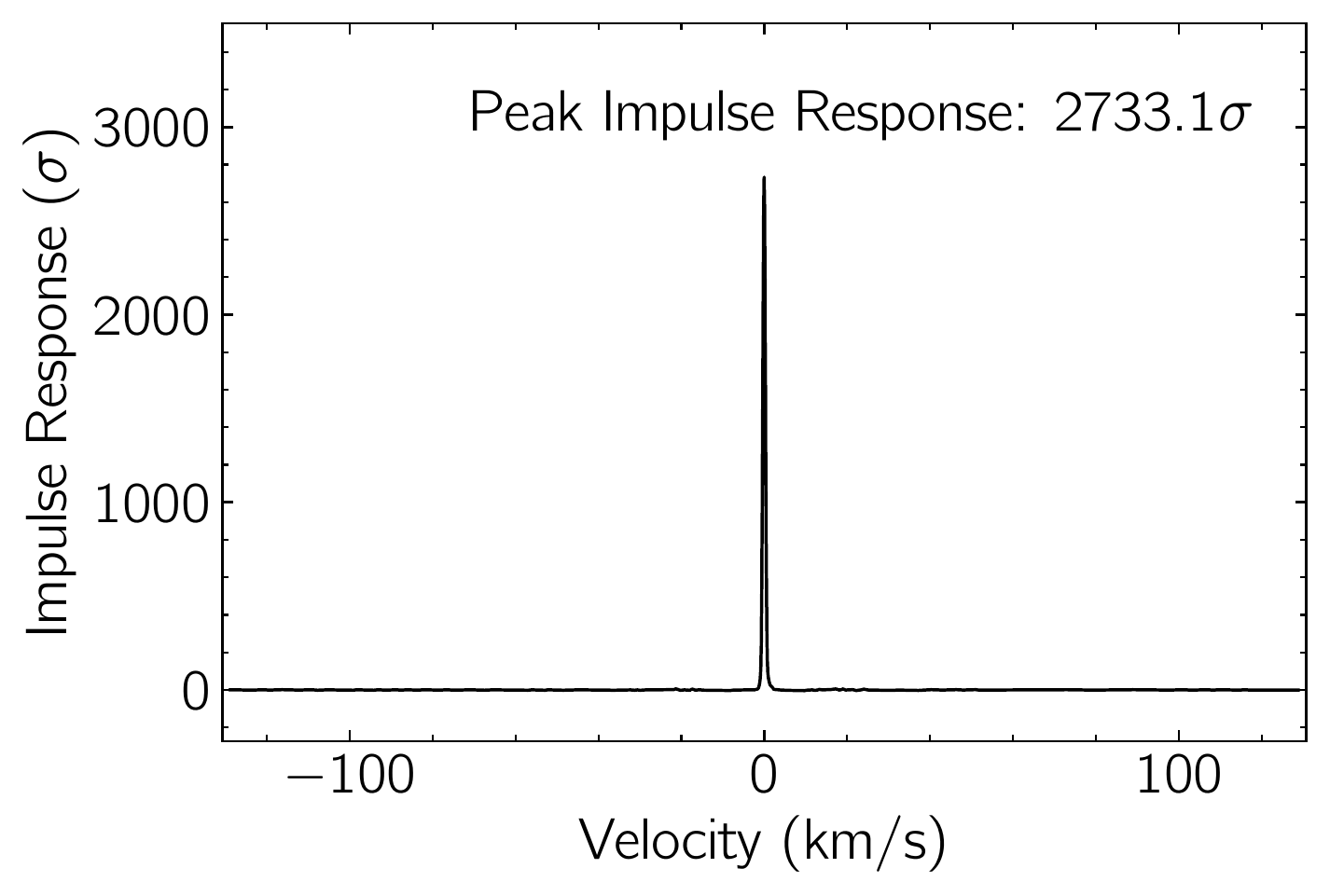}
    \caption{\emph{Left:} Velocity-stacked spectra of \ce{HC5N} in black, with the corresponding stack of the simulation using the best-fit parameters to the individual lines in red.  The data have been uniformly sampled to a resolution of 0.02\,km\,s$^{-1}$.  The intensity scale is the signal-to-noise ratio of the spectrum at any given velocity. \emph{Right:} Impulse response function of the stacked spectrum using the simulated line profile as a matched filter.  The intensity scale is the signal-to-noise ratio of the response function when centered at a given velocity.  The peak of the impulse response function provides a minimum significance for the detection of 2733.1$\sigma$.}
    \label{Fig_A5}
\end{figure*}

\begin{figure*}
    \centering
    \includegraphics[width=\textwidth]{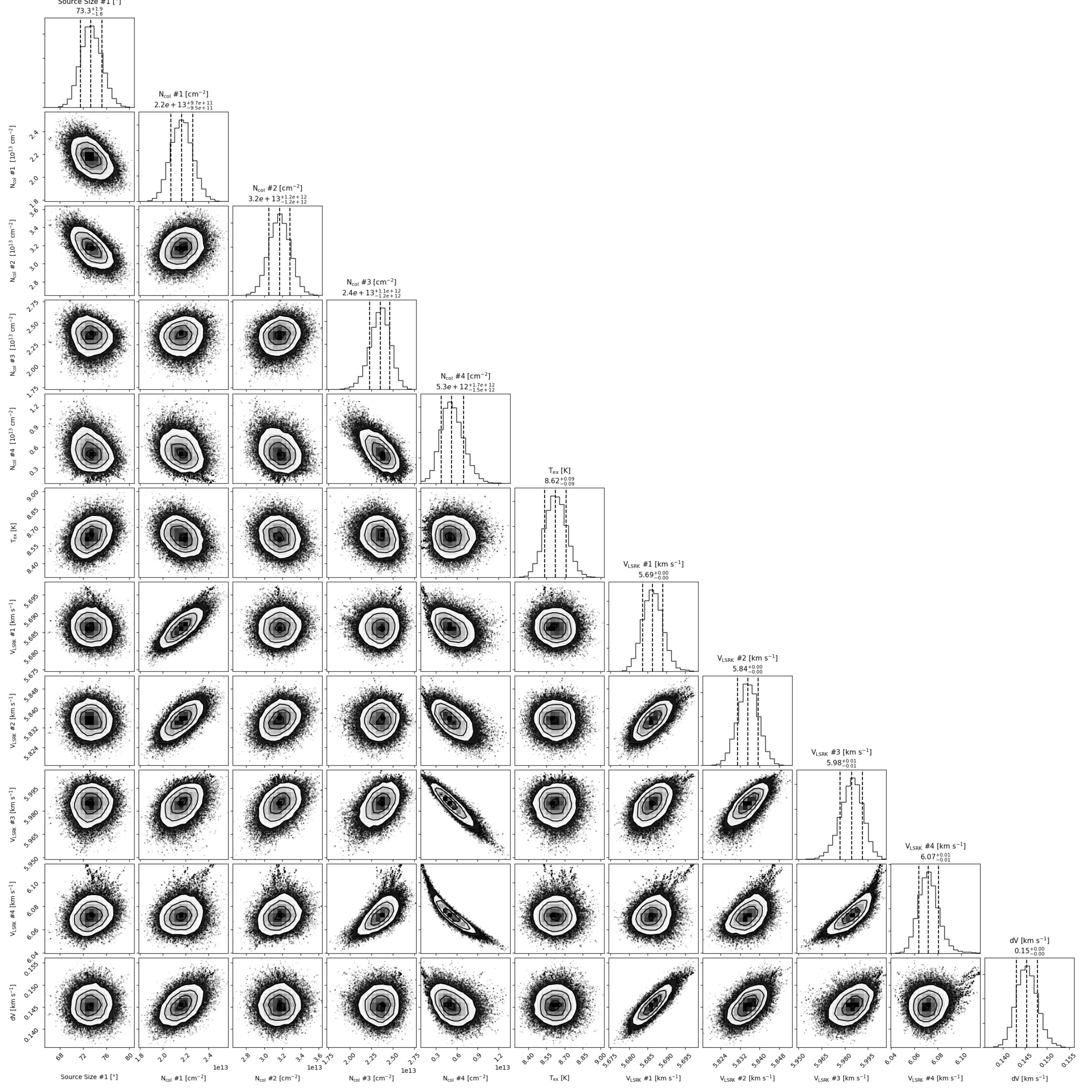}
    \caption{Parameter covariances and marginalized posterior distributions for the `co-spatial' HC$_{5}$N MCMC fit. 16$^{th}$, 50$^{th}$, and 84$^{th}$ confidence intervals (corresponding to $\pm$1 sigma for a Gaussian posterior distribution) are shown as vertical lines.}
    \label{Fig_A6}
\end{figure*}

The best-fit parameters for the MCMC analysis of \ce{HC5N}, under the `co-spatial' approximation, are given in Table~\ref{Table_A2}.  The individual detected lines are shown in Figure~\ref{Fig_A4}, while the stacked spectrum and matched filter results are shown in Figure~\ref{Fig_A5}. A corner plot of the parameter covariances for the \ce{HC5N} MCMC fit is shown in Figure~\ref{Fig_A6}.

\begin{table*}[!h]
\centering
\caption{\ce{HC5N} `co-spatial' best-fit parameters from MCMC analysis}
\begin{tabular}{c c c c c c}
\toprule
\multirow{2}{*}{Component}&	$v_{lsr}$					&	Size					&	\multicolumn{1}{c}{$N_T^\dagger$}					&	$T_{ex}$							&	$\Delta V$		\\
			&	(km s$^{-1}$)				&	($^{\prime\prime}$)		&	\multicolumn{1}{c}{(10$^{14}$ cm$^{-2}$)}		&	(K)								&	(km s$^{-1}$)	\\
\midrule
\hspace{0.1em}\vspace{-0.5em}\\
C1	&	$5.663^{+0.002}_{-0.001}$	&	\multirow{6}{*}{$128^{+8}_{-7}$}	&	$0.14^{+0.00}_{-0.00}$	&	\multirow{6}{*}{$8.7^{+0.0}_{-0.0}$}	&	\multirow{6}{*}{$0.137^{+0.002}_{-0.002}$}\\
\hspace{0.1em}\vspace{-0.5em}\\
C2	&	$5.817^{+0.002}_{-0.002}$	&		&	$0.26^{+0.01}_{-0.01}$	&		&	\\
\hspace{0.1em}\vspace{-0.5em}\\
C3	&	$5.935^{+0.005}_{-0.005}$	&		&	$0.13^{+0.01}_{-0.01}$	&		&	\\
\hspace{0.1em}\vspace{-0.5em}\\
C4	&	$6.065^{+0.002}_{-0.002}$	&		&	$0.14^{+0.00}_{-0.00}$	&		&	\\
\hspace{0.1em}\vspace{-0.5em}\\
\midrule
$N_T$ (Total)$^{\dagger\dagger}$	&	 \multicolumn{5}{c}{$6.69^{+0.13}_{-0.13}\times 10^{13}$~cm$^{-2}$}\\
\bottomrule
\end{tabular}

\begin{minipage}{0.75\textwidth}
	\footnotesize
	{Note} -- The quoted uncertainties represent the 16$^{th}$ and 84$^{th}$ percentile ($1\sigma$ for a Gaussian distribution) uncertainties.\\
	$^\dagger$Column density values are highly covariant with the derived source sizes.  The marginalized uncertainties on the column densities are therefore dominated by the largely unconstrained nature of the source sizes, and not by the signal-to-noise of the observations.\\
	$^{\dagger\dagger}$Uncertainties derived by adding the uncertainties of the individual components in quadrature.
\end{minipage}

\label{Table_A2}
\end{table*}

\clearpage
\subsubsection{HC$_7$N}
\begin{figure*}[!b]
    \centering
    \includegraphics[width=0.8\textwidth]{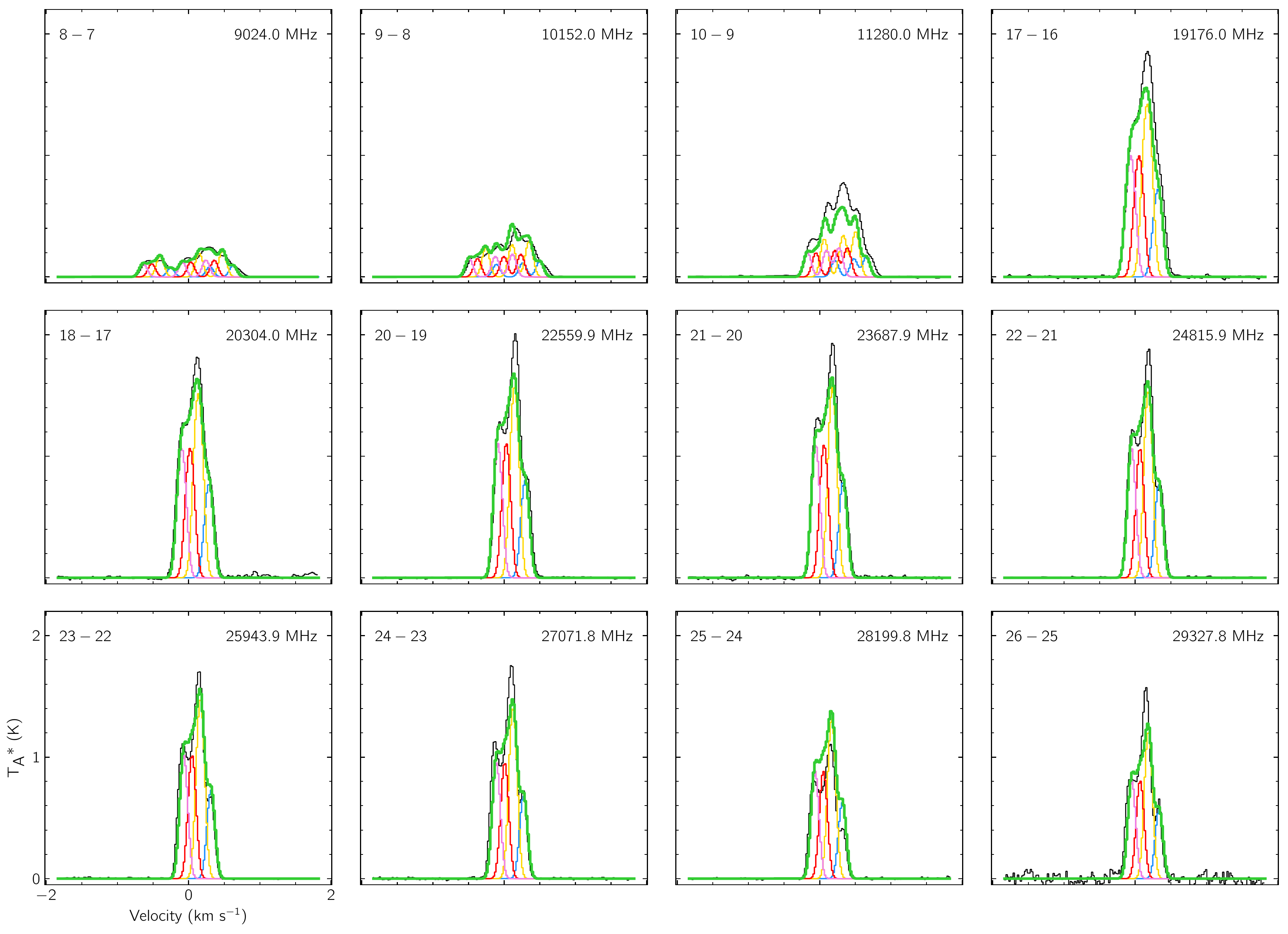}
    \caption{Individual line detections of \ce{HC7N} in the GOTHAM data.  The spectra (black) are displayed in velocity space relative to 5.8\,km\,s$^{-1}$, and using the rest frequency given in the top right of each panel. Quantum numbers are given in the top left of each panel, neglecting hyperfine splitting. The best-fit model to the data, including all velocity components, is overlaid in green.  Simulated spectra of the individual velocity components are shown in: blue (5.63\,km\,s$^{-1}$), gold (5.79\,km\,s$^{-1}$), red (5.91\,km\,s$^{-1}$), and violet (6.03\,km\,s$^{-1}$).  See Table~\ref{Table_A3}.}
    \label{Fig_A7}
\end{figure*}

\begin{figure*}[!b]
    \centering
    \includegraphics[width=0.4\textwidth]{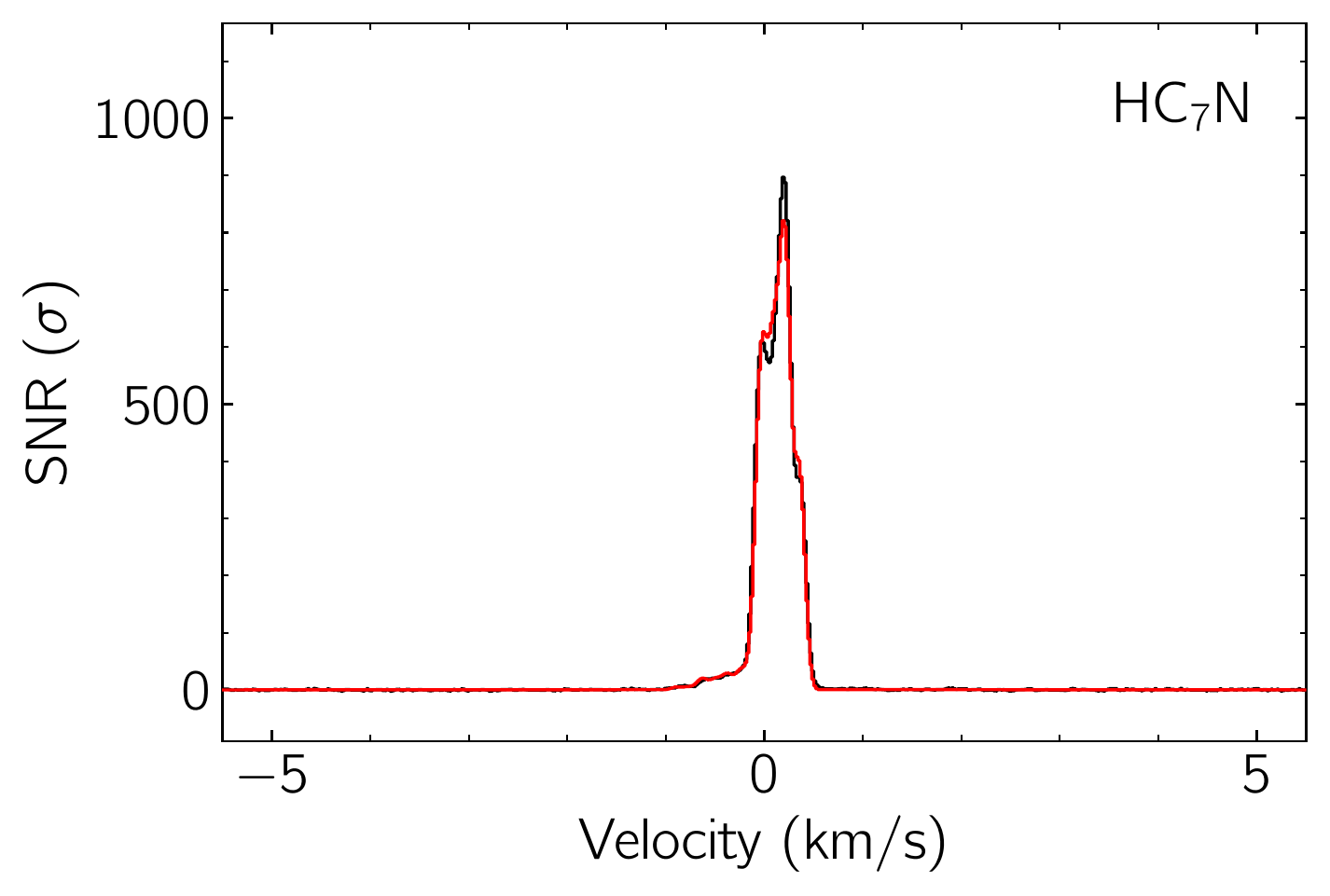}
    \includegraphics[width=0.4\textwidth]{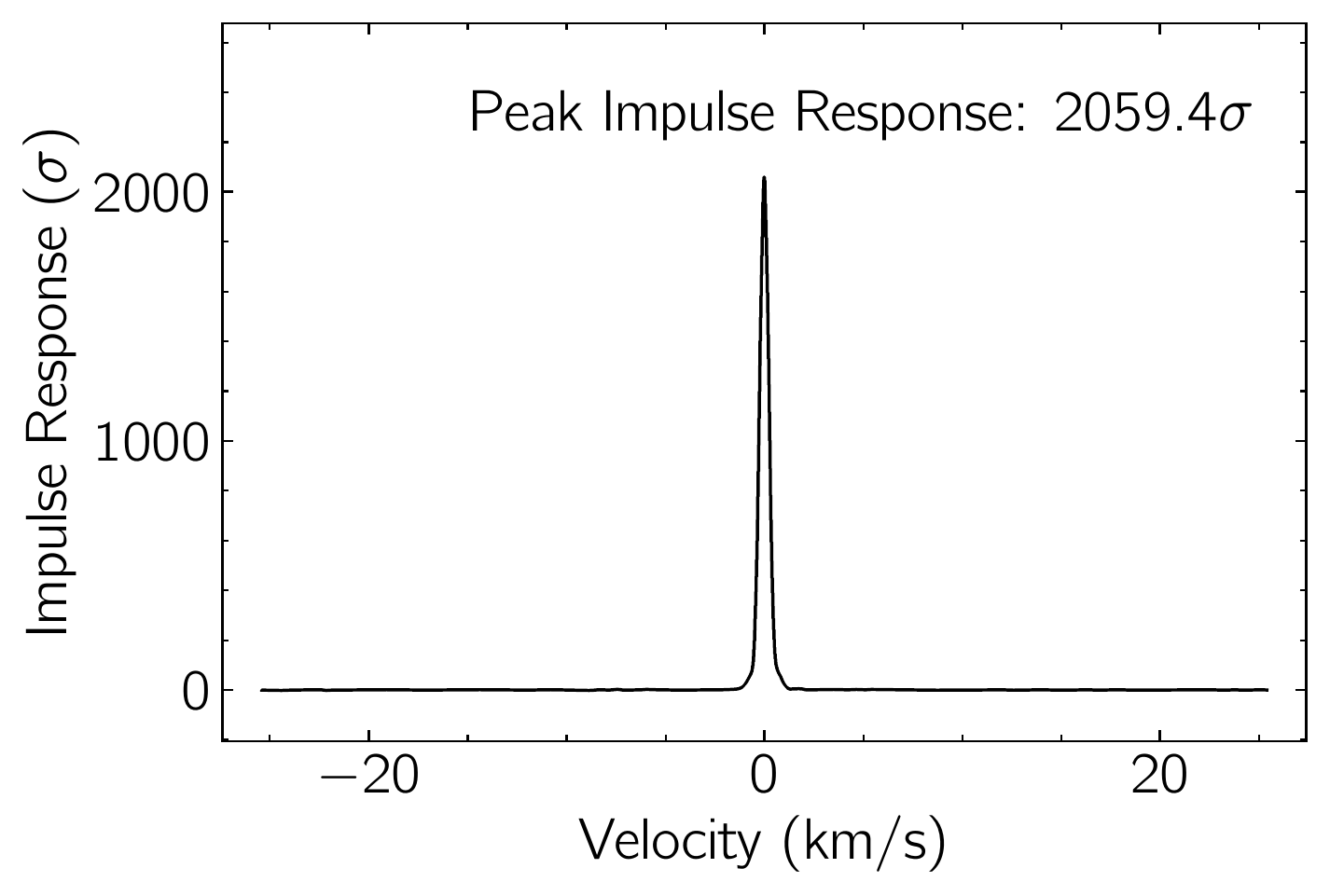}
    \caption{\emph{Left:} Velocity-stacked spectra of \ce{HC7N} in black, with the corresponding stack of the simulation using the best-fit parameters to the individual lines in red.  The data have been uniformly sampled to a resolution of 0.02\,km\,s$^{-1}$.  The intensity scale is the signal-to-noise ratio of the spectrum at any given velocity. \emph{Right:} Impulse response function of the stacked spectrum using the simulated line profile as a matched filter.  The intensity scale is the signal-to-noise ratio of the response function when centered at a given velocity.  The peak of the impulse response function provides a minimum significance for the detection of 2059.4$\sigma$.}
    \label{Fig_A8}
\end{figure*}

\begin{figure*}
    \centering
    \includegraphics[width=\textwidth]{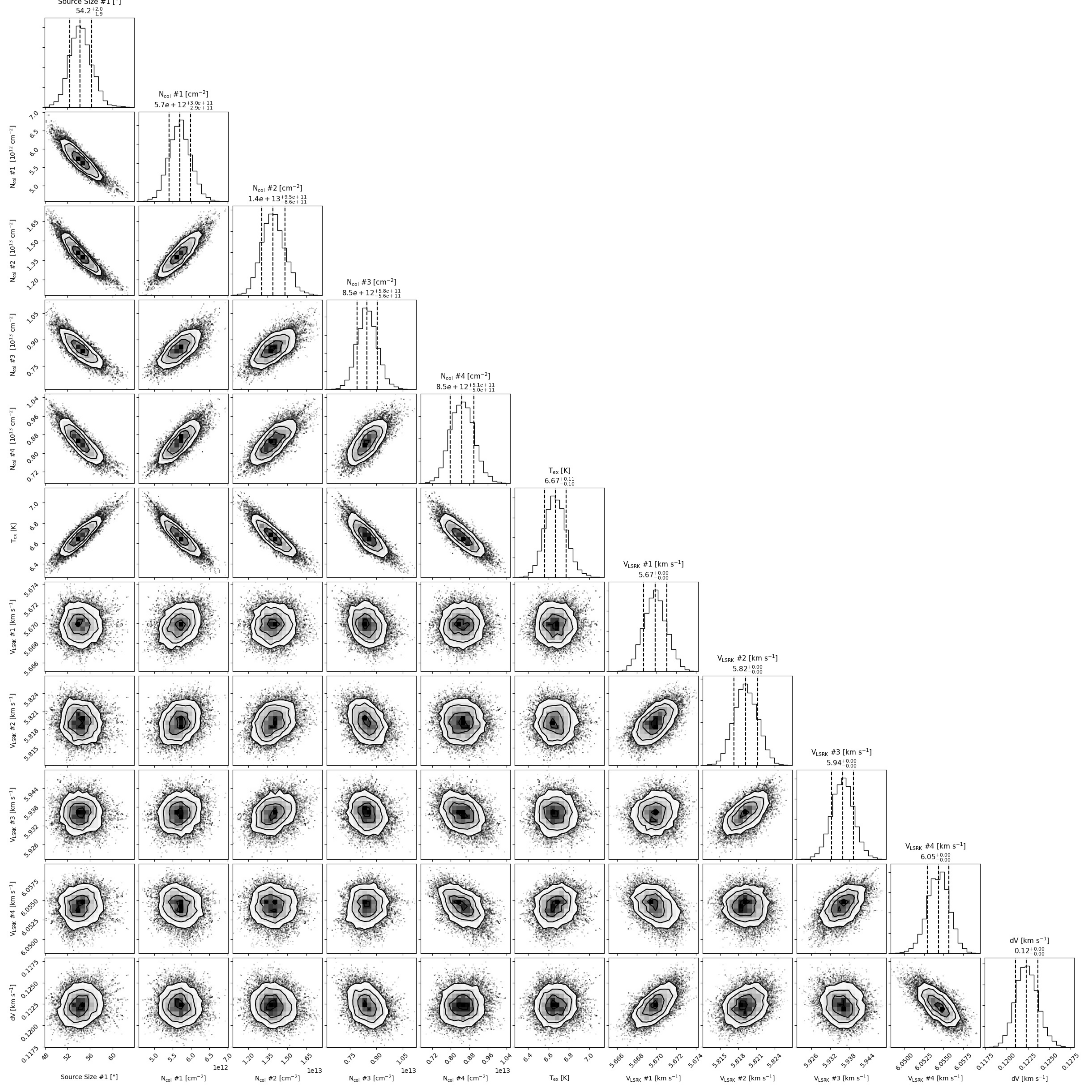}
    \caption{Parameter covariances and marginalized posterior distributions for the `co-spatial' HC$_{7}$N MCMC fit. 16$^{th}$, 50$^{th}$, and 84$^{th}$ confidence intervals (corresponding to $\pm$1 sigma for a Gaussian posterior distribution) are shown as vertical lines.}
    \label{Fig_A9}
\end{figure*}

The best-fit parameters for the MCMC analysis of \ce{HC7N}, under the `co-spatial' approximation, are given in Table~\ref{Table_A3}.  The individual detected lines are shown in Figure~\ref{Fig_A7}, while the stacked spectrum and matched filter results are shown in Figure~\ref{Fig_A8}. A corner plot of the parameter covariances for the \ce{HC7N} MCMC fit is shown in Figure~\ref{Fig_A9}.

\begin{table*}[!h]
\centering
\caption{\ce{HC7N} `co-spatial' best-fit parameters from MCMC analysis}
\begin{tabular}{c c c c c c}
\toprule
\multirow{2}{*}{Component}&	$v_{lsr}$					&	Size					&	\multicolumn{1}{c}{$N_T^\dagger$}					&	$T_{ex}$							&	$\Delta V$		\\
			&	(km s$^{-1}$)				&	($^{\prime\prime}$)		&	\multicolumn{1}{c}{(10$^{13}$ cm$^{-2}$)}		&	(K)								&	(km s$^{-1}$)	\\
\midrule
\hspace{0.1em}\vspace{-0.5em}\\
C1	&	$5.670^{+0.001}_{-0.001}$	&	\multirow{6}{*}{$54^{+2}_{-1}$}	&	$0.57^{+0.03}_{-0.03}$	&	\multirow{6}{*}{$6.7^{+0.1}_{-0.1}$}	&	\multirow{6}{*}{$0.122^{+0.001}_{-0.001}$}\\
\hspace{0.1em}\vspace{-0.5em}\\
C2	&	$5.819^{+0.002}_{-0.002}$	&		&	$1.39^{+0.09}_{-0.09}$	&		&	\\
\hspace{0.1em}\vspace{-0.5em}\\
C3	&	$5.936^{+0.003}_{-0.004}$	&		&	$0.85^{+0.06}_{-0.06}$	&		&	\\
\hspace{0.1em}\vspace{-0.5em}\\
C4	&	$6.054^{+0.001}_{-0.001}$	&		&	$0.85^{+0.05}_{-0.05}$	&		&	\\
\hspace{0.1em}\vspace{-0.5em}\\
\midrule
$N_T$ (Total)$^{\dagger\dagger}$	&	 \multicolumn{5}{c}{$3.65^{+0.13}_{-0.12}\times 10^{13}$~cm$^{-2}$}\\
\bottomrule
\end{tabular}

\begin{minipage}{0.75\textwidth}
	\footnotesize
	{Note} -- The quoted uncertainties represent the 16$^{th}$ and 84$^{th}$ percentile ($1\sigma$ for a Gaussian distribution) uncertainties.\\
	$^\dagger$Column density values are highly covariant with the derived source sizes.  The marginalized uncertainties on the column densities are therefore dominated by the largely unconstrained nature of the source sizes, and not by the signal-to-noise of the observations.\\
	$^{\dagger\dagger}$Uncertainties derived by adding the uncertainties of the individual components in quadrature.
\end{minipage}

\label{Table_A3}
\end{table*}

\clearpage
\subsubsection{HC$_9$N}
\begin{figure*}[!b]
    \centering
    \includegraphics[width=\textwidth]{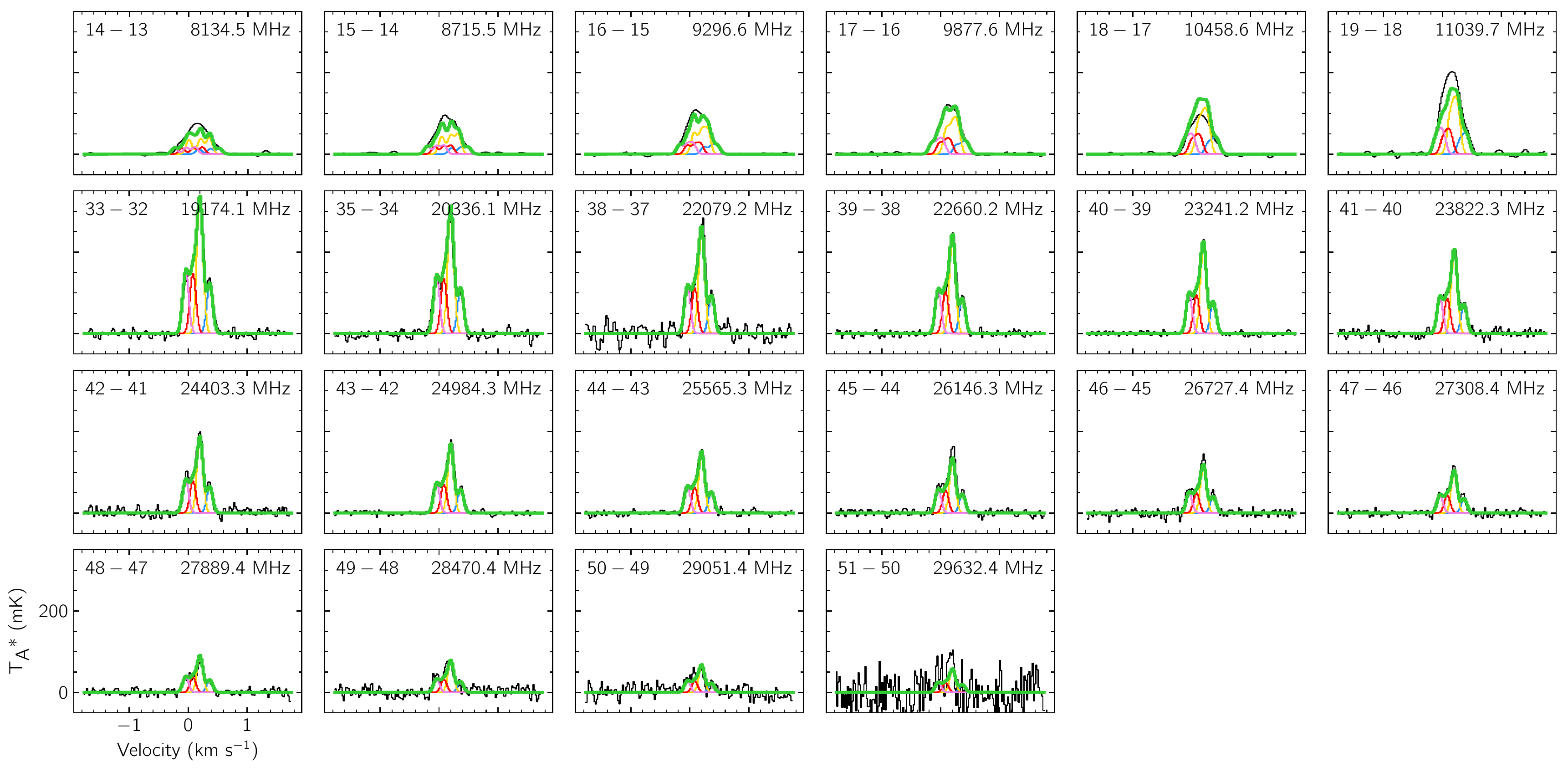}
    \caption{Individual line detections of \ce{HC9N} in the GOTHAM data.  The spectra (black) are displayed in velocity space relative to 5.8\,km\,s$^{-1}$, and using the rest frequency given in the top right of each panel. Quantum numbers are given in the top left of each panel, neglecting hyperfine splitting. The best-fit model to the data, including all velocity components, is overlaid in green.  Simulated spectra of the individual velocity components are shown in: blue (5.63\,km\,s$^{-1}$), gold (5.79\,km\,s$^{-1}$), red (5.91\,km\,s$^{-1}$), and violet (6.03\,km\,s$^{-1}$).  See Table~\ref{Table_A4}.}
    \label{Fig_A10}
\end{figure*}

\begin{figure*}[!b]
    \centering
    \includegraphics[width=0.4\textwidth]{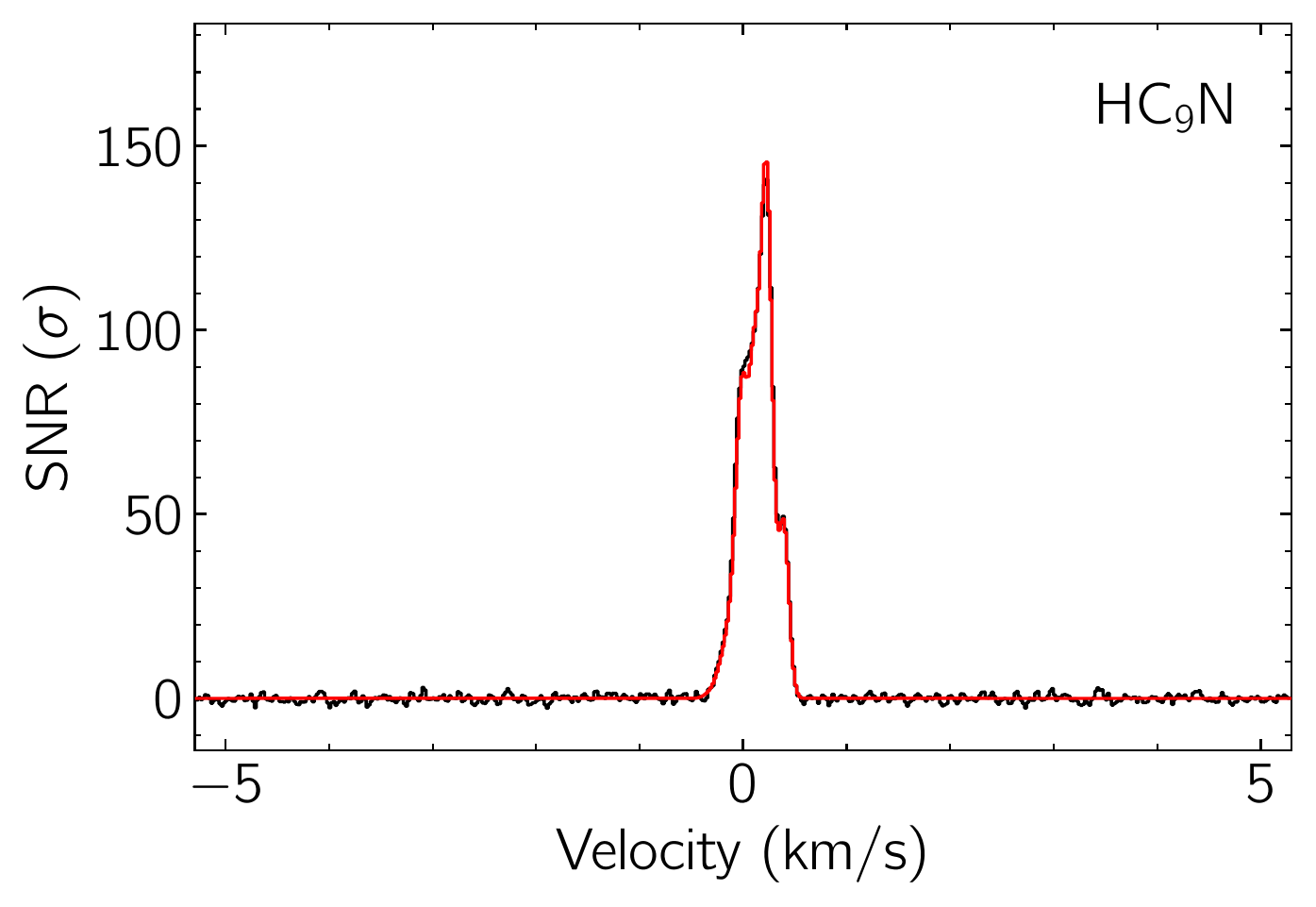}
    \includegraphics[width=0.4\textwidth]{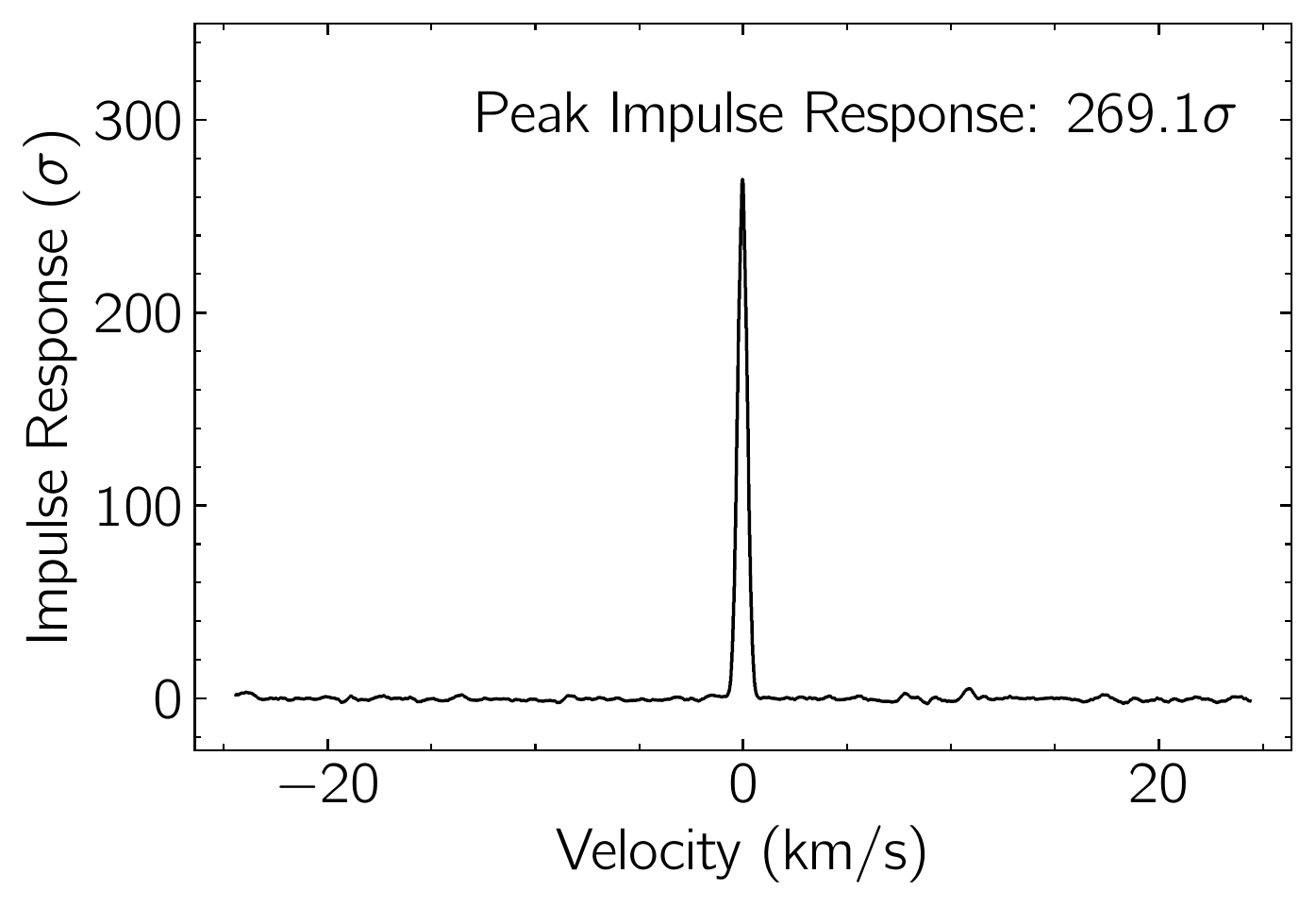}
    \caption{\emph{Left:} Velocity-stacked spectra of \ce{HC9N} in black, with the corresponding stack of the simulation using the best-fit parameters to the individual lines in red.  The data have been uniformly sampled to a resolution of 0.02\,km\,s$^{-1}$.  The intensity scale is the signal-to-noise ratio of the spectrum at any given velocity. \emph{Right:} Impulse response function of the stacked spectrum using the simulated line profile as a matched filter.  The intensity scale is the signal-to-noise ratio of the response function when centered at a given velocity.  The peak of the impulse response function provides a minimum significance for the detection of 269.1$\sigma$.}
    \label{Fig_A11}
\end{figure*}

\begin{figure*}
    \centering
    \includegraphics[width=\textwidth]{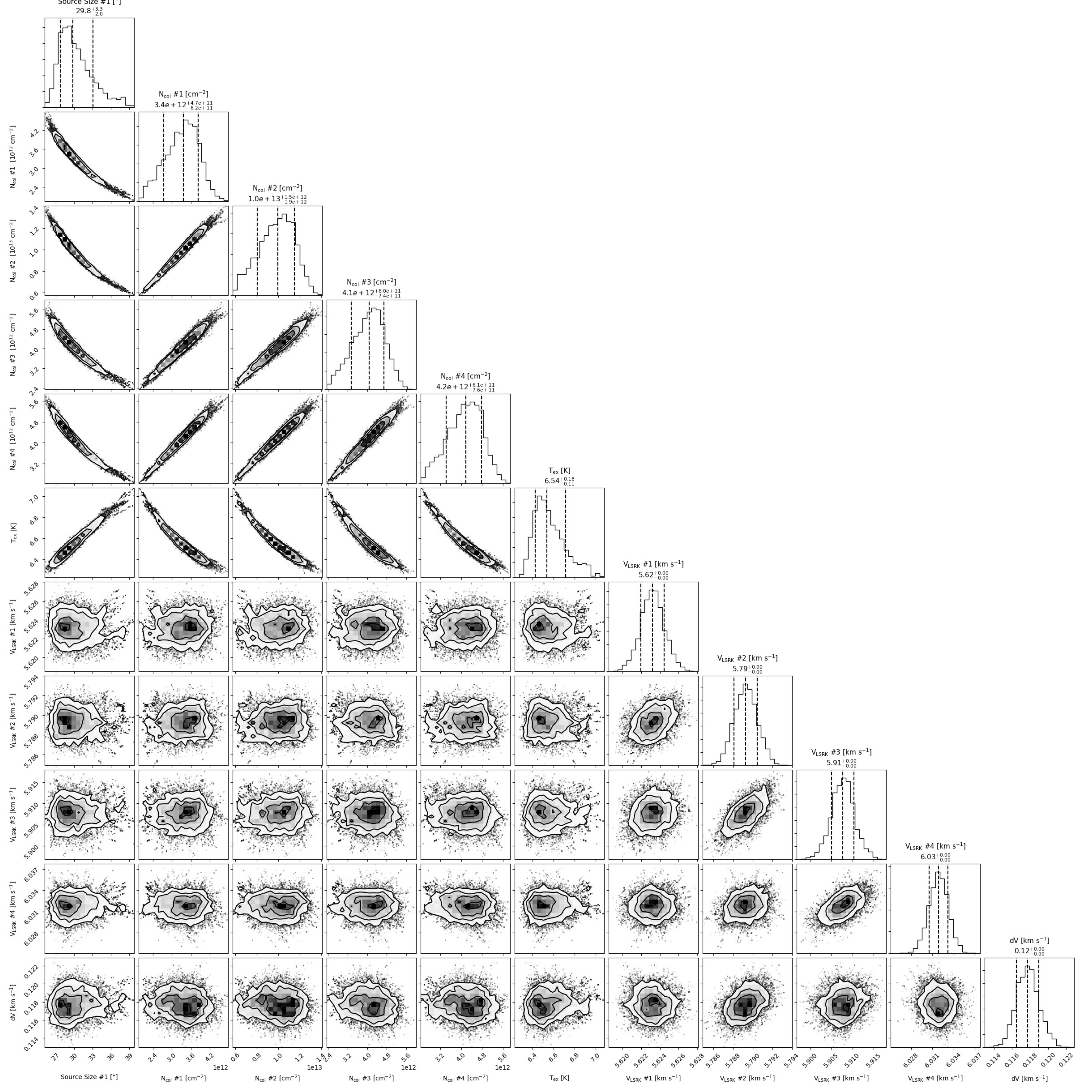}
    \caption{Parameter covariances and marginalized posterior distributions for the `co-spatial' HC$_{9}$N MCMC fit. 16$^{th}$, 50$^{th}$, and 84$^{th}$ confidence intervals (corresponding to $\pm$1 sigma for a Gaussian posterior distribution) are shown as vertical lines.}
    \label{Fig_A12}
\end{figure*}

The best-fit parameters for the MCMC analysis of \ce{HC9N}, under the `co-spatial' approximation, are given in Table~\ref{Table_A4}.  The individual detected lines are shown in Figure~\ref{Fig_A10}, while the stacked spectrum and matched filter results are shown in Figure~\ref{Fig_A11}. A corner plot of the parameter covariances for the \ce{HC9N} MCMC fit is shown in Figure~\ref{Fig_A12}.

\begin{table*}[!h]
\centering
\caption{\ce{HC9N} `co-spatial' best-fit parameters from MCMC analysis}
\begin{tabular}{c c c c c c}
\toprule
\multirow{2}{*}{Component}&	$v_{lsr}$					&	Size					&	\multicolumn{1}{c}{$N_T^\dagger$}					&	$T_{ex}$							&	$\Delta V$		\\
			&	(km s$^{-1}$)				&	($^{\prime\prime}$)		&	\multicolumn{1}{c}{(10$^{12}$ cm$^{-2}$)}		&	(K)								&	(km s$^{-1}$)	\\
\midrule
\hspace{0.1em}\vspace{-0.5em}\\
C1	&	$5.623^{+0.001}_{-0.001}$	&	\multirow{6}{*}{$29^{+3}_{-2}$}	&	$3.36^{+0.47}_{-0.62}$	&	\multirow{6}{*}{$6.5^{+0.2}_{-0.1}$}	&	\multirow{6}{*}{$0.118^{+0.001}_{-0.001}$}\\
\hspace{0.1em}\vspace{-0.5em}\\
C2	&	$5.789^{+0.001}_{-0.001}$	&		&	$9.96^{+1.53}_{-1.91}$	&		&	\\
\hspace{0.1em}\vspace{-0.5em}\\
C3	&	$5.908^{+0.003}_{-0.003}$	&		&	$4.09^{+0.60}_{-0.74}$	&		&	\\
\hspace{0.1em}\vspace{-0.5em}\\
C4	&	$6.032^{+0.001}_{-0.001}$	&		&	$4.18^{+0.61}_{-0.76}$	&		&   \\
\hspace{0.1em}\vspace{-0.5em}\\
\midrule
$N_T$ (Total)$^{\dagger\dagger}$	&	 \multicolumn{5}{c}{$2.16^{+0.18}_{-0.23}\times 10^{13}$~cm$^{-2}$}\\
\bottomrule
\end{tabular}

\begin{minipage}{0.75\textwidth}
	\footnotesize
	{Note} -- The quoted uncertainties represent the 16$^{th}$ and 84$^{th}$ percentile ($1\sigma$ for a Gaussian distribution) uncertainties.\\
	$^\dagger$Column density values are highly covariant with the derived source sizes.  The marginalized uncertainties on the column densities are therefore dominated by the largely unconstrained nature of the source sizes, and not by the signal-to-noise of the observations.\\
	$^{\dagger\dagger}$Uncertainties derived by adding the uncertainties of the individual components in quadrature.
\end{minipage}

\label{Table_A4}
\end{table*}

\clearpage
\subsubsection{HC$_{11}$N}
\begin{figure*}[!b]
    \centering
    \includegraphics[width=\textwidth]{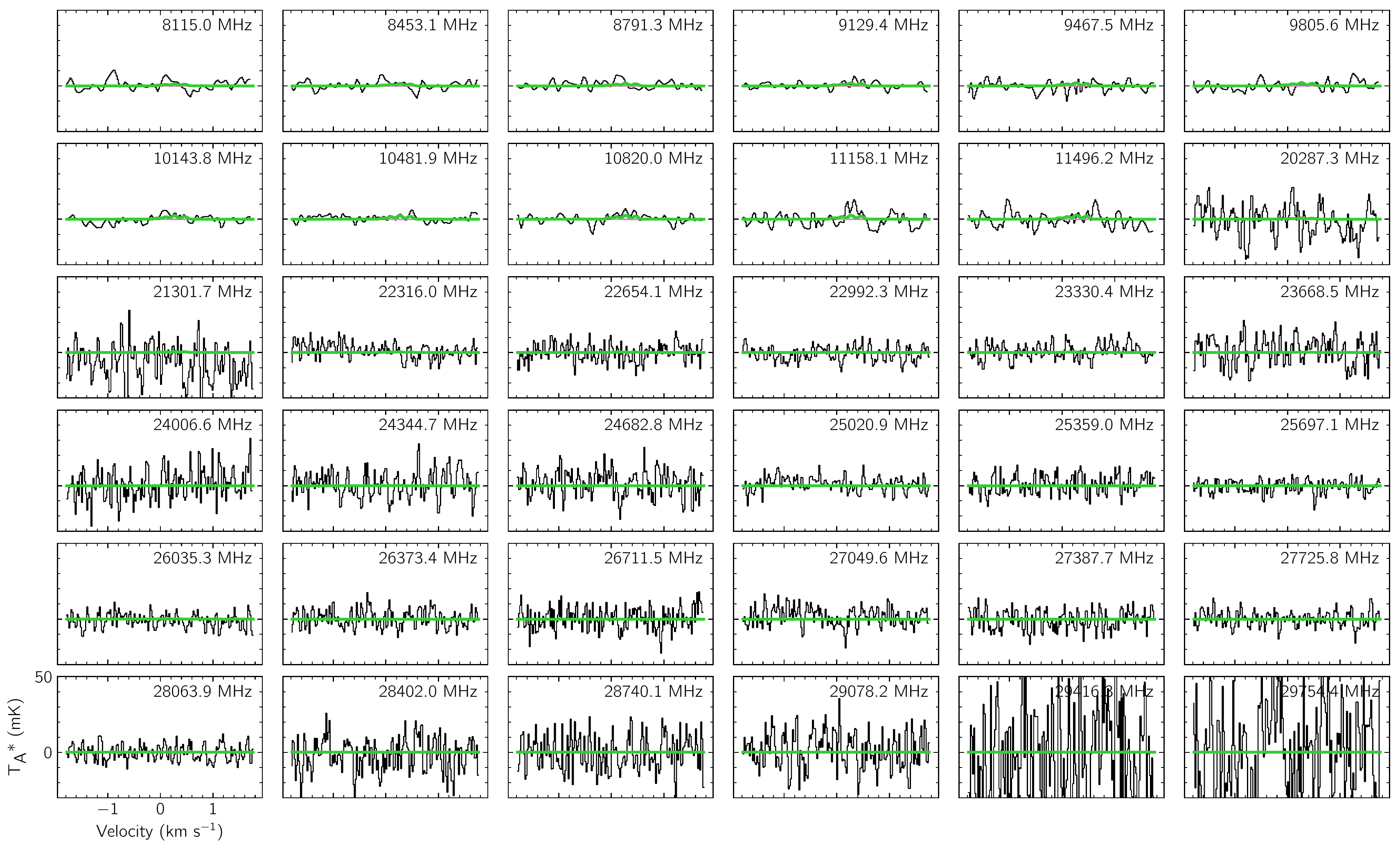}
    \caption{Individual line observations of \ce{HC11N} in the GOTHAM data.  The spectra (black) are displayed in velocity space relative to 5.8\,km\,s$^{-1}$, and using the rest frequency given in the top right of each panel. Quantum numbers are given in the top left of each panel, neglecting hyperfine splitting. The best-fit model to the data, including all velocity components, is overlaid in green.  Simulated spectra of the individual velocity components are shown in: blue (5.63\,km\,s$^{-1}$), gold (5.79\,km\,s$^{-1}$), red (5.91\,km\,s$^{-1}$), and violet (6.03\,km\,s$^{-1}$).  See Table~\ref{Table_A5}.}
    \label{Fig_A13}
\end{figure*}

\begin{figure*}[!b]
    \centering
    \includegraphics[width=0.4\textwidth]{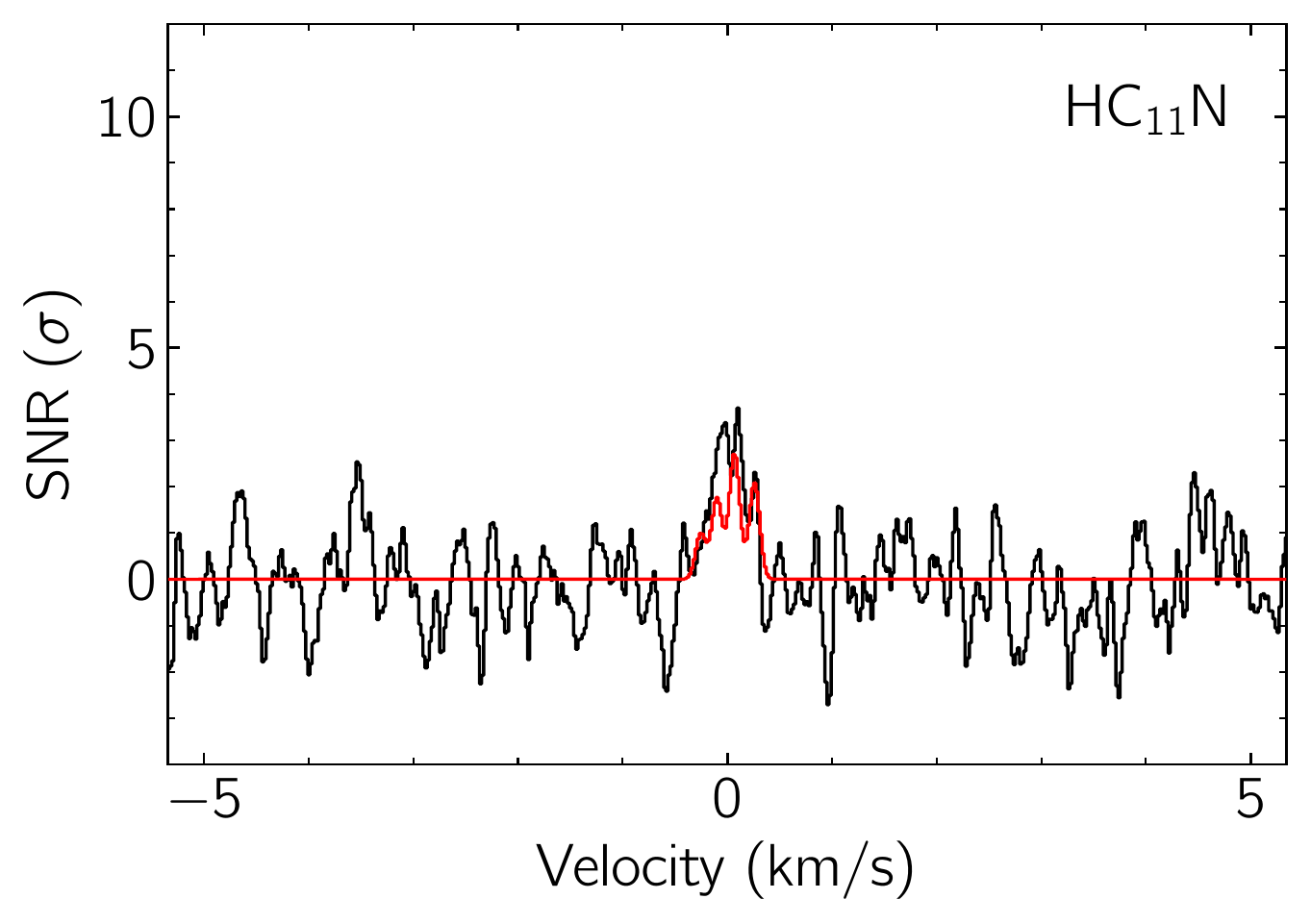}
    \includegraphics[width=0.4\textwidth]{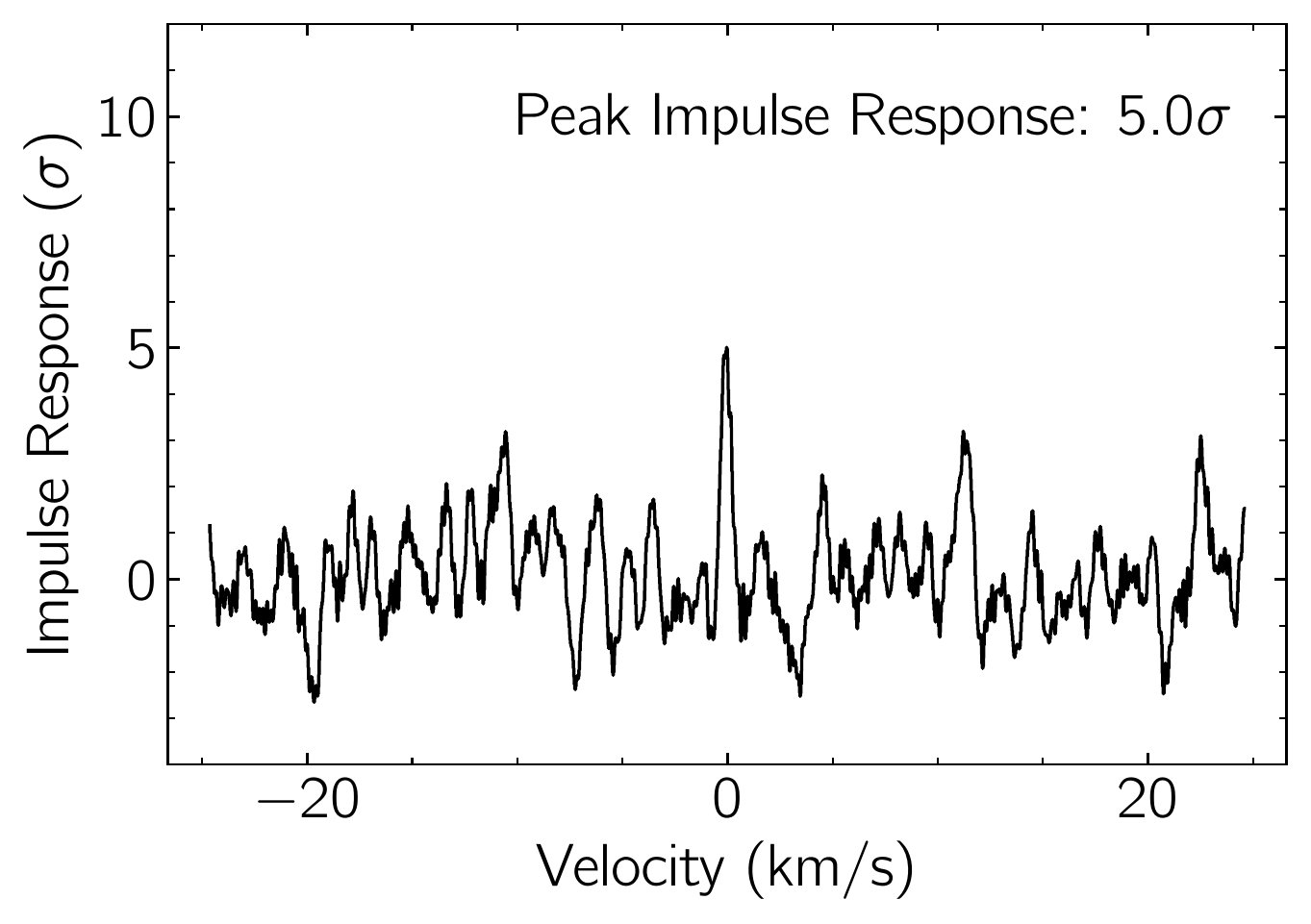}
    \caption{\emph{Left:} Velocity-stacked spectra of \ce{HC11N} in black, with the corresponding stack of the simulation using the best-fit parameters to the individual lines in red.  The data have been uniformly sampled to a resolution of 0.02\,km\,s$^{-1}$.  The intensity scale is the signal-to-noise ratio of the spectrum at any given velocity. \emph{Right:} Impulse response function of the stacked spectrum using the simulated line profile as a matched filter.  The intensity scale is the signal-to-noise ratio of the response function when centered at a given velocity.  The peak of the impulse response function provides a minimum significance for the detection of 5.0$\sigma$.}
    \label{Fig_A14}
\end{figure*}

\begin{figure*}
    \centering
    \includegraphics[width=\textwidth]{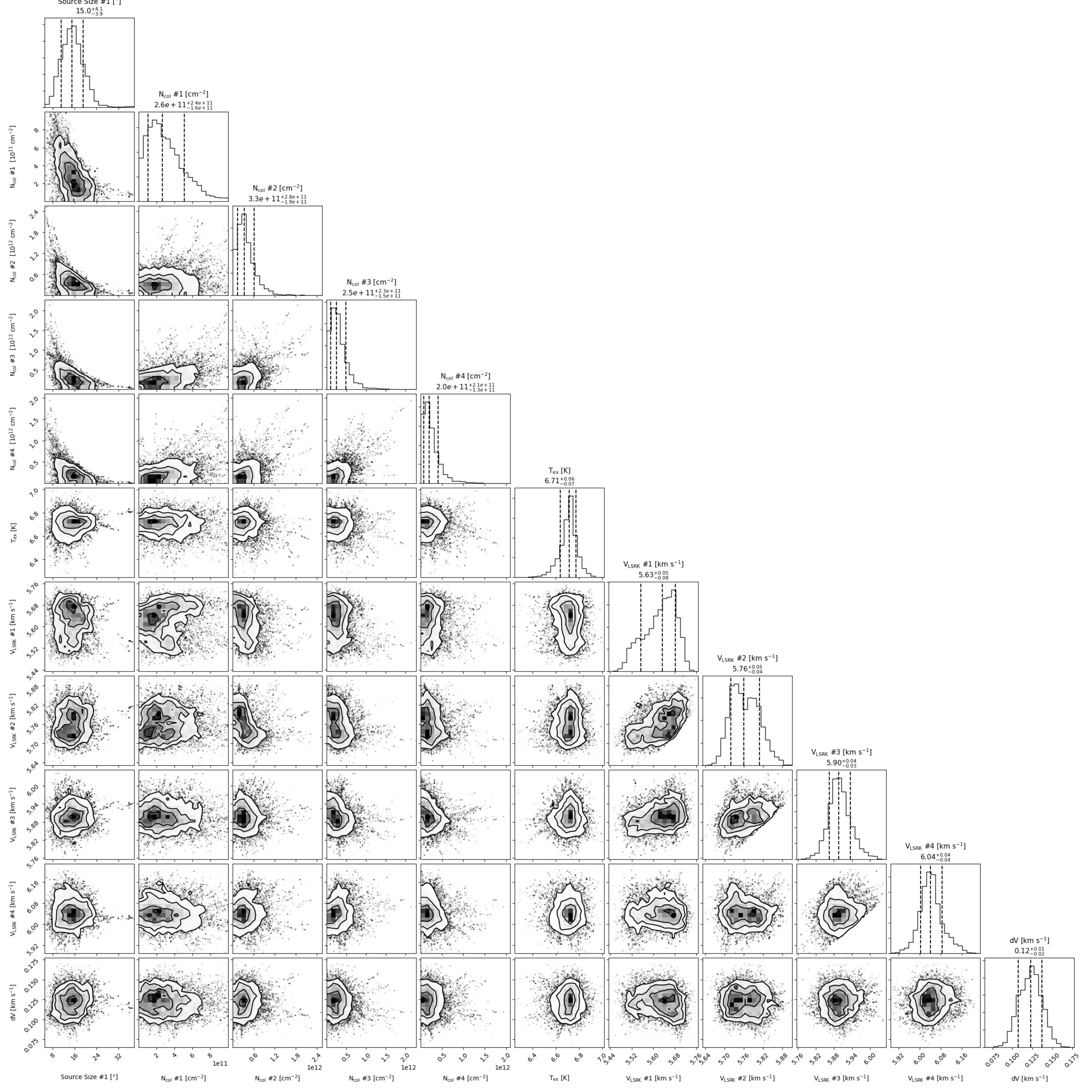}
    \caption{Parameter covariances and marginalized posterior distributions for the `co-spatial' HC$_{11}$N MCMC fit. 16$^{th}$, 50$^{th}$, and 84$^{th}$ confidence intervals (corresponding to $\pm$1 sigma for a Gaussian posterior distribution) are shown as vertical lines.}
    \label{Fig_A15}
\end{figure*}

The best-fit parameters for the MCMC analysis of \ce{HC11N}, under the `co-spatial' approximation, are given in Table~\ref{Table_A5}.  The individual observed lines are shown in Figure~\ref{Fig_A13}, while the stacked spectrum and matched filter results are shown in Figure~\ref{Fig_A14}. A corner plot of the parameter covariances for the \ce{HC11N} MCMC fit is shown in Figure~\ref{Fig_A15}.

\begin{table*}[!h]
\centering
\caption{\ce{HC11N} `co-spatial' best-fit parameters from MCMC analysis}
\begin{tabular}{c c c c c c}
\toprule
\multirow{2}{*}{Component}&	$v_{lsr}$					&	Size					&	\multicolumn{1}{c}{$N_T^\dagger$}					&	$T_{ex}$							&	$\Delta V$		\\
			&	(km s$^{-1}$)				&	($^{\prime\prime}$)		&	\multicolumn{1}{c}{(10$^{11}$ cm$^{-2}$)}		&	(K)								&	(km s$^{-1}$)	\\
\midrule
\hspace{0.1em}\vspace{-0.5em}\\
C1	&	$5.634^{+0.048}_{-0.080}$	&	\multirow{6}{*}{$14^{+4}_{-3}$}	&	$2.64^{+2.45}_{-1.60}$	&	\multirow{6}{*}{$6.7^{+0.1}_{-0.1}$}	&	\multirow{6}{*}{$0.123^{+0.014}_{-0.015}$}\\
\hspace{0.1em}\vspace{-0.5em}\\
C2	&	$5.760^{+0.049}_{-0.041}$	&		&	$3.29^{+2.80}_{-1.86}$	&		&	\\
\hspace{0.1em}\vspace{-0.5em}\\
C3	&	$5.896^{+0.040}_{-0.032}$	&		&	$2.53^{+2.31}_{-1.51}$	&		&	\\
\hspace{0.1em}\vspace{-0.5em}\\
C4	&	$6.041^{+0.044}_{-0.037}$	&		&	$1.96^{+2.14}_{-1.26}$	&		&	\\
\hspace{0.1em}\vspace{-0.5em}\\
\midrule
$N_T$ (Total)$^{\dagger\dagger}$	&	 \multicolumn{5}{c}{$1.04^{+0.49}_{-0.31}\times 10^{12}$~cm$^{-2}$}\\
\bottomrule
\end{tabular}

\begin{minipage}{0.75\textwidth}
	\footnotesize
	{Note} -- The quoted uncertainties represent the 16$^{th}$ and 84$^{th}$ percentile ($1\sigma$ for a Gaussian distribution) uncertainties.\\
	$^\dagger$Column density values are highly covariant with the derived source sizes.  The marginalized uncertainties on the column densities are therefore dominated by the largely unconstrained nature of the source sizes, and not by the signal-to-noise of the observations.\\
	$^{\dagger\dagger}$Uncertainties derived by adding the uncertainties of the individual components in quadrature.
\end{minipage}

\label{Table_A5}
\end{table*}

\subsection{Chemical modeling of cyanopolyynes in TMC-1}
Astrochemical models were run in order to better understand how similar the chemistry of \ce{HC11N} is to the smaller cyanopolyynes. The \texttt{NAUTILUS}-v1.1 code \citep{Ruaud_2016} was used along with a version of the KIDA 2014 network \citep{Wakelam_2014}, modified as in \citet{Shingledecker_2018, Xue:2020aa} and including the \ce{HC11N} reactions previously described in \citet{Loomis_2016}, and available as supporting material for that work. The \ce{HC11N} reactions added in that work were based on those of \ce{HC9N} included in the KIDA 2012 network \citep{wakelam_kinetic_2012}. For example, consider the following reaction:

\begin{equation}
    \ce{C8H2 + CN -> H + HC9N}
    \label{hc9nformation}
\end{equation}

\noindent
{in the network of \citet{Loomis_2016}, the corresponding formation route is included for \ce{HC11N}:}

\begin{equation}
    \ce{C10H2 + CN -> H + HC11N}
    \label{hc11nformation}
\end{equation}

\noindent
{where the above reaction utilizes the same rate coefficient parameters and branching fractions as in reaction \eqref{hc9nformation}.}

{As is currently the case with, e.g. \ce{HC9N}, the role of grain-chemistry for \ce{HC11N} is extremely limited. Adsorption onto grains serves mainly as a destruction pathway for the gas-phase molecule, after which it can be photodissociated {via secondary photons} into \ce{C6H} and \ce{C5N} with the same rate coefficient as the analogous \ce{HC9N} process. There is one formation route for \ce{HC11N} involving the barrierless association of H and \ce{C11N} that is of negligible importance on the overall gas-phase abundances.}

\begin{figure*}
    \centering
    \includegraphics[width=0.5\textwidth]{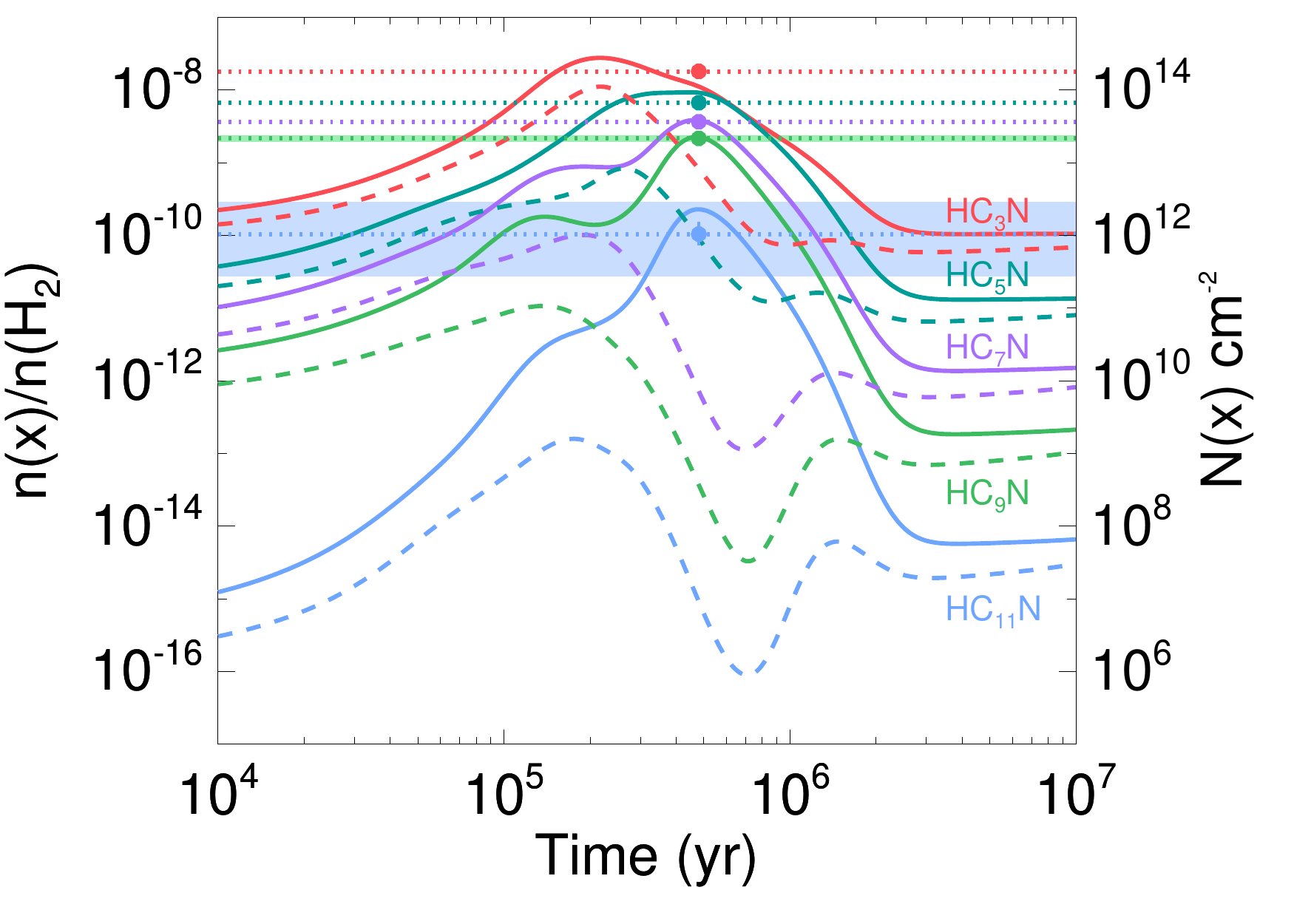}
    \caption{Calculated abundances (solid lines), abundances from the `co-spatial' MCMC analysis (dotted lines), and best-fit times (dots) for the cyanopolyynes HC$_n$N, $n\in[3,5,7,9,11]$ using C/O=1.1 (solid curves) and C/O=0.7 (dashed curves). Abundance ranges from the `separate components' MCMC analysis for \ce{HC9N} and \ce{HC11N} are shown by the green and blue bars, respectively. Equivalent column densities assuming $N(\mathrm{H_2})=10^{22}$ cm$^{-2}$ are shown on the right axis.}
    \label{colow}
\end{figure*}

For physical conditions, typical TMC-1 values were used, including $T_\mathrm{gas}=T_\mathrm{dust}=10$ K, a gas density of $10^4$ cm$^{-3}$, and a standard cosmic ray ionization rate of $1.3\times10^{-17}$ s$^{-1}$. Initial elemental abundances were, with the exception of oxygen, taken from \citet{Hincelin_2011}. {To obtain our initial oxygen abundances, we ran simulations using a range of values using $x(\mathrm{O})\in[1.0\times10^{-5},1.0\times10^{-3}]$. From this analysis, we found that the best agreement between calculated cyanopolyyne abundances and our observational results was obtained using $x(\mathrm{O})_{t=0}\approx1.55\times10^{-4}$, which implies a slightly carbon rich C/O$\sim1.1$. In Fig. \ref{colow}, model results are shown in which an initial C/O=0.7 was used, a more typical value in simulations of TMC-1 \citep{Hincelin_2011}. From a comparison of that figure with Fig. \ref{Fig_21}, one can see that the resulting agreement is significantly poorer. A detailed comparison between observations and our calculated abundances, as in \citet{agundez_chemistry_2013}, using our carbon-rich C/O=1.1 is beyond the scope of this work and will be explored in more detail in a subsequent study. Nevertheless, a comparison between our observationally derived abundance for propargyl cyanide (\ce{HCCCH2CN}), another species whose first detection we report in a companion work \citep{McGuire:2020bb}, and calculated abundances using the same model were found to be in excellent agreement.}

The results of our simulations are shown in Fig. \ref{Fig_21}. There, one can see that the calculated abundances of all species, with the exception of \ce{HC11N}, match the derived co-spatial MCMC results to within a factor of a few at a model time of $\sim5\times10^5$ yr. In our network, the chemistry of \ce{HC11N} largely follows that of the smaller cyanopolyynes, where the dominant formation routes for species of the general formula HC$_n$N ($n\in[3,5,7,9,11]$) are the dissociative recombination processes

\begin{equation}
    \mathrm{H_2C}_n\mathrm{N^+} + e^- \rightarrow \mathrm{H} + \mathrm{HC}_n\mathrm{N}
    \label{r1}
\end{equation}

\begin{equation}
    \mathrm{H_3C}_n\mathrm{N^+} + e^- \rightarrow \mathrm{H_2} + \mathrm{HC}_n\mathrm{N}
    \label{r2}
\end{equation}

\noindent
and the neutral-neutral reaction

\begin{equation}
    \mathrm{H_2C}_{n-1} + \mathrm{CN} \rightarrow \mathrm{H} + \mathrm{HC}_n\mathrm{N}.
    \label{r3}
\end{equation}

\noindent
Here, we find the dissociative recombination routes to dominate at times before $\sim10^5$ yr, with the neutral-neutral route becoming the major production pathway thereafter. At all simulation times, destruction occurs mainly via reaction with carbon atoms or ions such as C$^+$, H$_3^+$, and HCO$^+$. We note that the sharp decrease in all cyanopolyyne abundances observable in Fig. \ref{Fig_21} around $\sim$1~Myr corresponds approximately to the time at which most carbon in the source freezes out onto grains and mirrors the trend seen in the abundances of other carbon-bearing species that have formation routes dominated by purely gas-phase reactions.

{Though our calculated \ce{HC11N} abundances are within the errors of our `separate components' MCMC fit results, a more self-consistent comparison with the co-spatial MCMC abundances reveals that \ce{HC11N} is still overproduced relative to the smaller cyanopolyynes. This finding suggests that the assumptions we made previously in constructing our \ce{HC11N} network \citep{Loomis_2016}, i.e. that its chemistry largely follows that of \ce{HC9N} and the other cyanopolyynes, is somehow flawed.} One possibility is that the proposed reactions \eqref{r1}, \eqref{r2}, and \eqref{r3}, in addition to producing \ce{HC11N} from \ce{H2C11N+}, \ce{H3C11N+}, and \ce{H2C10}, respectively, may have other efficient product channels not included in our network, including perhaps the production of cyclic molecules. Additionally, recent studies have highlighted the importance of destruction pathways in understanding the abundances of interstellar molecules \citep{shingledecker_case_2019,shingledecker_isomers_2020}. {For example, it was found by \citet{shingledecker_case_2019} that propadienone exhibited a unique reactivity with atomic hydrogen. If \ce{HC11N} can similarly react efficiently with H, this} could help explain the sharp drop in its abundance relative to the other cyanopolyynes. {Another possibility suggested initially by \citet{herbst_negative_1981} and later confirmed by \citet{jerosimic_hcnn_2019} is that the longer the cyanopolyyne, the more stable the anion formed via electron association. The products of these associations, such as \ce{HC11N-}, lose their linear structures and are much more reactive than their neutral counterparts. Either of these} hypothetical destruction pathways might serve as a chemical link between linear carbon chain species and the cyclic aromatic molecules also observed in TMC-1 \citep{McGuire_2018,McGuire:2020aa,McGuire:2020aa}, providing an explanation for our chemical model's overproduction of \ce{HC11N} and underproduction of the newly detected aromatic species \citep{Burkhardt:2020aa,McCarthy:2020aa,McGuire:2020aa}.

\subsection{Lines Used in MCMC Fits}

Table~\ref{app:lines} shows the total number of transitions (including hyperfine components) of the molecules analyzed or discussed in this paper that were covered by GOTHAM observations at the time of analysis and were above our predicted flux threshold of 5\%, as discussed earlier.  Also included are the number of transitions, if any, that were coincident with interfering transitions of other species, and the total number of lines used after excluding interlopers.  Observational data windowed around these transitions, spectroscopic properties of each transition, and the partition function used in the MCMC analysis are provided in the Harvard Dataverse repository \citep{GOTHAMDR1}.

\begin{table*}[bht!]
    \centering
    \caption{Total number of transitions of a given species within the range of the GOTHAM data, number of interfering lines, and total number included in MCMC fit.}
    \begin{tabular}{l c c c}
    \toprule
                &   Transitions Covered &   Interfering Lines   &   Total Transitions   \\
    Molecule    &   By GOTHAM           &      In Data          &       Used in MCMC        \\
    \midrule
    \ce{HC3N}           &   6           &   0                   &   6     \\
    \ce{HC5N}           &   16          &   0                   &   16    \\
    \ce{HC7N}           &   36          &   0                   &   36     \\
    \ce{HC9N}           &   66          &   0                   &   66     \\
    \ce{HC11N}          &   19          &   0                   &   19     \\
    \ce{HC13N}          &   20          &   0                   &   20     \\
    propargyl cyanide   &   68          &   0                   &   68     \\
    benzonitrile        &   156         &   0                   &   156     \\
    1-CNN               &   1256        &   2                   &   1254    \\
    2-CNN               &   709         &   0                   &   709    \\
    \bottomrule
    \end{tabular}
    \label{app:lines}
\end{table*}


\begin{thebibliography}{}
\expandafter\ifx\csname natexlab\endcsname\relax\def\natexlab#1{#1}\fi
\providecommand{\url}[1]{\href{#1}{#1}}

\bibitem[{Agúndez \& Wakelam(2013)}]{agundez_chemistry_2013}
Agúndez, M., \& Wakelam, V. 2013, Chemical Reviews, 113, 8710.
\newblock \url{http://pubs.acs.org/doi/abs/10.1021/cr4001176}

\bibitem[{Bell {et~al.}(1982)Bell, Feldman, Kwok, \& Matthews}]{bell:389}
Bell, M.~B., Feldman, P.~A., Kwok, S., \& Matthews, H.~E. 1982, Nature, 295,
  389

\bibitem[{{Bell} {et~al.}(1997){Bell}, {Feldman}, {Travers}, {McCarthy},
  {Gottlieb}, \& {Thaddeus}}]{Bell_1997}
{Bell}, M.~B., {Feldman}, P.~A., {Travers}, M.~J., {et~al.} 1997, \apjl, 483,
  L61

\bibitem[{{Bell} \& {Matthews}(1985)}]{bell:l63}
{Bell}, M.~B., \& {Matthews}, H.~E. 1985, Astrophysical Journals Letters, 291,
  L63

\bibitem[{{Bell} {et~al.}(1998){Bell}, {Watson}, {Feldman}, \&
  {Travers}}]{Bell_1998}
{Bell}, M.~B., {Watson}, J.~K.~G., {Feldman}, P.~A., \& {Travers}, M.~J. 1998,
  \apj, 508, 286

\bibitem[{{Bujarrabal} {et~al.}(1981){Bujarrabal}, {Guelin}, {Morris}, \&
  {Thaddeus}}]{Bujarrabal_1981}
{Bujarrabal}, V., {Guelin}, M., {Morris}, M., \& {Thaddeus}, P. 1981, \aap, 99,
  239

\bibitem[{Burkhardt {et~al.}(2020)Burkhardt, Loomis, Shingledecker, Lee,
  Remijan, McCarthy, \& McGuire}]{Burkhardt:2020aa}
Burkhardt, A.~M., Loomis, R.~A., Shingledecker, C.~N., {et~al.} 2020, Nature
  Astronomy, in revision

\bibitem[{{Churchwell} {et~al.}(1978){Churchwell}, {Winnewisser}, \&
  {Walmsley}}]{Churchwell_1978}
{Churchwell}, E., {Winnewisser}, G., \& {Walmsley}, C.~M. 1978, \aap, 67, 139

\bibitem[{Cordiner {et~al.}(2017)Cordiner, Charnley, Kisiel, McGuire, \&
  Kuan}]{Cordiner_2017}
Cordiner, M.~A., Charnley, S.~B., Kisiel, Z., McGuire, B.~A., \& Kuan, Y.-J.
  2017, The Astrophysical Journal, 850, 187

\bibitem[{Crabtree {et~al.}(2016)Crabtree, Martin-Drumel, Brown, Gaster, Hall,
  \& McCarthy}]{Crabtree:2016fj}
Crabtree, K.~N., Martin-Drumel, M.-A., Brown, G.~G., {et~al.} 2016, The Journal
  of Chemical Physics, 144, 124201

\bibitem[{{Czekala}(2016)}]{Czekala_2016}
{Czekala}, I. 2016, {DiskJockey: Protoplanetary disk modeling for dynamical
  mass derivation}, , , ascl:1603.011

\bibitem[{{Dobashi} {et~al.}(2018){Dobashi}, {Shimoikura}, {Nakamura},
  {Kameno}, {Mizuno}, \& {Taniguchi}}]{Dobashi_2018}
{Dobashi}, K., {Shimoikura}, T., {Nakamura}, F., {et~al.} 2018, \apj, 864, 82

\bibitem[{{Dobashi} {et~al.}(2019){Dobashi}, {Shimoikura}, {Ochiai},
  {Nakamura}, {Kameno}, {Mizuno}, \& {Taniguchi}}]{Dobashi_2019}
{Dobashi}, K., {Shimoikura}, T., {Ochiai}, T., {et~al.} 2019, \apj, 879, 88

\bibitem[{{Drouin}(2017)}]{Drouin_2017}
{Drouin}, B.~J. 2017, Journal of Molecular Spectroscopy, 340, 1

\bibitem[{{Foreman-Mackey} {et~al.}(2013){Foreman-Mackey}, {Hogg}, {Lang}, \&
  {Goodman}}]{Foreman-Mackey_2013}
{Foreman-Mackey}, D., {Hogg}, D.~W., {Lang}, D., \& {Goodman}, J. 2013, \pasp,
  125, 306

\bibitem[{{Fuente} {et~al.}(2019){Fuente}, {Navarro}, {Caselli}, {Gerin},
  {Kramer}, {Roueff}, {Alonso-Albi}, {Bachiller}, {Cazaux}, {Commercon},
  {Friesen}, {Garc{\'\i}a-Burillo}, {Giuliano}, {Goicoechea}, {Gratier},
  {Hacar}, {Jim{\'e}nez-Serra}, {Kirk}, {Lattanzi}, {Loison}, {Malinen},
  {Marcelino}, {Mart{\'\i}n-Dom{\'e}nech}, {Mu{\~n}oz-Caro}, {Pineda},
  {Tafalla}, {Tercero}, {Ward-Thompson}, {Trevi{\~n}o-Morales},
  {Rivi{\'e}re-Marichalar}, {Roncero}, {Vidal}, \& {Ballester}}]{Fuente_2019}
{Fuente}, A., {Navarro}, D.~G., {Caselli}, P., {et~al.} 2019, \aap, 624, A105

\bibitem[{{Gelman} \& {Rubin}(1992)}]{Gelman_1992}
{Gelman}, A., \& {Rubin}, D.~B. 1992, Statistical Science, 7, 457

\bibitem[{{GOTHAM Collaboration}(2020)}]{GOTHAMDR1}
{GOTHAM Collaboration}. 2020, {Spectral Stacking Data for Phase 1 Science
  Release of GOTHAM}, v4.0,  Harvard Dataverse, doi:10.7910/DVN/PG7BHO.
\newblock \url{https://doi.org/10.7910/DVN/PG7BHO}

\bibitem[{{Gratier} {et~al.}(2016){Gratier}, {Majumdar}, {Ohishi}, {Roueff},
  {Loison}, {Hickson}, \& {Wakelam}}]{Gratier_2016}
{Gratier}, P., {Majumdar}, L., {Ohishi}, M., {et~al.} 2016, \apjs, 225, 25

\bibitem[{{Herbst}(1981)}]{herbst_negative_1981}
{Herbst}, E. 1981, \nat, 289, 656

\bibitem[{{Hincelin} {et~al.}(2011){Hincelin}, {Wakelam}, {Hersant},
  {Guilloteau}, {Loison}, {Honvault}, \& {Troe}}]{Hincelin_2011}
{Hincelin}, U., {Wakelam}, V., {Hersant}, F., {et~al.} 2011, \aap, 530, A61

\bibitem[{Jerosimić {et~al.}(2019)Jerosimić, Wester, \&
  Gianturco}]{jerosimic_hcnn_2019}
Jerosimić, S.~V., Wester, R., \& Gianturco, F.~A. 2019, Physical Chemistry
  Chemical Physics, 21, 11405.
\newblock
  \url{https://pubs.rsc.org/en/content/articlelanding/2019/cp/c9cp00877b}

\bibitem[{{Langston} \& {Turner}(2007)}]{Langston_2007}
{Langston}, G., \& {Turner}, B. 2007, \apj, 658, 455

\bibitem[{{Liu} {et~al.}(2001){Liu}, {Mehringer}, \& {Snyder}}]{Liu_2001}
{Liu}, S.-Y., {Mehringer}, D.~M., \& {Snyder}, L.~E. 2001, \apj, 552, 654

\bibitem[{{Loomis} {et~al.}(2018){Loomis}, {{\"O}berg}, {Andrews}, {Walsh},
  {Czekala}, {Huang}, \& {Rosenfeld}}]{Loomis_2018}
{Loomis}, R.~A., {{\"O}berg}, K.~I., {Andrews}, S.~M., {et~al.} 2018, \aj, 155,
  182

\bibitem[{{Loomis} {et~al.}(2016){Loomis}, {Shingledecker}, {Langston},
  {McGuire}, {Dollhopf}, {Burkhardt}, {Corby}, {Booth}, {Carroll}, {Turner}, \&
  {Remijan}}]{Loomis_2016}
{Loomis}, R.~A., {Shingledecker}, C.~N., {Langston}, G., {et~al.} 2016, \mnras,
  463, 4175

\bibitem[{{Loomis} {et~al.}(2020){Loomis}, {{\"O}berg}, {Andrews}, {Bergin},
  {Bergner}, {Blake}, {Cleeves}, {Czekala}, {Huang}, {Le Gal}, {M{\'e}nard},
  {Pegues}, {Qi}, {Walsh}, {Williams}, \& {Wilner}}]{Loomis_2020}
{Loomis}, R.~A., {{\"O}berg}, K.~I., {Andrews}, S.~M., {et~al.} 2020, \apj,
  893, 101

\bibitem[{{Mangum} \& {Shirley}(2015)}]{Mangum_Shirley_2015}
{Mangum}, J.~G., \& {Shirley}, Y.~L. 2015, \pasp, 127, 266

\bibitem[{McCarthy {et~al.}(2020)McCarthy, Lee, Loomis, Burkhardt,
  Shingledecker, Charnley, Cordiner, Herbst, Kalenskii, Willis, Xue, Remijan,
  \& McGuire}]{McCarthy:2020aa}
McCarthy, M.~C., Lee, K. L.~K., Loomis, R.~A., {et~al.} 2020, Nature Astronomy,
  accepted.

\bibitem[{{McGuire}(2018)}]{McGuire_2018_census}
{McGuire}, B.~A. 2018, \apjs, 239, 17

\bibitem[{{McGuire} {et~al.}(2018){McGuire}, {Burkhardt}, {Kalenskii},
  {Shingledecker}, {Remijan}, {Herbst}, \& {McCarthy}}]{McGuire_2018}
{McGuire}, B.~A., {Burkhardt}, A.~M., {Kalenskii}, S., {et~al.} 2018, Science,
  359, 202

\bibitem[{McGuire {et~al.}(2020{\natexlab{a}})McGuire, Burkhardt, Loomis,
  Shingledecker, Lee, Charnley, Cordiner, Herbst, Kalenskii, Momjian, Willis,
  Xue, Remijan, \& McCarthy}]{McGuire:2020bb}
McGuire, B.~A., Burkhardt, A.~M., Loomis, R.~A., {et~al.} 2020{\natexlab{a}},
  Astrophysical Journal Letters, 900, L10.

\bibitem[{McGuire {et~al.}(2020{\natexlab{b}})McGuire, Loomis, Burkhardt, Lee,
  Shingledecker, Charnley, Cordiner, Herbst, Kalenskii, Willis, Xue, Remijan,
  \& McCarthy}]{McGuire:2020aa}
McGuire, B.~A., Loomis, R.~A., Burkhardt, A.~M., {et~al.} 2020{\natexlab{b}},
  Science, in revision.

\bibitem[{North(1963)}]{North_1963}
North, D.~O. 1963, Proceedings of the IEEE, 51, 1016

\bibitem[{{Ohishi} \& {Kaifu}(1998)}]{Ohishi_1998}
{Ohishi}, M., \& {Kaifu}, N. 1998, Faraday Discussions, 109, 205

\bibitem[{Oka(1978)}]{oka:1982}
Oka, T. 1978, Journal of Molecular Spectroscopy, 72, 172

\bibitem[{{Olano} {et~al.}(1988){Olano}, {Walmsley}, \& {Wilson}}]{Olano_1988}
{Olano}, C.~A., {Walmsley}, C.~M., \& {Wilson}, T.~L. 1988, \aap, 196, 194

\bibitem[{{Pickett}(1991)}]{Pickett_1991}
{Pickett}, H.~M. 1991, Journal of Molecular Spectroscopy, 148, 371

\bibitem[{{Portillo} {et~al.}(2017){Portillo}, {Lee}, {Daylan}, \&
  {Finkbeiner}}]{Portillo_2017}
{Portillo}, S. K.~N., {Lee}, B. C.~G., {Daylan}, T., \& {Finkbeiner}, D.~P.
  2017, \aj, 154, 132

\bibitem[{{Remijan} {et~al.}(2005){Remijan}, {Hollis}, {Lovas}, {Plusquellic},
  \& {Jewell}}]{Remijan_2005}
{Remijan}, A.~J., {Hollis}, J.~M., {Lovas}, F.~J., {Plusquellic}, D.~F., \&
  {Jewell}, P.~R. 2005, \apj, 632, 333

\bibitem[{{Remijan} {et~al.}(2006){Remijan}, {Hollis}, {Snyder}, {Jewell}, \&
  {Lovas}}]{Remijan_2006}
{Remijan}, A.~J., {Hollis}, J.~M., {Snyder}, L.~E., {Jewell}, P.~R., \&
  {Lovas}, F.~J. 2006, \apjl, 643, L37

\bibitem[{{Ruaud} {et~al.}(2016){Ruaud}, {Wakelam}, \& {Hersant}}]{Ruaud_2016}
{Ruaud}, M., {Wakelam}, V., \& {Hersant}, F. 2016, \mnras, 459, 3756

\bibitem[{Shingledecker {et~al.}(2020)Shingledecker, Molpeceres, Rivilla,
  Majumdar, \& Kaestner}]{shingledecker_isomers_2020}
Shingledecker, C.~N., Molpeceres, G., Rivilla, V.~M., Majumdar, L., \&
  Kaestner, J. 2020, ApJ, submitted

\bibitem[{{Shingledecker} {et~al.}(2018){Shingledecker}, {Tennis}, {Le Gal}, \&
  {Herbst}}]{Shingledecker_2018}
{Shingledecker}, C.~N., {Tennis}, J., {Le Gal}, R., \& {Herbst}, E. 2018, \apj,
  861, 20

\bibitem[{Shingledecker {et~al.}(2019)Shingledecker, Álvarez Barcia, Korn, \&
  Kästner}]{shingledecker_case_2019}
Shingledecker, C.~N., Álvarez Barcia, S., Korn, V.~H., \& Kästner, J. 2019,
  The Astrophysical Journal, 878, 80

\bibitem[{{Siemiginowska} {et~al.}(2019){Siemiginowska}, {Eadie}, {Czekala},
  {Feigelson}, {Ford}, {Kashyap}, {Kuhn}, {Loredo}, {Ntampaka}, {Stevens},
  {Avelino}, {Borne}, {Budavari}, {Burkhart}, {Cisewski-Kehe}, {Civano},
  {Chilingarian}, {van Dyk}, {Fabbiano}, {Finkbeiner}, {Foreman-Mackey},
  {Freeman}, {Fruscione}, {Goodman}, {Graham}, {Guenther}, {Hakkila},
  {Hernquist}, {Huppenkothen}, {James}, {Law}, {Lazio}, {Lee},
  {L{\'o}pez-Morales}, {Mahabal}, {Mandel}, {Meng}, {Moustakas}, {Muna},
  {Peek}, {Richards}, {Portillo}, {Scargle}, {de Souza}, {Speagle}, {Stassun},
  {Stenning}, {Taylor}, {Tremblay}, {Trimble}, {Yanamand ra-Fisher}, \&
  {Young}}]{Astrostats_2020}
{Siemiginowska}, A., {Eadie}, G., {Czekala}, I., {et~al.} 2019, \baas, 51, 355

\bibitem[{Staff(2020)}]{GBTObsGuide}
Staff, G.~S. 2020, Green Bank Observatory Proposer's Guide for the Green Bank
  Telescope.
\newblock \url{https://www.gb.nrao.edu/scienceDocs/GBTpg.pdf}

\bibitem[{{Toelle} {et~al.}(1981){Toelle}, {Ungerechts}, {Walmsley},
  {Winnewisser}, \& {Churchwell}}]{Tolle_1981}
{Toelle}, F., {Ungerechts}, H., {Walmsley}, C.~M., {Winnewisser}, G., \&
  {Churchwell}, E. 1981, \aap, 95, 143

\bibitem[{Travers {et~al.}(1996)Travers, McCarthy, Kalmus, Gottlieb, \&
  Thaddeus}]{travers:l65}
Travers, M.~J., McCarthy, M.~C., Kalmus, P., Gottlieb, C.~A., \& Thaddeus, P.
  1996, The Astrophysical Journal Letters, 469, L65

\bibitem[{Wakelam {et~al.}(2012)Wakelam, Herbst, Loison, Smith, Chandrasekaran,
  Pavone, Adams, Bacchus-Montabonel, Bergeat, Béroff, Bierbaum, Chabot,
  Dalgarno, van Dishoeck, Faure, Geppert, Gerlich, Galli, Hébrard, Hersant,
  Hickson, Honvault, Klippenstein, Le~Picard, Nyman, Pernot, Schlemmer, Selsis,
  Sims, Talbi, Tennyson, Troe, Wester, \& Wiesenfeld}]{wakelam_kinetic_2012}
Wakelam, V., Herbst, E., Loison, J.-C., {et~al.} 2012, The Astrophysical
  Journal Supplement Series, 199, 21.
\newblock \url{http://adsabs.harvard.edu/abs/2012ApJS..199...21W}

\bibitem[{{Wakelam} {et~al.}(2015){Wakelam}, {Loison}, {Herbst}, {Pavone},
  {Bergeat}, {B{\'e}roff}, {Chabot}, {Faure}, {Galli}, {Geppert}, {Gerlich},
  {Gratier}, {Harada}, {Hickson}, {Honvault}, {Klippenstein}, {Le Picard},
  {Nyman}, {Ruaud}, {Schlemmer}, {Sims}, {Talbi}, {Tennyson}, \&
  {Wester}}]{Wakelam_2014}
{Wakelam}, V., {Loison}, J.~C., {Herbst}, E., {et~al.} 2015, \apjs, 217, 20

\bibitem[{{Walsh} {et~al.}(2016){Walsh}, {Loomis}, {{\"O}berg}, {Kama}, {van 't
  Hoff}, {Millar}, {Aikawa}, {Herbst}, {Widicus Weaver}, \&
  {Nomura}}]{Walsh_2016}
{Walsh}, C., {Loomis}, R.~A., {{\"O}berg}, K.~I., {et~al.} 2016, \apjl, 823,
  L10

\bibitem[{Woodward(1953)}]{Woodward_1953}
Woodward, P. 1953, Probability and Information Theory: With Applications to
  Radar, Electronics and Waves No. v. 3 (Elsevier Science \& Technology).
\newblock \url{https://books.google.com/books?id=RA0nAAAAMAAJ}

\bibitem[{Xue {et~al.}(2020)Xue, Willis, Loomis, Lee, Burkhardt, Shingledecker,
  Charnley, Cordiner, Kalenskii, McCarthy, Herbst, Remijan, \&
  McGuire}]{Xue:2020aa}
Xue, C., Willis, E.~R., Loomis, R.~A., {et~al.} 2020, Astrophysical Journal
  Letters, 900, L9.

\end{thebibliography}
\end{document}